\documentclass[%
reprint,
superscriptaddress,
amsmath,amssymb,
aps,
twocolumn,
prx,
]{revtex4-2}

\usepackage{graphicx}
\usepackage{dcolumn}
\usepackage{bm}
\usepackage{float}
\usepackage{subfigure} 
\usepackage{enumitem}
\usepackage{soul}
\usepackage{xcolor}
\usepackage{hyperref}
\hypersetup{colorlinks=true,citecolor=blue,linkcolor=blue,urlcolor=blue}

\def\XXint#1#2#3{{\setbox0=\hbox{$#1{#2#3}{\int}$}
     \vcenter{\hbox{$#2#3$}}\kern-.5\wd0}}

\usepackage{booktabs} 
\usepackage{array}    

\makeatletter
\def\switch@array{}
\makeatother

\definecolor{myred}{RGB}{205, 75, 60}
\definecolor{mypurple}{RGB}{125, 120, 185}
\definecolor{myblue}{RGB}{75, 135, 205}

\newcommand{\be}{\begin{equation}}
\newcommand{\ee}{\end{equation}}
\newcommand{\bea}{\begin{eqnarray}}
\newcommand{\eea}{\end{eqnarray}}
\newcommand{\bef}{\begin{eqnarray}
                                   \begin{array}{lll}}
\newcommand{\befr}{\begin{eqnarray*}
                                   \begin{array}{lll}}
\newcommand{\eef}{\end{array}
                                    \end{eqnarray}}
\newcommand{\eefr}{\end{array}
                                    \end{eqnarray*}}

\newcommand{\mr}[1]{\mathrm{#1}}

\begin{document}
\title{Tunable Chaos in the Finite Mean SYK Model}
\author{Arkaprava Mukherjee}%
 \email{mukherjee.210@osu.edu}
\affiliation{%
Department of Physics, The Ohio State University, Columbus, OH 43210, USA}%
\author{Sumilan Banerjee}
\email{sumilan@iisc.ac.in}
\affiliation{%
Centre for Condensed Matter Theory, Department of Physics, Indian Institute of Science, Bangalore 560012, India
}
\author{Sandip P. Trivedi}
\email{sandip@theory.tifr.res.in}
\affiliation{%
Department of Theoretical Physics, Tata Institute of Fundamental Research, 
Colaba, Mumbai 400 005, India}
\author{Nandini Trivedi}
\email{trivedi.15@osu.edu}
\affiliation{%
Department of Physics, The Ohio State University, Columbus, OH 43210, USA}%
\date{\today}

\begin{abstract}
The complex Sachdev-Ye-Kitaev (SYK) model, featuring fermions with all-to-all interactions, serves as a dual paradigm for understanding non-Fermi liquid behavior and the holographic nature of charged black holes. Two defining characteristics of the standard SYK model are its maximal chaos (Lyapunov exponent $\lambda_{\mr{L}}=2\pi T$ at temperature $T$), and its finite zero-temperature residual entropy. While previous studies have largely focused on couplings drawn from a zero-mean Gaussian distribution, we investigate a generalized model with a finite mean-to-standard-deviation ratio, $g\equiv J_{0}/\delta J$ of the coupling distribution in order to get deeper insight into the evolution of chaos. We find that increasing $g$ yields the following effects:
(i) The system remains a fast scrambler with $\lambda_{\mr{L}}=A~T$, but with a suppressed coefficient $A<2\pi$.
(ii) In the limit $g\to \infty$, out-of-time-ordered correlators (OTOCs) no longer exhibit
exponential 
growth with $\lambda_{\mr{L}}\simeq 0$.
(iii) The spectral correlations indicative of late-time chaos maintain Wigner-Dyson level spacing 
statistics for all values of $g$.
(iv) The system preserves a finite residual entropy, albeit with reduced magnitude, for all $g$ values. 
We conclude that in this generalized SYK model, there is a chaotic to non-chaotic crossover. Moreover different measures of chaos decouple, demonstrating that the presence of finite residual entropy does not strictly imply maximal chaos.

\end{abstract}

\maketitle

\section{\label{sec:SYK_AB}Introduction}

In recent years, the Sachdev-Ye-Kitaev (SYK) model~\cite{kitaev2015simple,Maldacena:2016hyu,PhysRevX.5.041025,SY1992} and its extensions \cite{RevModPhys.94.035004,Gu2017,Banerjee:2016ncu,Jian2017,Song2017,Davison2017,Zhang2017,Chowdhury2018,Haldar2018,Haldar2018a,Jian2018,Esterlis2019,Kim2021,Patel2023,PhysRevX.9.021043} have attracted considerable attention as solvable toy models to describe various strongly correlated states, such as strange metals ~\cite{
RevModPhys.94.035004} in condensed matter physics, as well as to understand black holes in quantum gravity ~\cite{Maldacena:2015waa,Maldacena:2016upp,Maldacena:2016hyu,Gu:2019jub,Nayak:2018qej,Moitra:2018jqs,Sachdev:2019bjn,Moitra:2019bub}. The SYK model is a $(0+1)$ dimensional model of fermions with $N$ sites or \emph{flavors} having infinite-range or all-to-all random four fermion interactions that are drawn from a real or complex Gaussian distribution with zero mean and finite standard deviation. The model has several interesting and unusual properties. It has a non-Fermi liquid (NFL) ground state that lacks a Landau quasiparticle description ~\cite{kitaev2015simple,Maldacena:2016hyu,PhysRevX.5.041025,RevModPhys.94.035004,PhysRevB.94.035135,PhysRevB.103.075141
}, and exhibits a non-zero residual entropy at zero temperature in the large-$N$ or thermodynamic ($N\to\infty$) limit. Also, and of special interest for this paper,  is the fact that the model is maximally chaotic or the fastest scrambler, saturating the Maldacena-Shenker-Stanford (MSS) bound with a Lyapunov exponent $\lambda_{\mr{L}}=2\pi T$ ($k_B=1$, $\hbar=1$) for temperature $T\to 0$. 

The similarities  this model exhibits with the  behaviour of near-extremal black holes are remarkable and make it  of great interest in the study of quantum gravity. In particular, the correspondence between quantum gravity and condensed matter systems opens the possibility to connect
with experiments in the near-future allowing us to study and test  non-trivial properties of black hole and wormhole physics in the laboratory. The prospect of  experiments leading  to further insights  and  cross-fertilisation between  the fields of quantum gravity, quantum information theory,  and condensed matter physics is a  truly exciting one!

Before we proceed, it is worth discussing, at least briefly, several  interesting and promising realizations of the model that have already been proposed in both solid state and cold atomic systems ~\cite{PhysRevB.96.121119,PhysRevX.7.031006,Franz:2018cqi,PhysRevLett.121.036403,Danshita:2016xbo}.
In these platforms, the random SYK interactions originate from the randomness of the degenerate single-particle wavefunctions. E.g., it has been shown that the irregular boundary of a graphene flake \cite{PhysRevLett.121.036403,PhysRevResearch.2.013254, PhysRevLett.131.036503} can give rise to a distribution of four-fermion couplings closely resembling a Gaussian with zero mean.
Given such experimental realizations, it is  a natural  to ask whether the random nature of the couplings can be changed, by altering various parameters in the experimental realisations,  so that   the four fermi couplings remains  Gaussian,  but  with a non-zero mean. And whether the striking features of the SYK model, such as the maximal chaos and $T=0$ non-zero residual entropy, survive in the presence of a finite mean?


\begin{figure*}[t]
\renewcommand{\arraystretch}{1.5}
        \begin{tabular}{@{} >{\bfseries}p{3.42cm} | p{4.6cm} | p{4.8cm} | p{4.6cm} @{}}
        \hline       
        Property & 
        \centering{$\displaystyle \boldsymbol{g=0}$} & 
        \centering{$\boldsymbol{g}>0$ (within some range)
        } & 
        $\boldsymbol{\quad\quad\quad\quad\quad g \to \infty}$\\
        \hline
         Lyapunov Exponent & 
       Fastest Scrambler: $\displaystyle \lambda_{\mr{L}} = 2\pi T$ & 
       Fast Scrambler:  $\displaystyle \lambda_{\mr{L}}= A T$ \newline  \normalsize{($A < 2\pi$, decreases with increasing $g$)} & 
       No Scrambling: $\displaystyle \lambda_{\mr{L}}= 0$ \\

        Spectral Form Factor & 
        Linear ramp with slope $=1$ (large-N result)& 
        Shorter ramp with slope $<1$  & 
        Vanishing ramp \\

       Level repulsion & 
       RMT type for each $\mathcal{Q}$ & 
       RMT type for each $\mathcal{Q}$ & 
       RMT type for each $\mathcal{Q}$  \\[0.2em]

        Residual Entropy $\displaystyle \frac{S_0}{N}$ & 
         $S_0/N = 0.47$ (= large$-N$ result)
        & 
$S_0/N < 0.47$, decreases with increasing $g$ & 
$S_0/N = 0.38$
\\[0.3em]

        Entanglement \newline Entropy & 
       Follows volume law and Page curve $S_{m}\sim \text{$m\log(2)$}$ & 
       Deviates from volume law \newline $S_{m}<\text{$m\log(2)$}$ & 
       Significant deviation \newline $S_{m}\ll\text{$m\log(2)$}$ \\
        \hline
        \end{tabular}
\caption{Evolution of chaotic properties 
for the generalized Sachdev-Ye-Kitaev (SYK) model driven by tuning the ratio mean-to-standard deviation $g = J_0 / \delta J$. At $g=0$, the system reduces to the standard SYK model, exhibiting  chaotic behavior across all conventional diagnostics, including maximal scrambling with Lyapunov exponent $\lambda_\mathrm{L}=2\pi T$ at low temperatures. In the opposite limit as $g \to \infty$ ($J_0=1, \delta J=0$), the couplings become completely uniform. Here the early-time diagnostic of chaos, the Lyapunov exponent, vanishes in the absence of randomness at low temperatures. For  finite values of $g$, within some range,  the system remains a fast scrambler with $\lambda_\mr{L}=AT$ and $A<2\pi$. Similarly, an intermediate-time chaos indicator, the spectral form factor, exhibits suppression of linear ramp with increasing $g$ at low temperatures, even indicating a vanishing of the ramp at finite $g$ for finite $N$. The linear ramp is characteristic of random-matrix theory (RMT) and hence its disappearance at finite $g$ suggests a MBL-like transition \cite{Altshuler1997,Micklitz2019} at finite $N$ and low temperature. In contrast, the late-time chaotic signature, level repulsion, persists for mid-spectrum states for all $g$ in all  fermion-number sectors. Notably, a finite residual entropy also persists across the entire range of $g$, even when the system ceases to be a  fast scrambler. The ground-state entanglement entropy ($\propto m$) for a subsystem with $m<N/2$ sites (out of $N$) sites also decreases with $g$ staring from close to the Page value $\sim m\log(2)$ at $g=0$.
}
\label{PHASE}
\end{figure*}

With this motivation, we generalize the standard SYK model in which the couplings are random variables, as was mentioned above, drawn from a Gaussian distribution,  with vanishing mean values. Here we study a variation of the SYK model, in which the complex couplings are drawn from a Gaussian distribution with both a finite mean $J_0$ and a finite standard deviation $\delta J$, defining the ratio $g=J_0/\delta J$. Note that  $g=0$   is the conventional SYK model, on the other hand when $g\rightarrow \infty$, the  randomness in the couplings disappears. We show that the SYK model with complex mean undergoes a chaotic-to-non-chaotic crossover in terms of scrambling or Lyapunov chaos as a function of $g$ at low temperatures, while retaining a finite residual entropy from $g=0$ to $g\to \infty$ limit. However, the model exhibits Wigner-Dyson random-matrix level spacing statistics for mid-spectrum states for the entire range of $g$. In the Discussion section we compare our results on the effect on chaos of tuning $g$ to previous works that added additional two-body interactions to the SYK model.  
\par

\par
One of our main motivations is to study how the chaotic properties of the model change as the  couplings go from being random to  ordered.  By adding a non-vanishing mean to the coupling distribution, we tune the model without altering the $N$-scaling of the model or introducing any chemical potential or two fermion interaction terms. From the gravity perspective one would like to know whether systems with more ordered couplings continue to behave like gravitational ones? Intuitively, one might expect that the system becomes less chaotic when the couplings become more ordered.  We are interested in asking whether this  expectation is indeed borne out  and in understanding precisely what  changes occur in the behaviour of the system.  In particular, we would like to know if   there are continuous changes in the system's properties, or if  these changes  are abrupt, suggesting a phase transition, as $g$ varies? 

 \par
To address these questions we  study  the chaotic nature  in  three different ways, corresponding to  three distinct time regimes: (1) The statistics of nearest-neighbour energy eigenstates, which corresponds to the late-time regime. (2) The spectral form factor (SFF), which corresponds to an intermediate-time regime. (3) The Out of Time Correlators (OTOCs) that probe the early-time regime. 

Some of our key findings are as follows (summarized in fig. \ref{PHASE}): 
We find that the mid-spectrum\footnote{For a precise definition of mid-spectrum  states see section \ref{sec:level2}.} nearest-neighbour many-body energy level spacings  continue to follow the Wigner-Dyson distribution, as $g$ changes. In contrast,  the  ramp region of the SFF at low temperature becomes increasingly smaller, as $g$ increases,  suggesting that the eigenvalue repulsion becomes stronger for eigenvalues which are somewhat apart.  The OTOCs, we find,  also show   evidence that the system becomes more ordered and less chaotic,  with increasing $g$. Further, as $g$ increases, starting from zero, the Lyapunov exponent decreases from its maximal value, so that the system  is still a fast scrambler, but not maximally chaotic,  with $\lambda_L\propto T<2\pi T$, at low temperature. This fast scrambling nature persists over a finite range of $g$, eventually crossing over to a non-chaotic phase for larger values of $g$  where the Lyapunov growth regime can no longer be detected. Finally, subject to the limitations of our analysis, we find that these changes- away from chaos towards greater order with increasing  $g$ are  continuous, and not marked by abrupt changes. 

In addition to properties that characterize chaos, we also investigate several other characteristics. We find, by taking  ED results for the entropy at  finite $N$ and then extrapolating to $N\to\infty$, that a non-zero  residual entropy in the ground state  is present when $g\ne 0$. While this entropy decreases with increasing $g$, we find, surprisingly,  that it is non-zero even for the fully ordered system with  $g\to\infty$.  We  also calculate  the bipartite entanglement in the ground state,  and find that it decreases with increasing $g$.

Before closing this section, let us make one final comment. As far as we can tell, the model we study here, with a complex mean,  does not admit an analytic solution  in the $N\to \infty$ limit. In contrast, when the mean takes a real value, we have been able to solve the model analytically for $N\to  \infty$. \cite{MukherjeeFuture}. In the analysis below we will use exact diagonalisation (ED) to study how the  behaviour of the model changes as $g$ is varied. In Appendix \ref{app:FINN} we discuss the size limitations, and the use of charge conservation to extend the simulations to larger sizes.

\section {Summary of results}

The paper is organized as follows:
In Section \ref{sec:level2}, we begin with a basic description of the model, including a brief introduction to its symmetry properties. We then discuss how the density of states changes   when a finite mean is introduced in the Gaussian distribution of random couplings. This includes  the extreme case of $g\to\infty$ where the randomness vanishes, due to the  standard deviation going to zero, at  a finite value of the mean. 

In section \ref{CHAOS} we turn to examining the chaotic properties of the model. We being by reviewing some general characteristics of chaos pertaining to three distinct time regimes. With this background in hand, we turn to examining  the chaotic behaviour of our    model  as a function of the  coupling $g$.   



In the late-time regime, chaos is primarily governed by nearest-neighbor level repulsion. Depending on the symmetries of the Hamiltonian, we find that the level-spacing distribution aligns with the predictions from Gaussian ensembles for all values of $g$. We elaborate on this in Section~\ref{LEV_STAT}.

In Section~\ref{SFF}, we analyze the intermediate-time behavior using the Spectral Form Factor. This analysis captures various aspects of level repulsion and nearest-neighbor level spacing, along with residual entropy at zero temperature. Here, we observe a controlled suppression of chaotic behavior with increasing mean.

In the early-time regime, chaos is diagnosed using the Out-of-Time-Order Correlator (OTOCs) (Sec.~\ref{OTOC}). From the early growth of the OTOCs, we extract the Lyapunov exponent and study its temperature dependence to assess the system's fast scrambling behavior. As $g$ is increased, within some range, we find that the system retains its fast scrambling nature  but   the value of the Lyapunov exponent decreases.

In section~\ref{RES_ENT} we turn our attention to the  zero-temperature entropy of the model as $g$ is varied.   From our  analysis, we find that the residual entropy also decreases with increasing mean. To understand this trend, we study the density of low-energy states and the corresponding level spacing near the ground state, which reveals an exponential scaling with $N$, consistent with  the presence of non-zero residual entropy.

In Section~\ref{GS_EE}, we study how the ground-state entanglement entropy changes with finite mean. We observe a consistent reduction in entanglement entropy as the mean increases. 

We end in section \ref{discupaper} with some  discussion and comments on the implications of our results. 

In Appendix \ref{SYMM_LEVEL}, we discuss the symmetries of the different particle number sectors of the model and their corresponding random matrix classifications.Appendix \ref{app:ENTRP_EXTRA} discusses an alternative analysis of the residual entropy, derived from both the disorder-averaged mean many-body level spacing and the total number of states within an energy window near the ground state.
Appendix \ref{app:ENTRP_EXTRA} discusses an alternative analysis of the residual entropy, derived from both the disorder-averaged mean many-body level spacing and the total number of states within an energy window near the ground state.
In Appendix \ref{app:SpHeat}, we discuss the temperature dependence of the specific heat obtained from ED in the finite-mean SYK model at low temperature.
Appendix \ref{app:FINN} contains details regarding our numerical simulation methods, along with multiple benchmarks against previous results discussed throughout the paper.




\section{\label{sec:level2}Complex SYK with finite mean}
The Sachdev-Ye-Kitaev model~\cite{PhysRevB.94.035135} describes a system of interacting fermions in 0+1 dimensions with all-to-all random coupling:
\bef
\displaystyle
\mathcal{H}=\frac{1}{(2N)^{3/2}}\sum_{i,j,k,l=1}^NJ_{ijkl}c^{\dagger}_{i}c^{\dagger}_{j}c_{k}c_{l}-\mu \sum c^{\dagger}_{i}c_{i}
\label{SIM_HAM}
\eef
with the conditions that
\bef
\displaystyle 
J_{ijkl}=-J_{jikl}=-J_{ijlk}~~~~~\& ~~~~~J_{ijkl}=J^{*}_{klij} \   \cdot
\label{Conditions_SYK}
\eef
Note that the sum in eq.(\ref{SIM_HAM}) is taken with all four indices, $i,j,k,l$ ranging from $1$ to $N$.

Using the conditions of eq.(\ref{Conditions_SYK}) we can take the independent components in $J_{ijkl}$  to be the ones with:
 $i<j$,  $k<l$ and $i\le k$. 
When $i=k$ we specify  the $j\le l$ component. 
For $i=k, j=l$,  $J_{ijkl}$ must be real, with no imaginary parts. 
The other components of $J_{ijkl}$ can then be obtained using the relations in eq.(\ref{Conditions_SYK}).

In general the couplings are complex. We denote the real and imaginary parts by 
\be\label{reim}
\displaystyle
J_{ijkl}=Re(J_{ijkl}) + i  Im(J_{ijkl})
\ee
In the complex SYK model one takes $Re(J_{ijkl})$, $Im(J_{ijkl})$,  for the independent couplings as  specified above, to be drawn from independent  Gaussian random distribution with a standard deviation $\delta J$ and with no mean value.  
It is worth being  explicit about this, especially since we will be adding a mean value below.

We take $Re(J_{ijkl})$ to be drawn from the distribution
\bea
\displaystyle 
\langle Re(J_{ijkl})Re(J_{i'j'k'l'})\rangle  =& & \delta J^2[\delta_{i,i'}\delta_{j,j'}\delta_{k,k'}\delta_{l,l'}] \\ \nonumber
\forall &&(i<j, k<l, i\le k); \\ \nonumber 
&&  (i'<j', k'<l', i'\le k')
\label{recoupling}
\eea
And similarly for  the imaginary parts 
\bea
\label{imcoupling}
\displaystyle
\langle Im(J_{ijkl})Im(J_{i'j'k'l'})\rangle=&& \delta J^2(\delta_{i,i'}\delta_{j,j'}\delta_{k,k'}\delta_{l,l'}) \\ \nonumber
\forall && (i<j; k<l; i\le k); \\ \nonumber 
&&  (i'<j', k'<l', i'\le k')
\eea
The only exception is when $i=k; j=l$, and $i'=k'; j'=l'$,  in which case the imaginary part vanishes as noted above, and we take the real part to be drawn from the  distribution, eq.(\ref{recoupling}).

To have a Hamiltonian which has nicer properties under particle-hole exchange we now add an additional term, following ~\cite{PhysRevB.94.035135}, so that the full Hamiltonian becomes,
\bef
\displaystyle
\mathcal{H}=\frac{1}{(2N)^{3/2}}\sum_{i,j,k,l=1}^NJ_{ijkl}[c^{\dagger}_{i}c^{\dagger}_{j}c_{k}c_{l} ]+{\mathcal \delta H} -\mu \sum c^{\dagger}_{i}c_{i}~~~~
\label{finalham}
\eef
where the additional term is
\bef
\displaystyle
\mathcal{\delta H} = 2 \frac{1}{(2N)^{3/2}}\sum_{i,j,k,l=1}^N J_{ijkl} [\delta_{i k} c_j^{\dagger} c_l]
\eef

Under the action of the  anti-unitary operator $S$ which exchanges particles and holes
\bef
\displaystyle
Sc_{i}S^{-1}=c^{\dagger}_{i},~~~~~~~Sc^{\dagger}_{i}S^{-1}=c_{i},~~~~J_{ijkl}\rightarrow J^{*}_{ijkl}\   \cdot
\label{SYM_CASE}
\eef
the Hamiltonian transforms as ${\cal H}\rightarrow S {\cal H}S^{-1}$. It is easy to see that the first two terms in  eq.(\ref{finalham}) are now invariant while the last term, proportional to the chemical potential, transforms as 
\be\label{lasterm}
\displaystyle
S (\mu \sum_i c_i^\dagger c_i) S^{-1}=\mu (N-\sum_i c_i^\dagger c_i)
\ee
At  the particle-hole symmetric point where the total fermion number $\displaystyle \sum_{i} c_i^\dagger c_i={N\over 2}$,  ${\cal H}$ is invariant. 

Next, we define the charge operator
\bef
\displaystyle
\mathcal{Q}=\sum_{i}c^{\dagger}_{i}c_{i}-\frac{N}{2},
\label{charge_op}
\eef
so that $\mathcal{Q}$ takes values from $-{N\over 2} $ to ${N\over 2}$, with the particle hole symmetric case corresponding to  $\mathcal{Q}=0$.
It is easy to see that $[{\mathcal Q},H]=0$, and that ${\mathcal Q}$ anti-commutes with the $S$ operator. 
As a result,  the eigenstates of energy come in pairs, with two eigenstates having  the same energy but opposite charge. 
Note that the Hamiltonian will be block diagonal in sectors of fixed charge, with charge sectors of  opposite value being connected by the operator $S$. 

We also define the Parity operator to be 
\be\label{defparity}
\displaystyle
P=(-1)^{{\cal Q}+{N\over 2}}
\ee

In our study of chaotic properties  we will find the transformation properties of different energy eigen sectors under $S$, ${\cal Q}$ and $P$ to be useful in  determining the relevant class of  random matrix theories.
\\

\medskip

\noindent{\textcolor{blue}{\it Introducing a Mean}}:
In this paper we study an altered  SYK model, where the random couplings are Gaussian distributed with a non-zero mean as well as a variance. 
We take the same mean value, for all the couplings (up to symmetry relations discussed above). More precisely, for the independent components, as specified above eq.(\ref{recoupling}), we take, 
\bef
\displaystyle
\langle J_{ijkl}\rangle=J_0 (1+ i) , \ \forall i<j, k<l, i<k
\label{meangn}
\eef
so that the real and imaginary parts have the same magnitude, 
and 
\be\label{meansp}
\displaystyle
\langle J_{ijkl}\rangle=J_0 \ \forall i<j, k<l, i=k, j=l
\ee

In our analysis, unless otherwise stated,  it will  be convenient  to set the standard deviation $\delta J=1$ to be unity and to measure energies and time scales in units of $\delta J$. The mean value $J_0$ which then enters the two-point correlator of couplings will in fact  denote the ratio 
\begin{equation}
    \label{defg}
    g={J_0/ \delta J}
\end{equation}
In our analysis we   will study the behavior of the altered SYK model as a function of $g$ and $N$.

Before proceeding let us mention one interesting extreme limit of the model, which we call the ``Clean SYK Model" where  the 
standard deviation, $\delta J$,  vanishes, with the mean value, $J_0$, being non-vanishing, so that $g\rightarrow \infty$. 
In this limit the randomness vanishes and the couplings take fixed values.  

Let us make three comments before we proceed. 
First, while we will not do so here, we note that more general   mean values  can also have been  considered. For example, 
 the magnitude of the real and imaginary mean values need  not be equal. Also, we have considered the Gaussian distribution for all the four -fermion couplings to be identical, more generally these could have been taken to be different. . 

Second, note that with $\mu=0$, upon  exchanging particles with  holes under the transformation,  $c_i\leftrightarrow c_i^\dagger$, we get  a Hamiltonian in which   the mean value, $\langle J_{ijkl}\rangle $,  takes its complex conjugate value,  
$\langle J_{ijkl}\rangle\rightarrow \langle J_{ijkl}\rangle^{*}$.
This follows from the fact that the antiunitary transformation $S$ is a symmetry when $\mu=0$. 
Note that  the charge ${\cal Q}$ sector in the original description is mapped to  the $-{\cal Q}$ sector under this exchange. As a result, the density of states, at energy $E$, $\rho(E)$, for the $\mu=0$ case, which is obtained by summing over all ${\cal Q}$ sectors, is unchanged under particle-hole transformation. In fact, the density of states at fixed energy $E$ and charge ${\cal Q}$, $\rho(E,{\cal Q})$ will also remain unchanged under the transformation, of the mean value to its complex conjugate,  since, as noted above, the energy eigenstates are paired with the partners  having equal   and  opposite values of ${\cal Q}$.

Third, it is difficult to solve the model analytically in general once a non-zero mean is introduced. We will therefore investigate its behaviour numerically below, using the method of exact diagonalisation (ED). 
ED  is difficult to carry out  for  large values of $N$ since the dimensionality of the Hilbert space grows like $2^N$. 
However since the Hamiltonian conserves the charge ${\cal Q}$, we can increase the range of $N$ that is handled, by working in fixed charge sectors, and then adding the results from all the charge sectors together. 

Two more points are worth mentioning before  we proceed. 
First, in some limits the SYK model, with a mean, can be solved exactly. These limits include the clean case with $J_0$ being real. 
A more detailed analysis of such limits, with the resulting changes in chaotic behaviour, will be presented in a subsequent paper, \cite{MukherjeeFuture}
.  Second, most of the analysis of chaos below will be carried out by setting $\mu=0$. It will also be worth studying how the chaotic behavior changes as $\mu$ is varied. 


\medskip

\begin{figure}[h] 
  \setlength{\unitlength}{1pt}
  \begin{picture}(\linewidth, 380)
    \put(0,0){\includegraphics[width=\columnwidth]{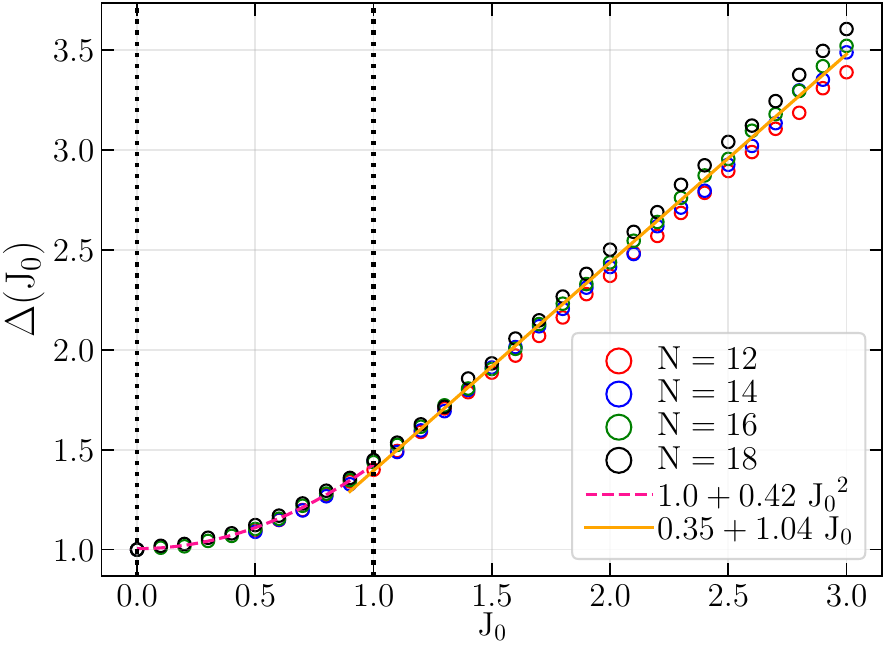}}
    \put(50,155){\large\textcolor{black}{\textbf{(b)}}}
    \put(9,190){\includegraphics[width=0.98\columnwidth]{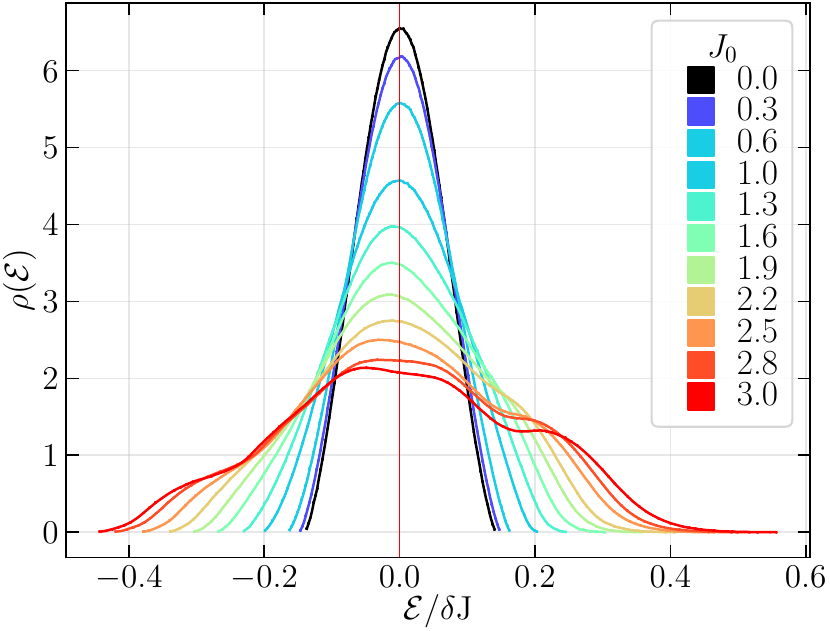}}
    \put(50,345){\large\textcolor{black}{\textbf{(a)}}}

  \end{picture}
\caption{a) Density of states ${\rho}({\cal E}) $  for  $N=18$, and $\delta J=1$, for  changing mean values $J_0$ (different colors). (b) Scaling of  the energy spread $\Delta$ as a function of $J_0$ for different $N$ (indicated by different colors). For $J_0 < 1$, a quadratic fit (dashed line) and for  $1 < J_0 < 3$, a linear fit (solid line) have been used.}
  
\label{SCALING_SYK}
\vspace*{-0.20in}
\end{figure}

\noindent{\textcolor{blue}{\it Density of States}}: The density of states for varying  values of  $J_0$,  with $N$ taking values from $12$ to $18$, and with $\delta J=1$,  are plotted in fig. \ref{SCALING_SYK} (a). 
The spread in energy, or the many-body energy bandwidth, $(E_{max}-E_{min})$,
scales linearly in $N$, it is therefore useful to rescale the energy, 
${\cal E}={E\over N}$.
The normalized density of states $\rho({\cal E})$ for various finite values is shown in fig. \ref{SCALING_SYK} and satisfy
\bef\label{conda}
\displaystyle
\int d{\cal E} {\rho}({\cal E}) =1 \  \cdot
\eef
As $J_0$  increases, we see from fig. \ref{SCALING_SYK}(a) that the spread in  rescaled energy ${\cal E}$ increases. In order to quantify the spread, we define $\Delta(J_0)$ 
\bef
\displaystyle
\label{defratio}
\Delta(J_0)={E_{max}(J_0)-E_{min}(J_0)\over E_{max}(J_0=0)-E_{min}(J_0=0)}
\label{DELTAJ}
\eef
the ratio of the spread in energies for mean value $J_0$ normalised by the spread when $J_0=0$. Here $E_{max}(J_0)$ and $E_{min}(J_0)$ are the maximum and minimum energies respectively for a fixed value of $N$.
From fig. \ref{SCALING_SYK}(b) we see that for the range of $N$ considered here, $\Delta(J_0)$ is well fit by the functional form 
\bef
\displaystyle
\Delta(J_0)=\begin{cases}
1+0.42 J_0^{2};~~J_0 < 1
\\
0.35+J_0~~;~~J_0\geq1
\end{cases}
\label{SCALE_FN}
\eef
so that $\Delta(J_0)$ varies quadratically, at small mean values,  for $J_0<1$,  and linearly for larger values of $J_0$. There is also a weak but observable dependence of $\Delta$ on $N$.
\\

\begin{figure*}
\hspace{-2.2cm}
\begin{picture}(0.9\linewidth,0.28\linewidth)
    \put(0,0){\includegraphics[width=0.33\linewidth]{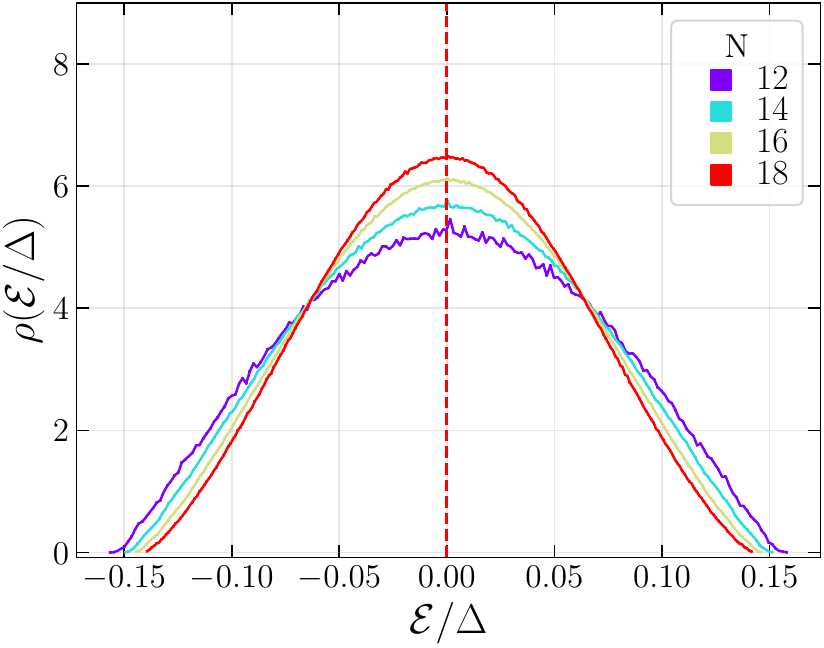}}
    \put(20,120){\normalsize{\textcolor{black}{$(a)$}}}
    
    \put(177,0){\includegraphics[width=0.34\linewidth]{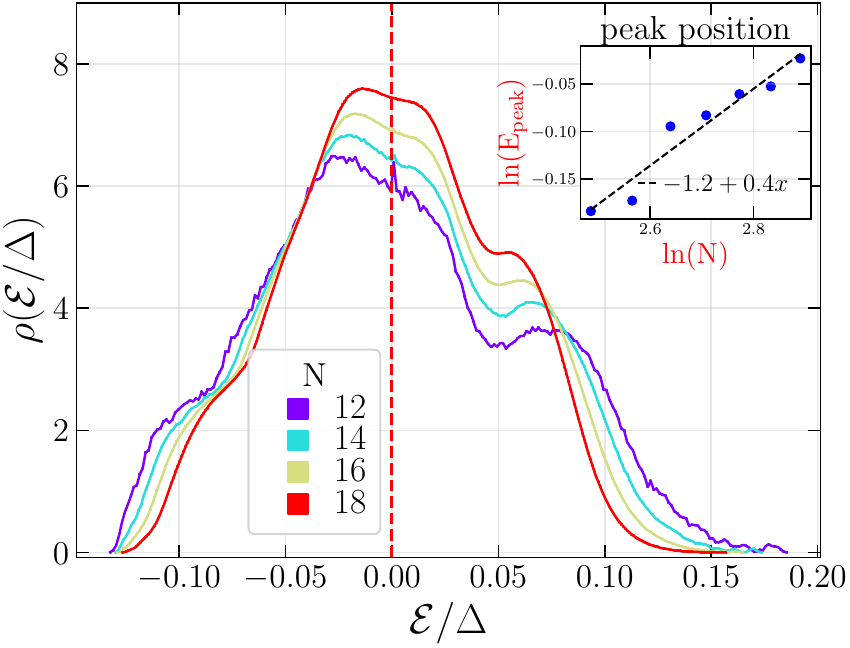}}
    \put(197,120){\normalsize{\textcolor{black}{$(b)$}}}
    
{
        \linethickness{1.1pt}
        \color{black}
        \multiput(283,95)(0,-9){5}{\line(0,-3){6}}
                \put(283,50){\vector(0,-3){2.5}} 
    }
    \thinlines 
    \put(347,0){\includegraphics[width=0.33\linewidth]{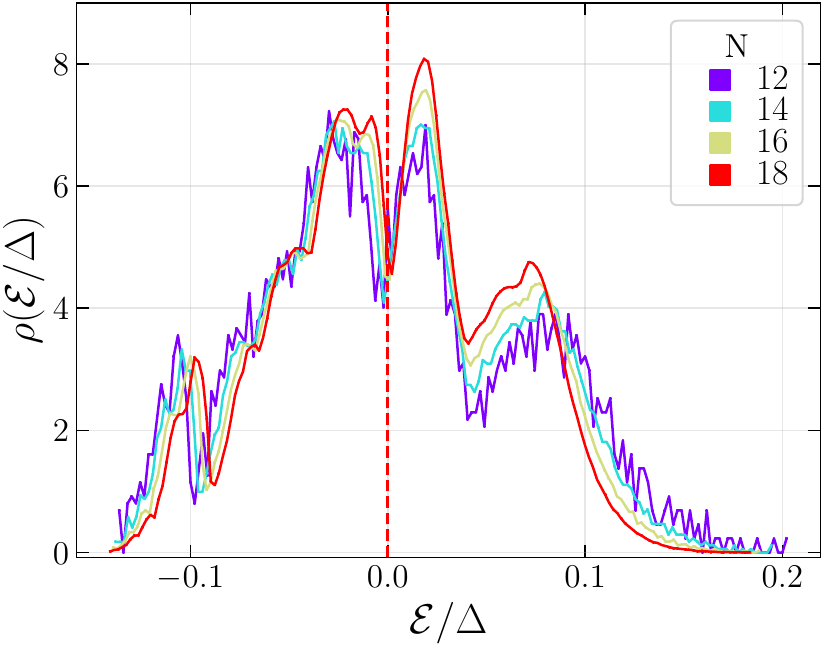}}
    \put(367,120){\normalsize{\textcolor{black}{$(c)$}}}
\end{picture}
\caption{Many-body density of states for different $N$ for a) $J_0=0, J=1$, b)  $J_0=3, J=1$, c) $J_0=1, J=0$.  Inset figure in b) shows the position of the positive peak energy as a function of $N$  in a \rm{log}-\rm{log} Plot.}
\label{DOS_PLOT_0}
\end{figure*}
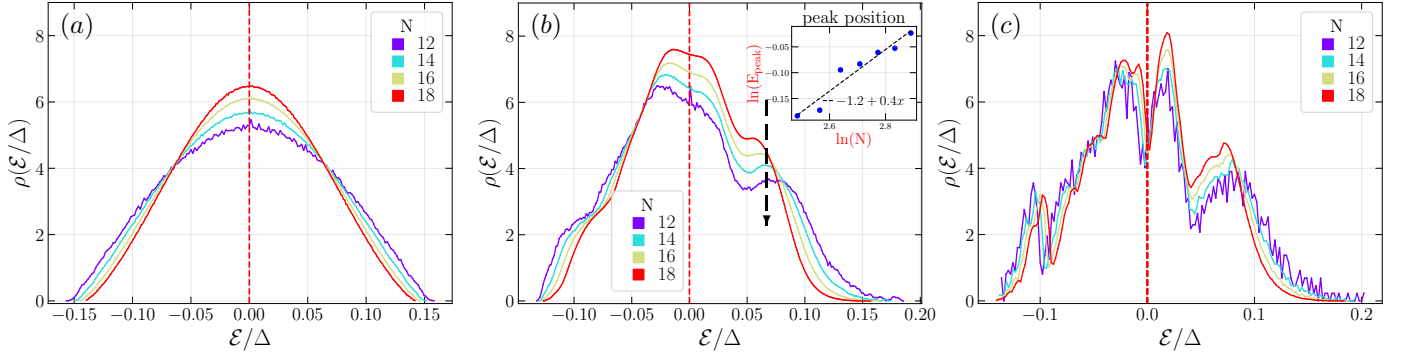

In fig:-\ref{DOS_PLOT_0} we consider the effect of varying mean values on the density of states by studying three cases: (a) $J_0=0, \delta  J=1$ ; (b) $J_0=3,\delta  J=1$; and (c) $J_0=1+\iota, \delta J=0$. The last case corresponds to the Clean SYK model referred to  above. 
From fig. \ref{DOS_PLOT_0}(a) we see that a weak dependence on $N$ in the spread of energies continues to persist, even after rescaling the energy with $N$. 

In fig. \ref{DOS_PLOT_0}(b) and \ref{DOS_PLOT_0}(c)  it is useful to consider rescaling the energy also by $\Delta$, the spread, eq.(\ref{defratio}), when   comparing the different cases. More precisely, for the cases (a), (b) above, where $\delta J=1$, $\Delta$ continues to be defined as given in eq.(\ref{defratio}). However for case (c) where $\delta J=0$, for uniformity,  we define
\be
\displaystyle
\label{defdeltac}
\Delta={E_{max}(J_0, \delta J=0)-E_{min}(J_0, \delta  J=0)\over E_{max}(J_{0}=0,\delta  J=1)-E_{min}(J_{0}=0,
\delta  J=1)}
\ee
with  the denominator being the spread $E_{max}-E_{min}$ for the case with a vanishing mean and  standard deviation  $\delta J=1$.

The density of states in fig. \ref{DOS_PLOT_0}(a) is symmetric under the exchange 
${\cal E}\rightarrow -{\cal E}$, with a central peak at ${\cal E}=0$. We see from fig. \ref{DOS_PLOT_0}(b) that once a mean value is introduced this symmetry  is no longer present, and additional peaks appear in the density of states ${\rho}({\cal E})$. In particular, there is a pronounced peak at a positive value of  ${\cal E}$ for positive $Re(J_0)$. We will refer to this peak as the ``positive peak" in our discussion below. 
Note that, as mentioned above, under the transformation, of the mean value to its complex conjugate
, $Re(J_{ijkl})$ remains unchanged, so $\rho(E)$ is unchanged.
Also under $\langle J_{ijkl}\rangle\rightarrow -\langle J_{ijkl}\rangle$, the density of states $\rho({\cal E})\rightarrow \rho(-{\cal E })$. 
These facts show that, more generally,  whether the pronounced  peak is in fact  at positive or negative ${\cal E}$,  will be determined by the sign of $Re(\langle J_{ijkl}\rangle)$.

From the inset \rm{log}-\rm{log} plot in fig. \ref{DOS_PLOT_0}(b) we see that the location of the positive peak scales like $E\propto N^{0.4}\sim \sqrt{N}$. We note that the positive peak appears within the otherwise \emph{negative temperature $T(\mathcal{E})$} regime, i.e., for $\mathcal{E}\gtrsim 0$ where $1/T(\mathcal{E})=\partial \log \rho(\mathcal{E})/\partial \mathcal{E}<0$, and leads to a range of energy $\mathcal{E}$ with a positive temperature for $\mathcal{E}\gtrsim 0$.
In addition to the positive peak there is a peak near ${\cal E}=0$, now slightly shifted to negative ${\cal E}$. 

These features in the spectrum persists for the range of $N$ we have analysed.

Finally in fig. \ref{DOS_PLOT_0}(c) we consider the Clean model with a fixed coupling. Since there is no averaging to be done on the couplings, the data looks ``more noisy" in this case, but the noise goes down as $N$ increases. We see that the central peak in this case gets split into two. 
There is, in addition,  a pronounced  peak for both positive and negative energies.


In summary, the study of the density of states shows interesting changes in the system, as one moves from randomness to order with increasing $g$. 
We turn to analysing the chaotic behaviour as a function of $g$ next. 

\section{Chaos}\label{CHAOS}
In this section  we  discuss the chaotic behaviour  of the  system. We will consider three  different  diagnostics, which probe  different  time scales of the chaotic dynamics. We begin with a general discussion of these three diagnostics before turning to analysing our system. 

\medskip

\noindent{\textcolor{blue}{\it {I) The nearest-neighbour eigenvalue spacing:}} The spacing of nearest neighbour energy levels is  a well known  characteristic of   chaos. It is widely believed, according to the Bohigas-Giannoni-Schmit (BGS) conjecture \cite{Bohigas1984},  that the statistics governing these spacings should exhibit universal features. In fact,  depending on the symmetry properties of the Hamiltonian, the statistics  should     match  that of an      appropriately  identified random matrix theory \cite{PhysRevD.101.066017,you2017sachdev,PhysRevX.12.021040}. 

In general, the spacing of energy levels, $\Delta E$, should be inversely proportional  to the time scale involved in the dynamics $t$,  
\be\label{relte}
\displaystyle
\Delta E\propto {1\over t}
\ee

  The nearest neighbour spacings should therefore correspond to the very long time dynamical properties of the system, as we will see shortly.  
\medskip

\noindent{\textcolor{blue}{\it {II) The spectral form factor (SFF):}}
This is defined,  as a function of time $t$,  to be 
\bef
\displaystyle
\text{SFF}(t, \beta) = \frac{\langle Z(\beta + i t) Z(\beta - i t) \rangle}{\langle Z(\beta) \rangle^2}.
\label{SFF_EQ}
\eef

Here $\beta$ is the inverse temperature, $Z(\beta +it)=\mathrm{Tr}(e^{-(\beta +it) H})$ is the partition function at complex temperature, and the angle brackets in $\langle Z(\beta+it)Z(\beta-it)\rangle$ denote   an ensemble average over the random  couplings.

For chaotic systems (including random matrix ensembles),  $\text{SFF}(t)$ exhibits three distinct behaviors  as a function of $t$:
\begin{enumerate}[label=(\roman*), itemsep=0pt, parsep=0pt]
\item a rapid  decay initially for $t<t_{th},$ where $t_{th}$ is the Thouless time, 
\item a linear ``ramp'' region at  intermediate times, $t_{th}<t <t_{H}$, in which it grows, 
\item a  ``plateau'' region at  long times, $t>t_{H}$,  where it attains a constant value.
\end{enumerate}
The time scale $t_H$   which  marks the onset of the ``plateau''  region is called the Heisenberg time, $t_H$. Since it is  of order  the longest time scale,  $t_H$ should be  determined by the spacing of the nearest neighbour eigenvalues. At low temperatures, 
more precisely, $t_{H}$ would be determined by  the spacing of energy levels among  the low lying states.  In contrast, the    ramp region, which corresponds to   intermediate time scales, should  probe the spectral characteristics of energy levels at intermediate spacing. 

The expectation that $t_H$ is determined by the spacing of nearest neighbour energy levels is indeed borne out for both RMTs and also the Majorana SYK model.
For a Random Matrix theory, with  $\mathcal{N}_\mathrm{H}\times \mathcal{N}_\mathrm{H}$  matrices, in the large $\mathcal{N}_\mathrm{H}$  limit, the spacing of nearest energy levels is of order 
$1/\mathcal{N}_\mathrm{H}$. It is known in this case  that, up to temperature dependent corrections,  $t_H\sim \mathcal{N}_\mathrm{H}$, in agreement with the discussion above.
In the Majorana SYK model it is known, \cite{Cotler:2016fpe}, again neglecting temperature dependent effects, that $t_H\sim e^{Ns_0}$ at low temperatures, where $Ns_0$ is the  $T=0$ residual entropy. Note that $Ns_0$ is defined in the large $N$ theory,  by first taking the $N\rightarrow \infty$ limit and then the $T\to 0$ limit. At finite and large $N$, one expects the system to have a near degeneracy close to the ground state, with the spacings being $\Delta E \sim e^{-Ns_0}$. So we see, using the relation eq.(\ref{relte}), that  $t_{H}$ is again of order the nearest neighbour spacings close to the ground state. 


Let us end this brief overview  of the SFF with a few more  comments. First, 
in RMT the Thouless time is known to be of order $t_{th}\sim \sqrt{\mathcal{N}_\mathrm{H}}$ 
. The ramp region therefore extends over the time $\sqrt{\mathcal{N}_\mathrm{H}}<t<\mathcal{N}_\mathrm{H}$. Interestingly,   the Majorana SYK model,  exhibits similar behaviour at low temperatures with $e^{N s_0}$ playing the role of $ \mathcal{N_H}$.  It is known, \cite{Cotler:2016fpe},  that at low temperatures, $\beta \delta J\ll 1$,  $t_{th} \sim e^{Ns_0/2}$, up to small temperature dependent corrections, so that the ramp region extends from $\displaystyle e^{Ns_0/ 2}<t<e^{Ns_0}$.

It is also worth mentioning that the behaviour of the SFF in the Majorana SYK model, and in the JT (Jackiw–Teitelboim) theory of gravity \cite{1985NuPhB.252..343J,1974AnPhy..88..286R,Saad:2019lba}
 agree with each other, in the low temperature limit. 
In JT theory $Ns_0$ enters as a parameter, determining the weight of different topological sectors.  Up to subdominant temperature corrections it turns out in this theory also,  
\bef
\label{JTuniv}
\displaystyle
t_{H} \sim e^{N s_0},~~
t_{th} \sim  e^{N s_0/ 2}.
\eef
 It is also worth noting here that JT theory arises quite universally  in the description of the low energy behaviour of near-extremal black holes, including fast rotating black holes which have been  observed in the sky \cite{Nayak:2018qej,Moitra:2018jqs,Sachdev:2019bjn,Moitra:2019bub}. 
These near-extremal black holes carry a charge, for example angular momentum, and their behaviour is in fact more directly related to the charged SYK model. 


Second, it is useful to define a disconnected component of  the SFF given by 
\be\label{dis}
\displaystyle
g_d(t)={\langle Z(\beta+i t\rangle \langle Z(\beta-it)\rangle \over \langle Z(\beta)\rangle^2}
\ee
and a connected component which is 
\be\label{conn}
\displaystyle
g_c(t)=\text{SFF}(t)-g_d(t)
\ee
During the  initial decay period $g_d$  gives the dominant contribution.  $t_{th}$ marks the time when $g_d$ and $g_c$ become comparable. And thereafter, in the ramp region, $g_c$ gives the dominant contribution to the SFF.  

Finally, the late time plateau value for the SFF, which to smooth out oscillations can be defined as follows, 
\be\label{ltv}
\displaystyle
\text{SFF}(t\rightarrow \infty) \simeq\lim_{T\to\infty}\frac{1}{T}\int_{t_H}^{T}\left |\frac{Z(\beta,t)}{Z(\beta)}\right |^{2}={\sum_m N_m^2 e^{-2\beta E_m}\over Z(\beta)^2}.
\ee

\medskip

\noindent{\textcolor{blue}{\it {III) Out of Time Correlators (OTOCs):}}
The four point out of time correlator is a useful diagnostic of chaos. For an operator ${\cal O}$, at temperature $T={1/ \beta}$,
it is defined to be 
\be\label{defO} 
\displaystyle
G_4(t)= {\langle Z\rangle \langle\mathrm{Tr}(y{\cal O}(t) y{\cal O}(0) y{\cal O}(t) y{\cal O}(0))\rangle\over   \langle\mathrm{Tr}( e^{-\beta H}{\cal O}(t) {\cal O}(t))\rangle \langle\mathrm{Tr}(e^{-\beta H}{\cal O}(0){\cal O}(0))\rangle}
\ee
where $y=e^{-\beta H/4}$, and $Z=\mathrm{Tr}(e^{-\beta H})$.
With random couplings, as in the system at hand, one takes an ensemble average of both the numerator and denominator, as denoted  by the angle brackets above. Note that  the average is taken separately for  the numerator and denominator. 

For a system in the large $N$ limit, like the SYK model it is useful to define $F_4(t)$ through the relation: 
\be\label{deff4}
\displaystyle
G_4(t)=1-{1\over N} F_4(t).
\ee
The function, $F_4$,  arises from the connected component of the $4$ point function in the numerator in eq.(\ref{defO}), and is, to leading order, independent of $N$.

\begin{figure*}
\hspace{-2.3cm}
\centering
\begin{picture}(0.9\linewidth,0.28\linewidth)
    \put(0,0){\includegraphics[width=0.33\linewidth]{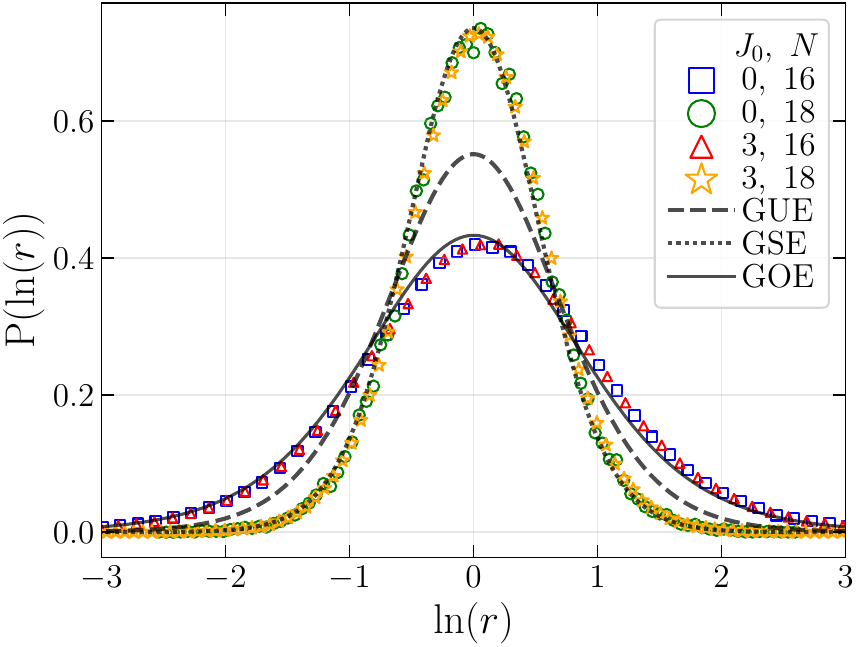}}
    \put(24,113){\normalsize{\textcolor{black}{$(a)$}}}
     \put(75,33){\fbox{$\mathcal{Q}=0$}}
    \put(175,0){\includegraphics[width=0.33\linewidth]{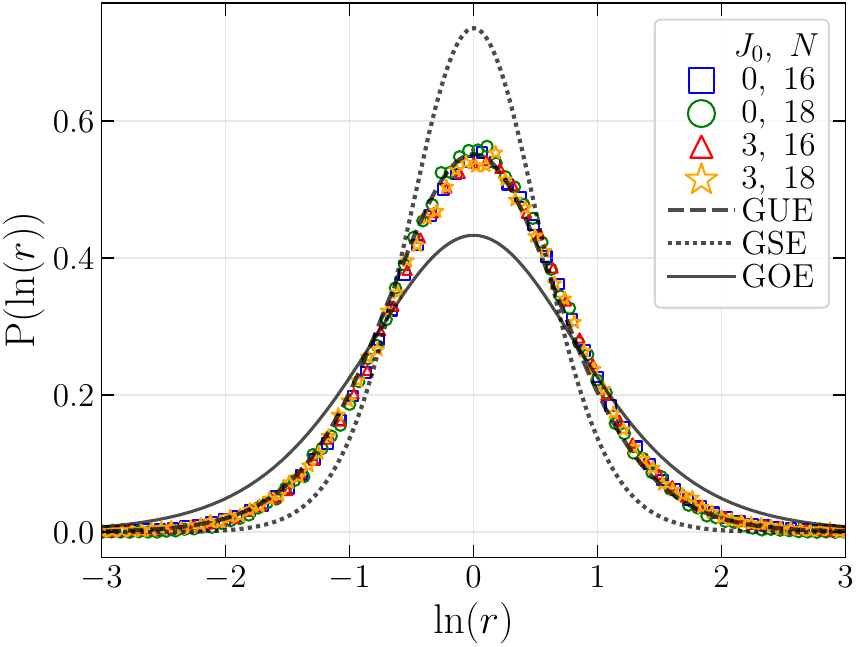}}
    \put(200,113){\normalsize{\textcolor{black}{$(b)$}}}
     \put(250,33){\fbox{$\mathcal{Q}=2$}}
    \put(345,0){\includegraphics[width=0.33\linewidth]{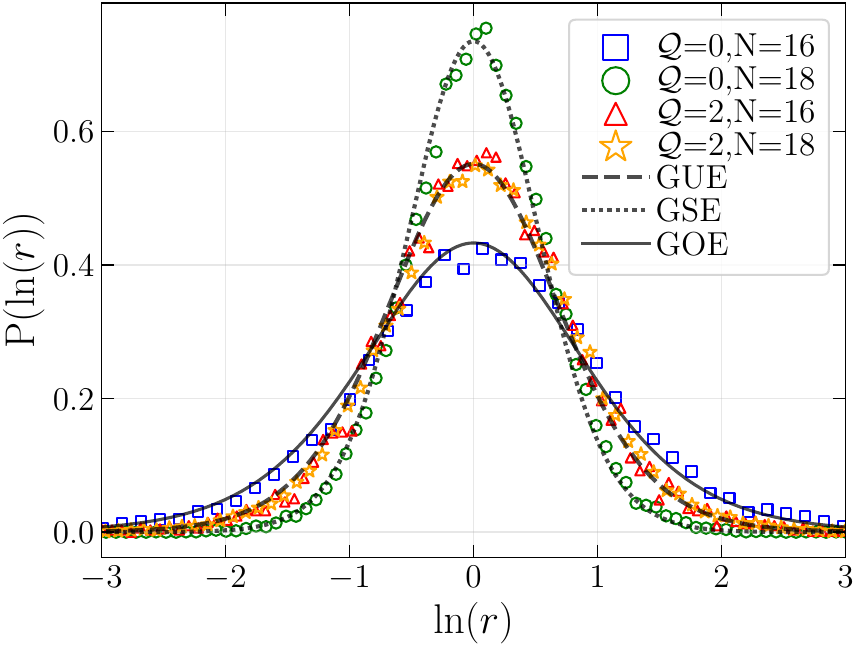}}
    \put(371,113){\normalsize{\textcolor{black}{$(c)$}}}
\end{picture}

\caption{Plot of the probability $P(r)$ where $r$ is the ratio of the energy separation of the ith energy level from the one below relative to the one above, as a function of $\ln(r)$, showing agreement with RMT ensembles (GUE, GOE, GSE). a) $\mathcal{Q}=0$ sector, for  $N=16,18$, and $J_0=0,3$, $\delta J=1$;  b)  ${\cal Q}=2$ sector for $N=16,18$ and $J_0=0,3$ $\delta J=1$;  c) Fixed coupling model with $\delta J=0$, for ${\cal Q}=0,2$. In all the plots two values of $N=16,18$ are shown and $J_0=1$. } 
\label{LEVEL_SP_SYK}
\end{figure*}
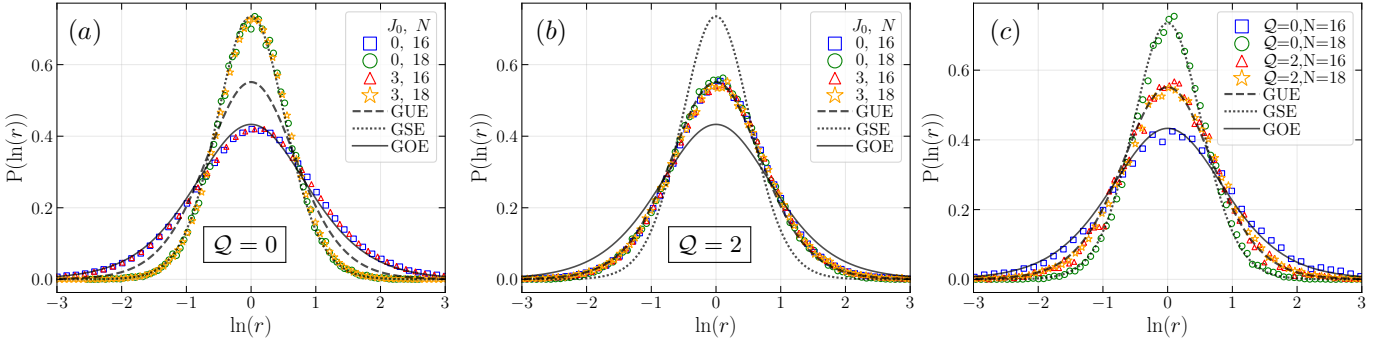

For a fast  scrambler 
\be\label{fscr}
F_4(t)\sim e^{\lambda_\mr{L}t},
\ee
where the Lyapunov exponent, $\lambda_\mr{L}$, is linear in  $T$, 
\be\label{lexp}
\lambda_\mr{L}\propto  T.
\ee
The Lyapunov exponent, and associated exponential growth, is the quantum analogue of the classical butterfly effect, making OTOCs a useful diagnostic of quantum chaos. 

A famous result,  \cite{Maldacena:2015waa},  says that $\lambda_\mr{L}$ satisfies the chaos bound ($k_\mr{B}=1$, $\hbar=1$)
\bef
\label{cbound}
\lambda_\mr{L} \le {2\pi T}.
\eef
The SYK model, with zero mean,  saturates this bound. It is also a remarkable fact that all black holes, in two derivative gravity, at finite temperature, saturate this bound. 

The exponential growth, eq.(\ref{fscr}), continues till the two terms in eq.(\ref{deff4}) become comparable, which happens at a scrambling time of order 
\be
\label{tscramb}
\displaystyle
t_{sc}\sim {\log(N)\over \lambda_\mr{L}}\ee
For $t>t_{sc}$ the exponential growth stops and $F_4$ saturates to a finite value. 

Comparing with the discussion of the SFF   above we see that $t_{sc}$, which goes like $\log(N)$, is parametrically much smaller at large $N$, than  both $t_{th}$ and $t_{H}$. As mentioned above in the SYK model, for example, $t_{th}$ and $t_H$   are exponential in $N$,  going like $e^{N s_0/ 2}$ and $e^{Ns_0}$ respectively. 
Thus, we see that the OTOCs are a diagnostic of chaos on much smaller time scales than the SFF,  or the nearest -neighbour statistics.

 We now turn to examining these three diagnostics of chaos in the charged SYK system with a varying mean value for the couplings. As was mentioned above, unless explicitly stated, the numerical results   in this section presented below are  obtained  by setting $\mu=0$. 
 
\subsection{\label{LEV_STAT}Level-Spacing Statistics}
We start by  studying the statistics of nearest neighbour level spacings among energy eigenstates, in the middle of the many-body spectrum (mid-spectrum), corresponding to infinite temperature $\beta=0$.
As mentioned above, we expect the statistics of nearest neighbour spacings to be universal and determined by the symmetries of the Hamiltonian \cite{altland1997nonstandard}.  
The mid-spectrum states are defined to be those which lie in the range $-0.1<{{\mathcal E}\over \Delta}<0.1$,  see fig. \ref{DOS_PLOT_0}.

For the system we are studying, the following symmetries are important for determining which RMT is relevant:  Parity $P$, particle- hole exchange $S$,  and charge ${\cal Q}$. These are discussed in appendix.\ref{SYMM_LEVEL}.

Some key properties of these symmetries determining the map to RMTs are as follows.  A  sector with charge   ${\cal Q}$ is mapped to  the $-{\cal Q}$ sector by $S$. This leads to the conclusion that sectors with non-zero ${\cal Q}$   should be governed by the GUE. On the other hand, the ${\cal Q}=0$ sector, which can arise only when $N$ is an even integer,  is invariant under $S$.   The Parity symmetry  is important in determining the map in this case. When, 
$N=0, {\rm mod } \  4$,  $P=(-1)^{{\cal Q}+{N\over 2}}=1$, which leads to a map to the GOE, while when $N=2 \ {\rm mod} \ 4$,  $P=(-1)^{{\cal Q}+{N\over 2}}=-1$, which leads to a map to the GSE. 

We thus arrive at the following table \cite{PhysRevD.101.066017}:
\be
\begin{tabular}{ |c|c|c|c|c| } 
 \hline
N mod 4 & 0 & 1 & 2 & 3  \\ 
 \hline
$\mathcal{Q}=0$ & GOE&  ~& GSE& ~   \\ 
 \hline
$\mathcal{Q}\neq 0$ & GUE & GUE& GUE& GUE \\ 
 \hline
\end{tabular}
\label{RMTCLASS}
\ee
To  study the nearest neighbour spacings  a quantity often used is the level spacing ratios $(r)$
\bef
\displaystyle
r_i=\frac{\Delta \lambda_i}{\Delta \lambda_{i+1}}=\frac{E_{i}-E_{i-1}}{E_{i+1}-E_{i}}
\label{rnde}
\eef 
which measure the ratio between neighboring energy level spacings\footnote{Alternatively one can consider the nearest neighbour spacing itself, rather than the ratio $r$, but in that case one has to `unfold' the spectrum, \cite{PhysRevD.101.066017}}.

The probability, $P(r)$, for this ratio to take  values between  $r$ and $r+dr$, can be computed for the GUE, GOE and GSE ensembles,   \cite{PhysRevLett.110.084101,PhysRevD.101.066017,you2017sachdev,PhysRevD.94.126010}, and takes the form, 
\bef
\displaystyle
P(\ln r)= \frac{N_\beta~r^{\beta+1}(1+r)^\beta}{\left(1+r+r^2\right)^{1+\frac{3 \beta}{2}}};~\begin{cases}
GOE,~N_{\beta=1}=\frac{27}{8}\\
GUE,~N_{\beta=2}=\frac{81 \sqrt{3}}{4 \pi}\\
GSE,~N_{\beta=4}=\frac{729 \sqrt{3}}{4 \pi}
\end{cases}
\label{RMTcases}
\eef

It is convenient to plot $P(r)$ as a function of $\ln(r)$ as we have done  in fig. \ref{LEVEL_SP_SYK}. 
In fig. \ref{LEVEL_SP_SYK} (a) we consider the ${\cal Q}=0$ sector of the system, for the case with zero mean, $J_0=0$ and when $J_0=3$ (with the standard deviation  $\delta J=1$). We see that very good agreement is obtained with the GOE when $N=16$,  $N {\rm mod} 4=0$, and with the GSE when $N=18$, $N {\rm mod} 4 =2$, as expected from the table above.  Importantly, this agreement, which was known to be true for the complex SYK model, continues to hold when a mean value is turned on. 

In fig. \ref{LEVEL_SP_SYK}(b) we consider the ${\cal Q}=2$ sector, which, from table eq.(\ref{RMTCLASS}), should map to the GUE. We see that this expectation is indeed borne out by the numerics, for both the zero and non-zero mean cases, studied here for $N=16$ and $N=18$.

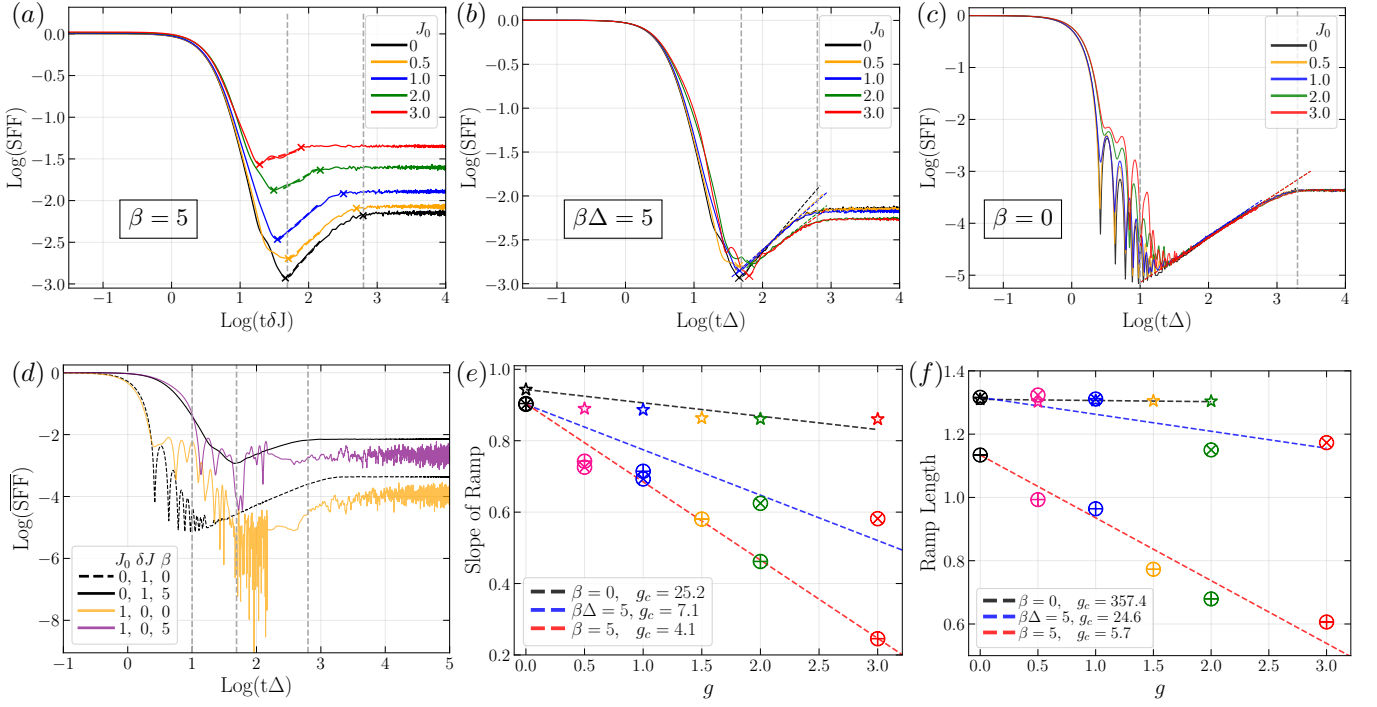
\begin{figure*}[t]
\centering
\setlength{\unitlength}{\linewidth}
\begin{picture}(1, 0.5)


    \put(-0.005, 0.27){\includegraphics[width=0.33\linewidth]{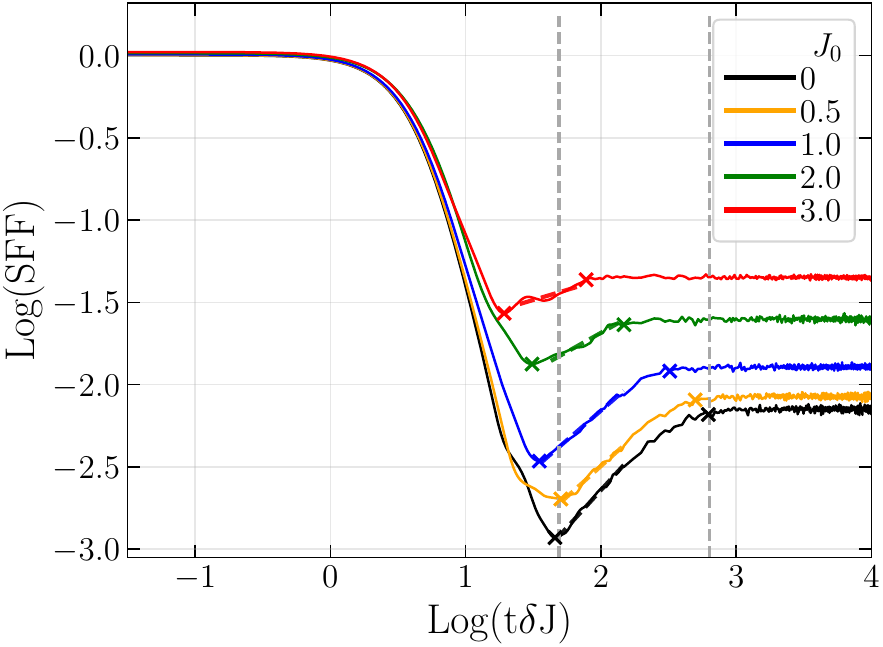}}
    \put(0.00, 0.5){\normalsize{\textcolor{black}{$(a)$}}}
    \put(0.08, 0.35){\fbox{$\beta=5$}}

    \put(0.330, 0.27){\includegraphics[width=0.33\linewidth]{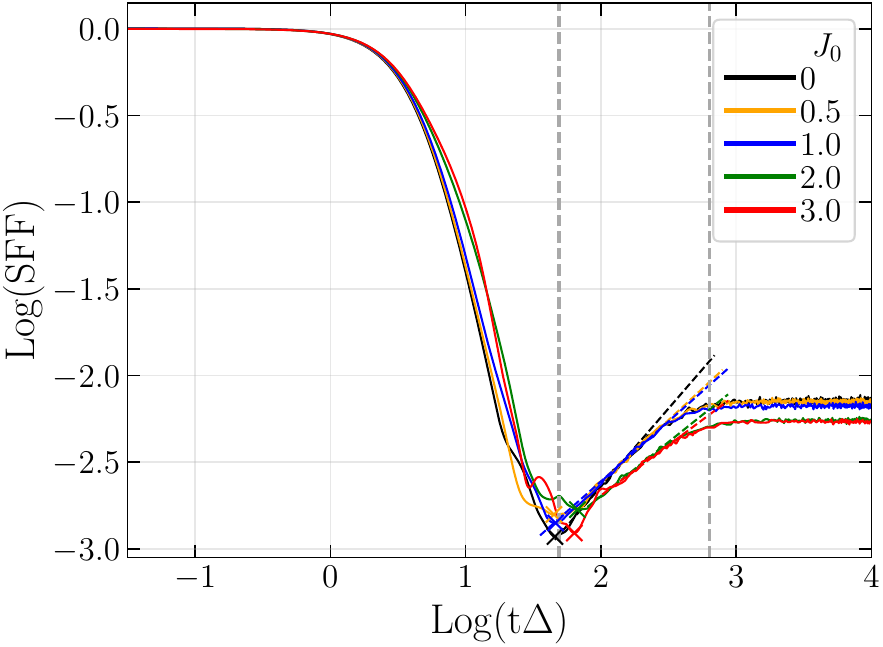}}
    \put(0.328, 0.5){\normalsize{\textcolor{black}{$(b)$}}}
    \put(0.408, 0.35){\fbox{$\beta\Delta=5$}}

    \put(0.669, 0.27){\includegraphics[width=0.32\linewidth]{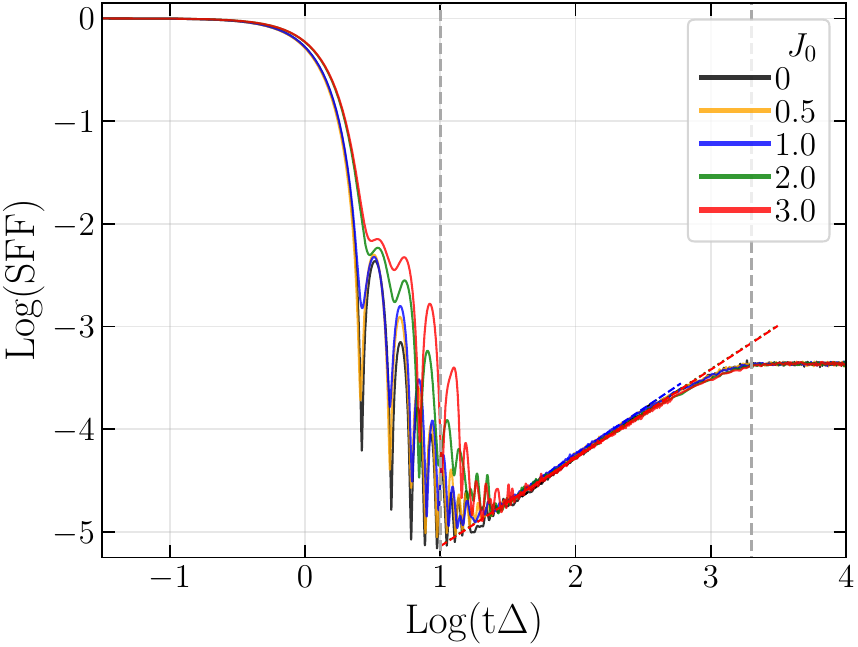}}
    \put(0.668, 0.5){\normalsize{\textcolor{black}{$(c)$}}}
    \put(0.718, 0.35){\fbox{$\beta=0$}}
    \put(-0.002, 0.00){\includegraphics[width=0.33\linewidth]{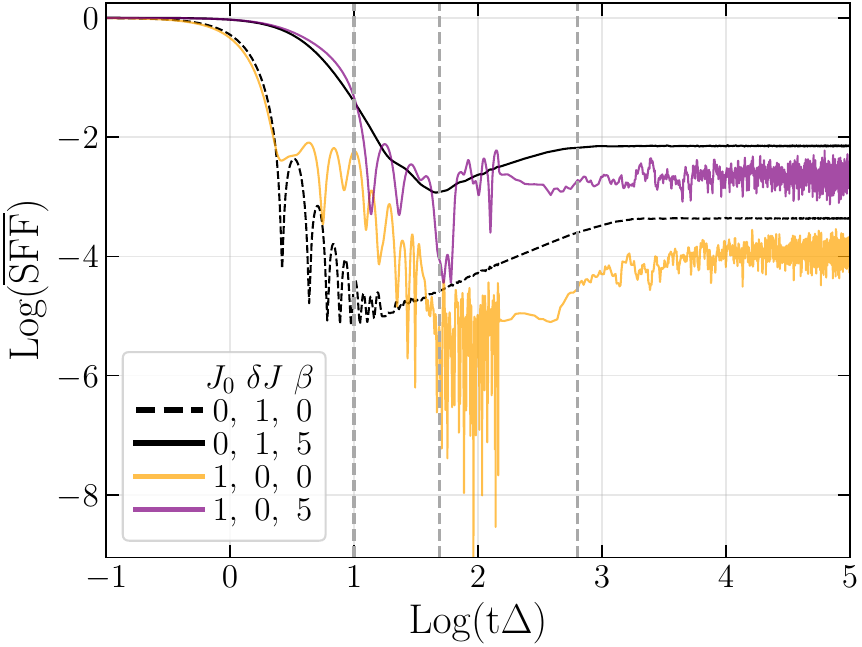}}
    \put(0.00, 0.235){\normalsize{\textcolor{black}{$(d)$}}}

    \put(0.334, 0.00){\includegraphics[width=0.325\linewidth]{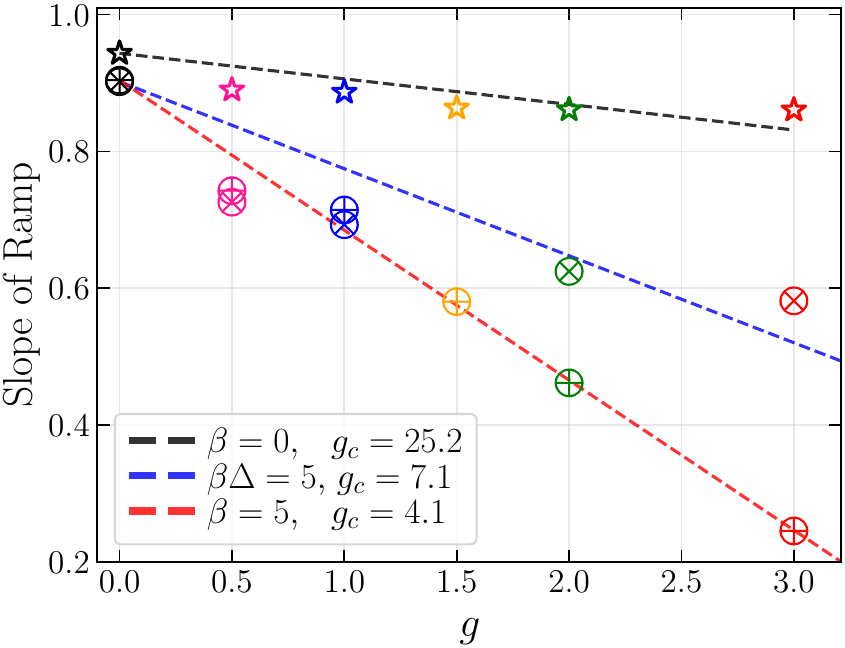}}
    \put(0.328, 0.235){\normalsize{\textcolor{black}{$(e)$}}}

    \put(0.67, 0.00){\includegraphics[width=0.32\linewidth]{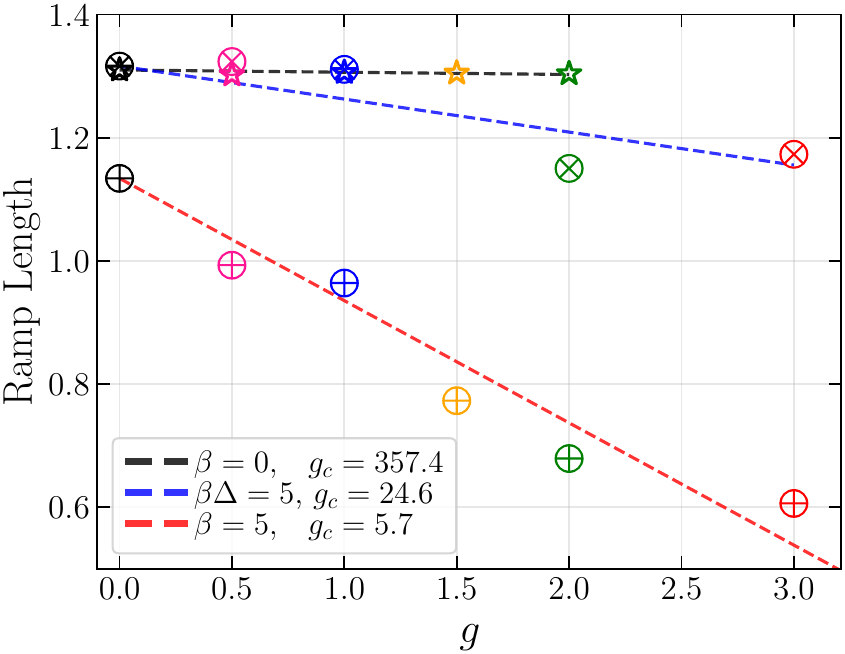}}
    \put(0.662, 0.235){\normalsize{\textcolor{black}{$(f)$}}}

\end{picture}

\caption{(a), (b), (c):   Spectral form factor SFF for different mean values, $J_0$, at $\beta=5$ and $\beta=0$. (d)  SFF for zero mean ($J_0=0, \delta J=1$), and zero standard deviation ($\delta J=0, J_0=1$).  (e) Slope of the ramp region extracted from (a), (b) as a fuunction of $g=J_{0}/\delta J$ (ratio of mean-to-standard deviation ). 
(f) The length of the ramp region obtained from (a) and (b) as a function of $g$. 
}
\label{SFF_SYK}
\end{figure*}
Finally, in fig. \ref{LEVEL_SP_SYK}(c) we consider the case where the mean value $J_0=1$, but the standard deviation vanishes, so that the couplings take fixed values. Here we study both ${\cal Q}=0, 2$ sectors,   with $N=16$ and $18$.
We see that in all cases that there is very good agreement with the expected RMT statistics. 

While we have only plotted the data for one non-zero value of the mean, in fig. \ref{LEVEL_SP_SYK}(a), fig. \ref{LEVEL_SP_SYK}(b), we have verified it numerically for other values of the mean as well, and  expect the agreement with RMT to hold generally. 
Let us also  note that the statistics of nearest neighbour eigenvalues does not change by changing the sign of ${\cal Q}$. This is because we have set $\mu=0$, and the operator $S$, defined in eq. \ref{SYM_CASE},  then  commutes with $H$, takes ${\cal Q}\rightarrow -{\cal Q}$. Nor does it change by changing the sign of the mean value,  or by complex conjugating  the mean value (since, as was noted above  
 this exchanges the ${\cal Q} $  and $-{\cal Q} $ sectors). As a result,  such changes in the sign of ${\cal Q}$ or the  mean value,  do not have to be considered anew.  

To summarise, our results are in agreement with the general expectations of random matrix universality. The symmetry characteristics of the Hamiltonian we are studying, regardless of its detailed nature,  in particular the ratio of the mean and standard deviation in the couplings, determines the nearest neighbour statistics. 
It is worth emphasising that even when the randomness vanishes, and the couplings take fixed values,  fig. \ref{LEVEL_SP_SYK}(c), the statistics is chaotic and described by RMT universality. In contrast to this universality in the nearest neighbour spacings, the density of states itself is   sensitive to the detailed nature of the Hamiltonian, and,  as we discussed in the previous section, changes quite significantly when a mean value is turned on. 

\subsection{\label{SFF}Spectral-Form-Factor}

Next we turn to the SFF, defined in eq.(\ref{SFF_EQ}) above. 
Note that we have set $\mu=0$ in obtaining the numerical plots below. Also, unless that standard deviation, $\delta J$, vanishes, for the clean case, we choose units where it is set to unity. 

In fig. \ref{SFF_SYK}) we plot the data obtained by varying the mean value $J_0$ for the $N=12$ case. 
In fig. \ref{SFF_SYK} (a) we consider the case when $\beta=5$ for five different values of $J_0$. 
We see that the qualitative features of a dip, ramp and plateau region, remain as $J_0$ changes, but the ramp region becomes smaller as $J_0$ increases. 
We also see that  $t_{th}$,  where the initial decay period ends and the ramp starts, decreases as $J_0$ increases (note that $x$ axis is $\log(t)$).
In addition,  $\text{SFF}(t_{th})$, increases  as $J_0$ increases.  And 
$t_{H}$ shifts to smaller values while  the plateau value,  $\text{SFF}(t_{H})$, increases.
as $J_0$ increases.

Next, let us turn to fig.\ref{SFF_SYK}  (b). 
Here we have rescaled the time direction with $\Delta$, which is  the spread in energy levels, eq. \ref{DELTAJ}, and also plotted the data keeping $\beta \Delta$ fixed, for various values of $J_0$. The purpose is to see if some of the features in fig. 5(a) can be accounted for by the fact that the spread in energies changes as $J_0$ changes. From eq. \ref{SCALE_FN} and fig \ref{SCALING_SYK} we see that $\Delta $ increases as $J_0$ increases, so this means we are also lowering $\beta \propto {1\over \Delta}$, as $J_0$ increases. 

We see from fig.\ref{SFF_SYK}(b) that the profile for the  SFF  in the initial decay region varies much less now, with varying $J_0$. We also see that $t_{H}\Delta $ is now relatively insensitive to changing $J_0$, showing the $t_{H}$  approximately scales like $1/\Delta$. 
Note also  that the ramp region persist in a more pronounced manner, compared to fig.\ref{SFF_SYK}(a), as $J_0$ increases. In effect, this means that lowering the value of $\beta$, as $J_0$ increases, restores the ramp region, which was decidedly getting smaller in fig.\ref{SFF_SYK}(a). We also see that some of the trends, after the rescaling, are different compared to  fig.\ref{SFF_SYK}(a). In particular,   $t_{th} \Delta $ shifts to larger values as $J_0$ increases, instead of smaller  values; and the plateau, value, $\text{SFF}(t_{H})$, decreases as $J_0$ increases. 

In fig.\ref{SFF_SYK}(c) we consider the case with $\beta\rightarrow 0$, i.e. $T\rightarrow \infty$. 
We continue to rescale the time $t$ by $\Delta$ axis, so  the  $x$ axis is given by  $\log(t \Delta)$. 
We see that the initial decay phase is more ``noisy", but the ramp region is, interestingly, universal. This means, as a function of $t$, that the ramp would be shifted to increasing $t$,  with  $t_{th}$ and $t_{H}$ being shifted $\propto   \frac{1}{\Delta}$ . This behaviour is not unexpected, after noting the relation, eq.(\ref{relte}) and the fact that the total spread in energies is given by $\Delta$. 

The universal value of $\text{SFF}$ in the plateau region is related to the dimensionality of the Hilbert space. From the numerical analysis we find that $\log (\text{SFF})$ in this region is $\simeq -3.312$. This agrees with theoretical expectations. As discussed above the late time value is given by eq.(\ref{ltv}),  and at $\beta\rightarrow 0$ this becomes, 
\be\label{ltvalue}
\displaystyle
{Z(2 \beta=0)\over Z(\beta=0)^2}= 2\times \ 2^{-N},
\ee
after noting that 
 for $N=12$ the spectrum  is two fold degenerate.  The  RHS of eq.(\ref{ltvalue}) evaluates to  $\log(2\times 2^{-12})\simeq -3.312$. 
~ Note that in the  numerical simulation, instead of taking the infinite-time limit, we implement an ensemble averaging over disorder realizations to suppress the oscillatory terms and yield a smooth function of time. 

Next, we turn to  fig.\ref{SFF_SYK}(d). Here we compare the behaviour of the system with vanishing mean,  $J_0=0$, and non-vanishing standard deviation, $\delta J=1$, to approach the behaviour of the clean system, with vanishing standard deviation,  $\delta J=0$ and  non-vanishing mean, $J_0=1$. We study two different temperatures, $\beta=0,5$.  


From fig.\ref{SFF_SYK}(d) we see that  when 
 $\delta J=1$, the two curves for $\beta=5,0$ clearly show the initial decay, ramp and plateau stages. But when $J=0$ there is indeed  no discernible ramp.  For the $\delta J=0, J_0=1$ cases,  fig.\ref{SFF_SYK}(d) shows that there is an initial decay, and, as best as one can tell from the numerical analysis,  also  a final asymptote, as $t\rightarrow \infty$,   with the asymptotic value of the SFF becoming smaller as $\beta$ decreases. The data for the $\delta J=0$ cases in fig.\ref{SFF_SYK}(d) is 
a  time average  obtained after binning the data for intervals $\Delta t=10$. This reduces the noisy nature of the data,   due to a lack of ensemble averaging, but only to some extent.  

In fig.\ref{SFF_SYK}(e) we show the slope  of the ramp region obtained from the previous plots of fig.\ref{SFF_SYK}. For $J_{0}=0,\delta J=1$ the slope is known to be 1. Here, from  our numerical calculation,  we get it to be $0.94$, for $\beta=0$,and  $0.9$ for $\beta=5$. With increasing $J_{0}$ the value of the  slope  decreases, suggesting an extrapolation to a \emph{critical value} $g_c$ exists, at which the slope vanishes.  From  linear fitting we get $g_{c}\approx 4.1 $ for $\beta=5$ and  $g_{c}\approx 7.1$ for $\beta \Delta=5$.
A putative critical value corresponding to the disappearance of SFF linear ramp, a characteristic of RMT behavior, suggests an MBL-like transition, that can exist at finite $N$ for a 0+1 dimensional model \cite{Altshuler1997,Micklitz2019} at low temperatures.
\be
\begin{minipage}{0.49\linewidth}
    \centering
    Slope of the Ramp \newline
    \begin{tabular*}{\linewidth}{@{\extracolsep{\fill}} |c| c |c |c |@{}}
        \hline
        $J_{0}$   &  $\beta\delta J=5$ & $\beta\Delta=5$ & $\beta=0$  \\
        \hline
        0   & 0.9  & 0.9  & 0.94\\
        \hline
        0.5   & 0.74  & 0.72    & 0.89 \\
        \hline
        1.0   & 0.71& 0 .69   & 0.88\\
        \hline
        2.0   & 0.46  & 0.62   & 0.862 \\
        \hline
        3.0   & 0.245  & 0.58    & 0.86\\
        \hline
    \end{tabular*}
\end{minipage}~~~
\begin{minipage}{0.49\linewidth}
    \centering
    Length of the Ramp \newline
    \begin{tabular*}{\linewidth}{@{\extracolsep{\fill}} | c |c |c |@{}}
        \hline
         $\beta\delta J=5$ & $\beta\Delta=5$ & $\beta=0$  \\
        \hline
        1.13  & 1.31    & 6.26 \\
                \hline
        0.99  & 1.32    & 6.26 \\
                \hline
        0.96  & 1.31    & 6.26 \\
                \hline
        0.68  & 1.15    & 6.26 \\
                \hline
        0.6 & 1.17   & 6.26 \\
        \hline
    \end{tabular*}
\end{minipage}
\ee

 In fig.\ref{SFF_SYK}(f) 
we take $N=12$. For a few different values of $\beta$, 
we  study the variation of the ramp 
length from its dip time to plateau time ($t_{H}-t_{th}$) as $g$ is varied, and plot it against increasing $J_{0}$.  In this case,  linear fitting suggests a $g_{c}\approx 6$ . We also find that when  $\beta=0$ the ramp length does not change appreciably  as  $g$ is varied.   

\subsection{\label{OTOC}Out-of-time-order correlators}
The $4$-point OTOC is defined in eq.\ref{defO}. More precisely we  will use a flavour averaged version of the OTOC  defined as follows. We first consider the OTOC for the two flavours $i,j$ 
\be
\label{defgij}
\displaystyle
\tilde{G}_{i j}(t)=\frac{<Z><\mathrm{Tr}\left(y \psi_i(t) y \psi_j(0) y \psi_i(t) y \psi_j(0)\right)>}{\left\langle \mathrm{Tr}(\psi_j(0) \psi_j(0))\right\rangle\left\langle \mathrm{Tr}(\psi_i(t) \psi_i(t))\right\rangle}
\ee
Then  define the flavour averaged version, by averaging over all flavours,  
\be
\displaystyle
\label{Gdef}
G_4(t)=\frac{1}{2N(2N-1)}\sum_{i,j=1}^{2N}\tilde{G}_{i j}(t)
\ee
$F_4$ is now  defined in terms of $G_4$ as given in eq.(\ref{deff4}). 

Note that the $\psi_i$ fermions which appear in eq.(\ref{defgij}) are Majorana fermions, defined by 
\bef
\displaystyle
    \label{defmajo}
    \psi_{i}=c^{\dagger}_{i}+ c_{i},\\
    \psi_{i+1}=\iota (c^{\dagger}_{i}- c_{i}),
    \eef
    There are a total of $N$ complex fermions, this means we get $2N$ Majorana fermions; the index $i$ specifying the Majorana fermion $\psi_i$ runs from   $i=1, \cdots 2N$. 

OTOCs have been extensively studied in the Majorana SYK model, \cite{PhysRevLett.126.030602,Maldacena:2015waa,PhysRevResearch.2.013254}. This prompted us to  use Majorana fermions in our investigation. We have verified that   the results for the Lyapunov exponents we obtain, in the numerics,  remain unchanged if we use complex fermions instead. 

The numerical data we present below has been obtained by taking  $N=12$, complex fermions.    
The flavour averaging we carry out, allows us to reduce  noise in the data substantially and plays a key role, we believe, in allowing us to   extract reasonable values for the Lyapunov exponents,  even at the  modest  value of $N=12$, as we will see below. 

\begin{figure*}[t]
\centering
\setlength{\unitlength}{\linewidth}

\begin{picture}(1, 0.74)
    \put(0.008, 0.51){\includegraphics[width=0.32\linewidth]{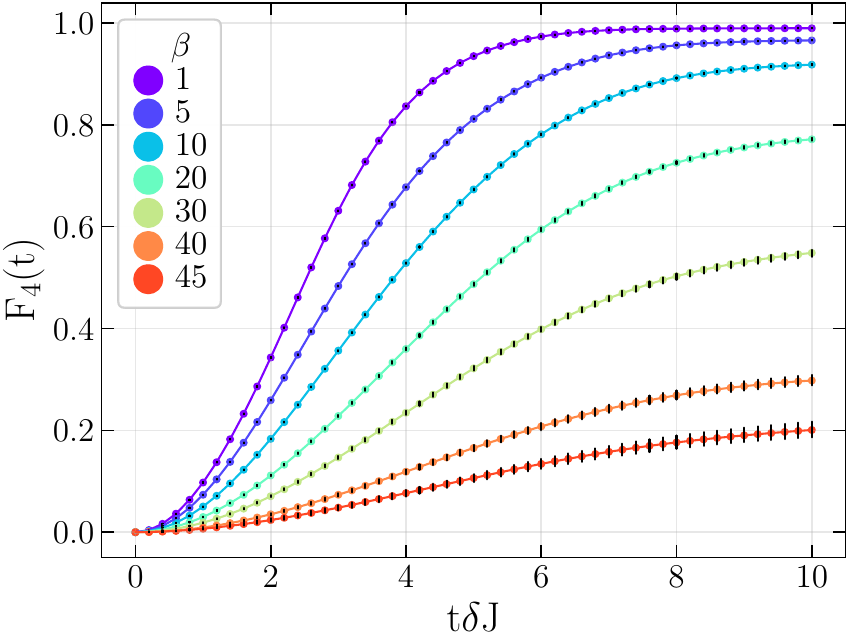}}
    \put(0, 0.74){\normalsize{\textcolor{black}{$(a)$}}}

    \put(0.34, 0.51){\includegraphics[width=0.32\linewidth]{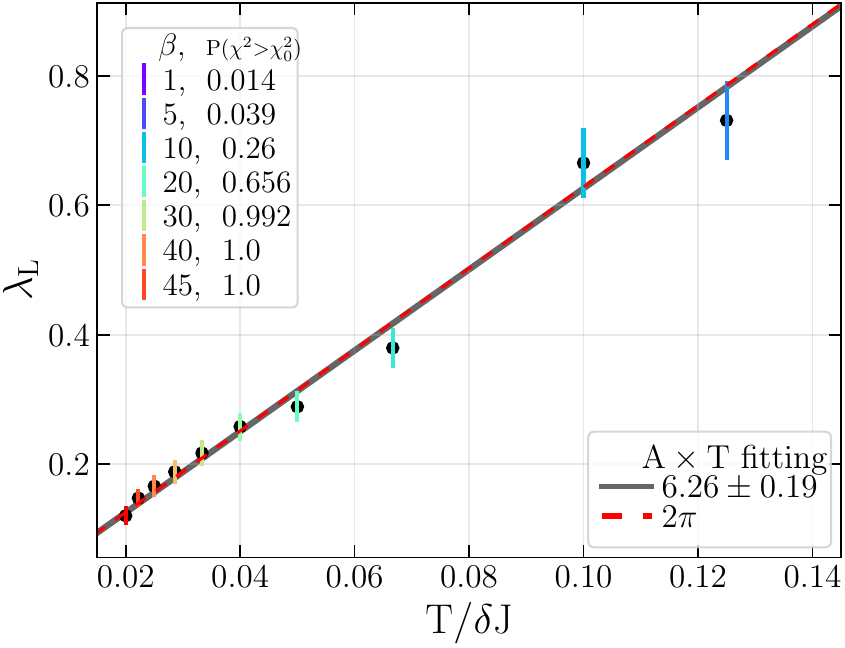}}
    \put(0.335, 0.74){\normalsize{\textcolor{black}{$(b)$}}}

    \put(0.677, 0.51){\includegraphics[width=0.32\linewidth]{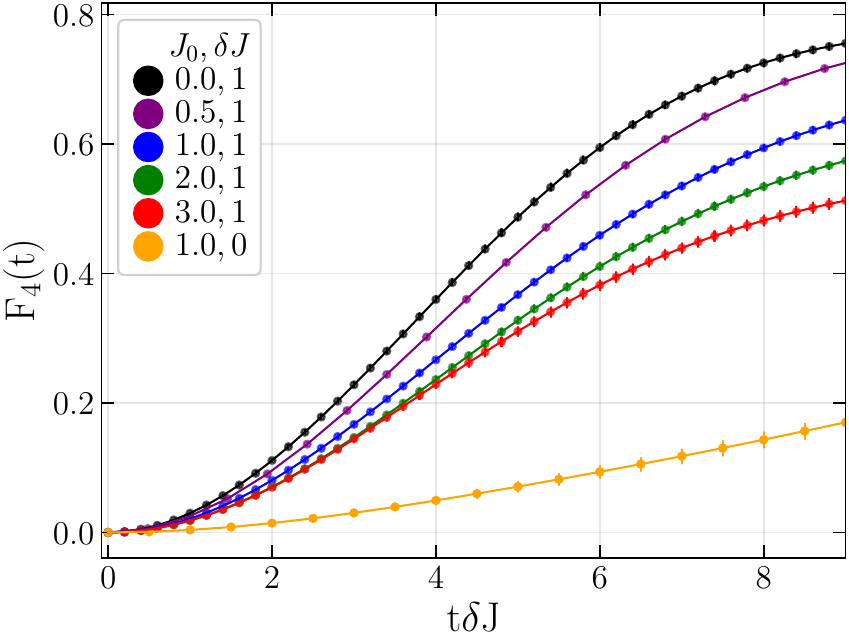}}
    \put(0.665, 0.74){\normalsize{\textcolor{black}{$(c)$}}}

    \put(0.008, 0.26){\includegraphics[width=0.33\linewidth]{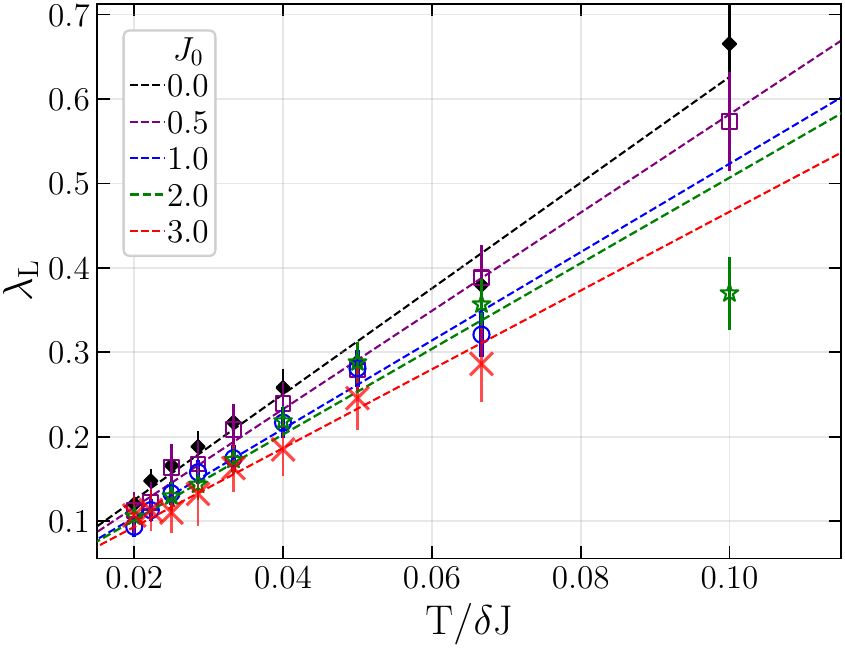}}
    \put(0.0, 0.5){\normalsize{\textcolor{black}{$(d)$}}}

    \put(0.34, 0.26){\includegraphics[width=0.33\linewidth]{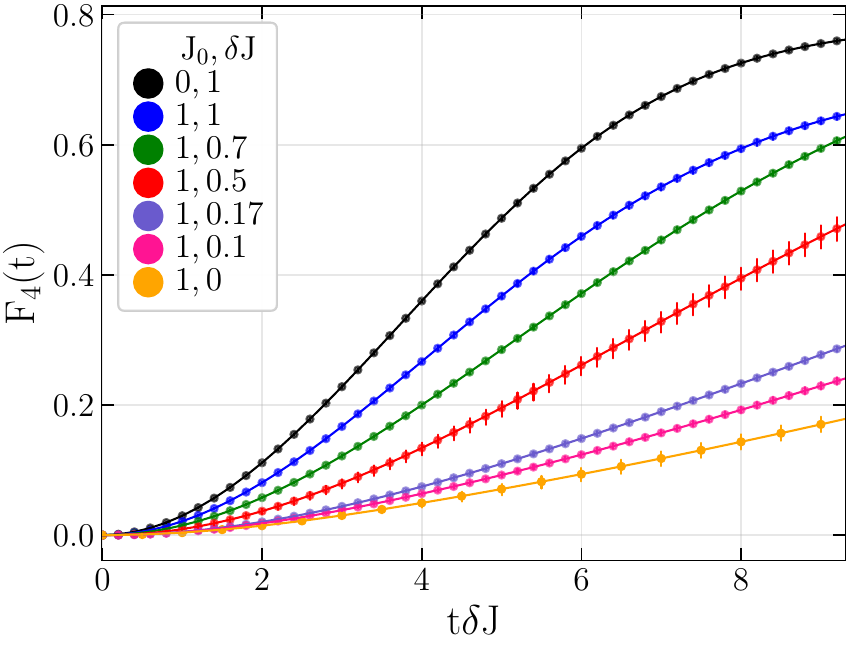}}
    \put(0.337, 0.5){\normalsize{\textcolor{black}{$(e)$}}}

    \put(0.677, 0.26){\includegraphics[width=0.33\linewidth]{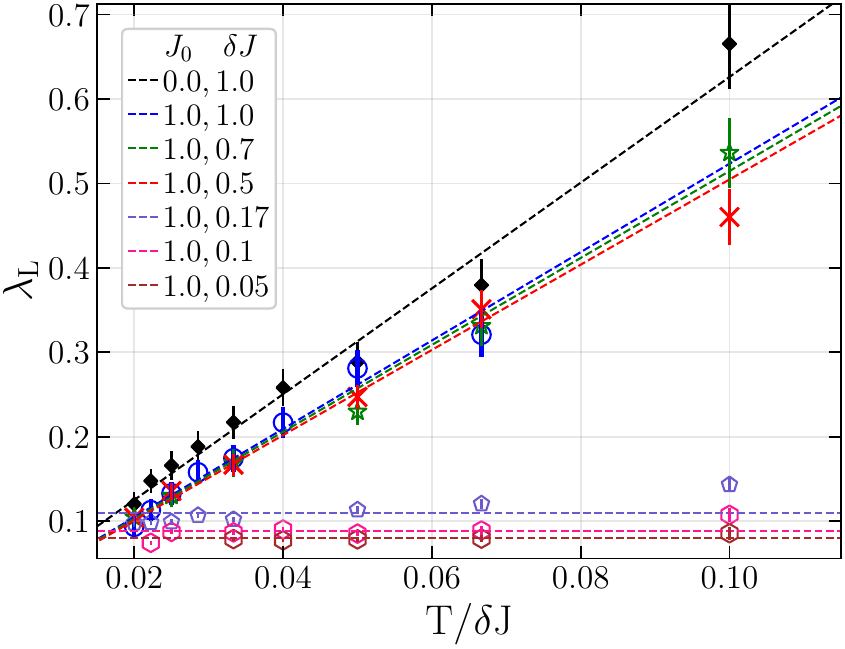}}
    \put(0.6699, 0.5){\normalsize{\textcolor{black}{$(f)$}}}
  \put(0.16, 0.23){\normalsize{\textcolor{black}{$(g)$}}}
    \put(0.2, 0.15){\begin{tabular*}{0.33\linewidth}{@{\extracolsep{\fill}} |c| c |c |c |c |@{}}
            \hline
            $J_{0}$ & $\delta J$ & $g$ & $A$ & $\pm\Delta A$ \\
            \hline
    0   & 1.0  & 0    & 6.26 & 0.19 \\
     \hline
    0.5 & 1.0  & 0.5  & 5.82 & 0.24 \\
     \hline
    1.0 & 1.0  & 1.0  & 5.23 & 0.17 \\
     \hline
    2.0 & 1.0  & 2.0  & 5.07 & 0.18 \\
     \hline
    3.0 & 1.0  & 3.0  & 4.66 & 0.24 \\
            \hline
    1.0 & 0.7  & 1.42 & 5.15 & 0.12 \\
\hline
    1.0 & 0.5  & 2    & 5.05 & 0.14 \\
    \hline
        \end{tabular*}}
    \put(0.556, 0.00){\includegraphics[width=0.32\linewidth]{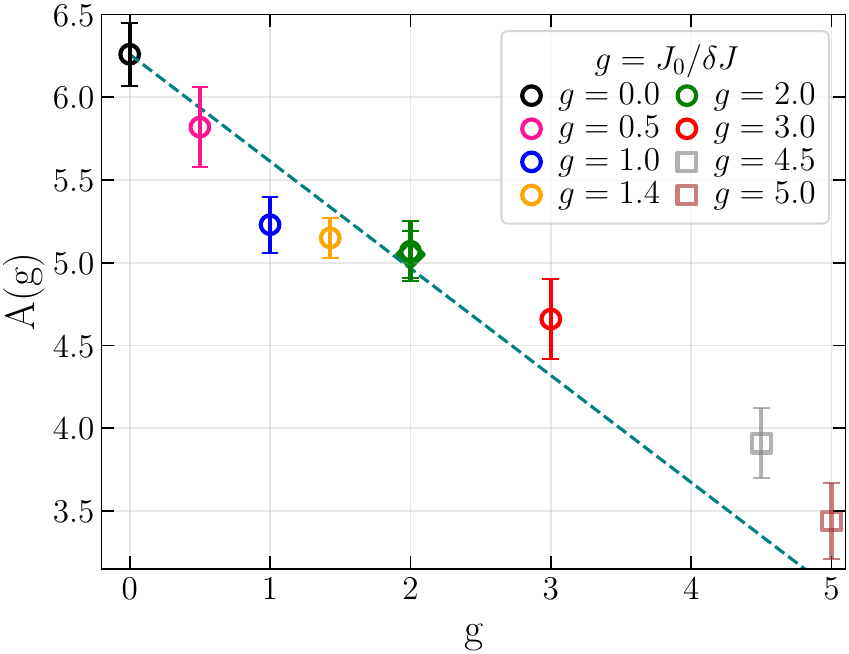}}
    \put(0.545, 0.23){\normalsize{\textcolor{black}{$(h)$}}}
\end{picture}
\caption{ a) OTOCs with $J_0=0$
at different values of $\beta$. b) Lyapunov exponent as a function of $\beta$, for $J_0=0$. 
Black curve from fitting data, and red curve for Lyapunov exponent saturating the  chaos bound, in good agreement.
c) OTOCs for varying  $J_0$ at $\beta=20$, and $\delta J=1$; showing that the rate of increase of $F_4$ slows down as $J_0$ increases. d)Lyapunov exponent $\lambda_{\mr{L}}$ vs $T/\delta J$ 
showing agreement with fast scrambling behaviour, eq.(\ref{fflyapunov}). The coefficient $A$ becomes smaller as $J_0$ increases. e) OTOCs for varying  $\delta J$, keeping $J_0=1$, and  $\beta=20$. Black curve is for  $J_{0}=0, \delta J=1, \beta=20$. f)  $\lambda_{\mr{L}}$ plotted against $T$
, with a linear fit confirming the fast scrambling behavior. A deviation from this behavior is observed for larger ratios $g=J_0/\delta J \geq 6$. (g) Tabulated values of the extracted coefficient $A$ and their associated errors $\Delta A$, derived from panels (d) and (e). (h) The coefficient $A(g)$ from panel (g) plotted as a function of  $g$. Note that by defining flavor-averaging of the OTOCs allows us to suppresses noise substantially and enables reliable extraction of Lyapunov exponents for modest $N=12$.
}

\label{OTOC_SYK}
\end{figure*}
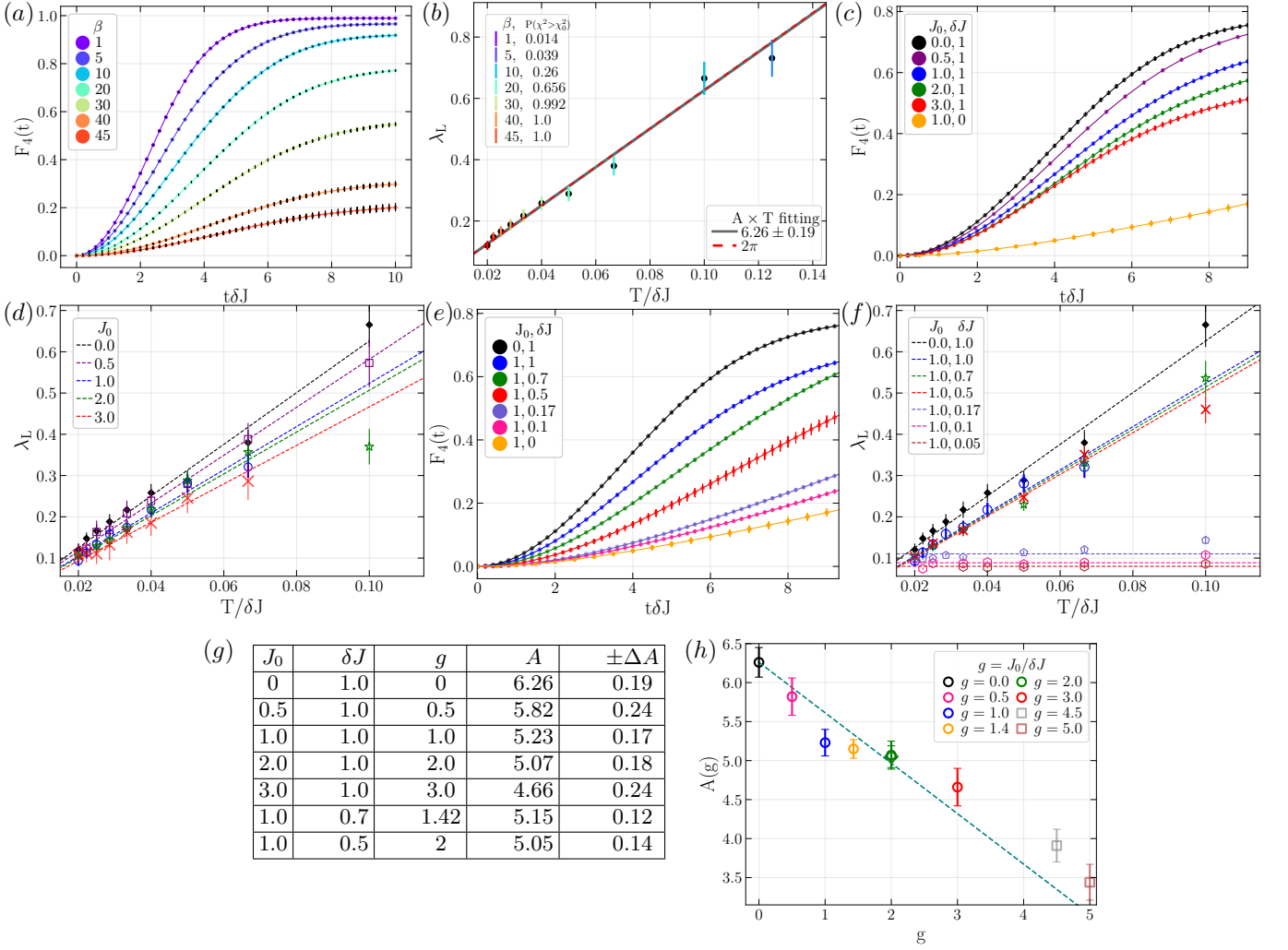
 We now turn to the data as shown in  fig. \ref{OTOC_SYK}.  
In fig. \ref{OTOC_SYK}(a) and (b) we consider the conventional, 
$J_0=0,\delta J=1$,  complex SYK model, 
 as a way of benchmarking our results before turning to the $J_0\ne 0$ case subsequently. fig. \ref{OTOC_SYK}(a) considers the model for varying $\beta$. 
We see that $F_4$, eq.(\ref{deff4}) rises rapidly with time and eventually saturates.
The error bars in the plot are obtained by random sampling over 70\% of the available data, while carrying out the average  over the species in eq.(\ref{Gdef}). 
 We see that the errors are small. 
 
 In fig. \ref{OTOC_SYK}(b) we plot the Lyapunov exponent as a function of $\beta$. For a given value of $\beta$ this exponent is obtained from fig. \ref{OTOC_SYK}(a) as follows. 
 We first identify an interval of time 
\([t_{\text{min}}, t_{\text{max}}]\),
over which $F_4$ is well fitted by a function of the form 
\be
\displaystyle
\label{expform}
F_4(t)=\frac{1}{N}e^{\lambda_\mr{L} t}\sim a+b e^{\lambda_\mr{L} t}
\ee
The values of $t_{\text{min}}, t_{\text{max}}$ are identified such that 
the corresponding \( P(\chi^2 > \chi^2_0) \) indicates a statistically reasonable fit.
This gives a value for the Lyapunov exponent. The error bar on this value is then obtained from the fitting procedure applied by \texttt{Scipy.curve.optimize}.

The resulting values of $\lambda_{\mr{L}}$ as a function of $T$ are shown in fig. \ref{OTOC_SYK}(b). The inset shows the values for $P(\chi^2>\chi_0^2) $, for the various $\beta$ values. The curve in fig. \ref{OTOC_SYK}(b) is obtained by fitting the $T$ dependence to be of the form 
\bef\label{fflyapunov}
\displaystyle
\lambda_\mr{L}= A T
\eef
In the large $N $ limit it is known analytically that $A$ takes the value,  \cite{Maldacena:2015waa}, 
\bef
\displaystyle
\label{valA}
A=2\pi,\eef
saturating the chaos bound. 
We see that the best fit we obtain, 
$A=6.26\pm 0.19$, agrees well with this value,
and we also see that the statistical uncertainty is about $3\%$ and small.

Fig. \ref{OTOC_SYK}(a) and (b) give us confidence that our numerical methods are working for computing the Lyapunov exponent. We remind the reader that the  analysis above has been carried out  for $N=12$. As mentioned above, we believe the flavour averaging we are doing is responsible, to some extent, for these methods producing reliable results, even at this modest value of $N$.

  It is worth mentioning here, that an earlier study of OTOCs in the Majorana SYK model at similar system sizes was presented in~\cite{Anegawa:2023vxq} for the Majorana SYK model. This study also employed the same methodology as we are doing in our current investigation, using  the  \texttt{Dynamite} package in conjunction with \texttt{PETSc} and \texttt{SLEPc}, following the method described in \cite{PhysRevLett.126.030602}. The results in \cite{PhysRevLett.126.030602} are at much larger values of $N$ and even allow for an extrapolation to $N\rightarrow \infty$. The agreement obtained in  ~\cite{Anegawa:2023vxq}, with the results in \cite{PhysRevLett.126.030602}
  (in the regime \(\beta > 20\)), gives us further confidence that despite the limitations of small system size, our method allows us to obtain the  OTOCs  reliably, using much less computational resources.
  
 In fig. \ref{OTOC_SYK}(c) we turn to cases with non-zero $J_0$. We take $\beta=20$ and obtain the dependence of $F_4(t)$  for various values of $J_0$.  We see that the error bars in 6(c), again obtained by random sampling over  $70\%$ of the data while doing the flavour averaging in eq.(\ref{Gdef}),  remains small. 
 A  clear important qualitative trend we see is that the rate of increase of $F_4$ becomes smaller, as $J_0$ increases. The standard deviation  for all the cases considered is $ \delta J=1$, except for the orange line, which pertains to  $J_0=1, \delta J=0$, i.e., ${J_0/\delta J}=\infty$. 
 We see that in this limiting case, the rate of increase slows down a great deal, and, not surprisingly,   $F_4$ cannot be fitted with an exponentially growing form, eq.(\ref{expform}), over a suitably identified  interval $[t_{\text min}, t_{\text max}]$. 

From fig. \ref{OTOC_SYK}(c) the value of the Lyapunov exponent for  $\beta=20$ can be obtained as we did in the $J_0=0$ case of  fig. \ref{OTOC_SYK}(a). This procedure can then be repeated for different values of $\beta$, to obtain $\lambda_{\mr{L}}$ for varying values of $J_0, \beta$.  We do this   for all cases except the one where $\delta J=0$, which cannot be fitted to the Lyapunov form eq.(\ref{fscr}). 

  Fig. \ref{OTOC_SYK}(d) is  a plot of the resulting values of $\lambda_{\mr{L}}$ vs $T/\delta J$. We see that the functional dependence, for the different values of $J_0$ considered, is well described by the  functional form, eq.(\ref{expform}). The resulting values of the coefficient $A$, for different values of $J_0$, are given in the table fig. \ref{OTOC_SYK} (g). It is clear, as we anticipated from the qualitative trend in fig. \ref{OTOC_SYK}(c), that  $A$ becomes smaller as $J_0$ increases.   

  To summarise, we learn from the analysis described above that as $J_0$ increases, starting from zero, the scrambling behaviour slows down. Up to $J_0=3$, the system is still a fast scrambler with a Lyapunov growth of the OTOC, but the Lyapunov coefficient, which saturates the chaos bound when $J_0=0$, becomes smaller as $J_0$ increases. From the orange curve in fig. \ref{OTOC_SYK}(c), corresponding to $J_0=1, \delta J=0$, we learn that eventually,  when ${J_0/\delta J}$ becomes big enough,  the fast scrambling behaviour goes away and the OTOC does not grow exponentially any more.

Finally in fig. \ref{OTOC_SYK} e) and f) , as a check on our previous results, we carry out a similar analysis, but instead of varying $J_0$, keeping $\delta J$ fixed, we now vary $\delta J$ keeping $J_0=1$ (except for the black curve in fig. \ref{OTOC_SYK}(e), included as a benchmark, where $J_0=0, \delta J=1$). In fig. \ref{OTOC_SYK}(e) $\beta$ is kept fixed at $20$. We see that as $J_0/\delta J$ increases the rate of increase of $F_4$ slows down, in qualitative agreement with fig. \ref{OTOC_SYK}(c). 

The Lyapunov exponents obtained from fig. \ref{OTOC_SYK}(e) are plotted against $T/\delta J$ in 
fig. \ref{OTOC_SYK}(f).  We see that the first $4$ cases  are well fitted by a straight line, showing a good fit to the fast scrambling form, eq.\ref{expform}, with the coefficient $A$ decreasing as ${J_0\over \delta J}$ increases. 
However a significant change occurs for the last two cases, $\delta J=0.17$ and $ \delta J=0.1$. For these a straight line is no longer a good fit, showing once again,  as was discussed above, that once the ratio ${J_0/\delta J}$ becomes big enough the fast scrambling behaviour stops. For the last three cases instead of fitting with $T$ we show the constant line at the average value of $\lambda_{\mr{L}}$. This proccedure suggest the following values of average $\lambda_{\mr{L}}$; 0.11 ($\delta J=0.17$), 0.09 ($\delta J=0.1$),0.08 ($\delta J=0.05$). Qualitatively this constant $\lambda_{\mr{L}}$ approximation holds for $T/\delta J<0.07$.
We also note, as a check on our analysis,  that the value for $\lambda_{\mr{L}}$ in  fig.\ref{OTOC_SYK}(f) for $J_0=1, \delta J=0.5$ is $A=5.05\pm 0.14$,  this is  in good agreement with what we got in fig.\ref{OTOC_SYK}(d) for $J_0=2, \delta J=1$, $A=5.07\pm 0.18$.  \\

To conclude this section, we have seen from the analysis above that as the ratio $g=J_0/\delta J$ increases, the different diagnostics of chaos behave differently. The nearest neighbour spacings continue to satisfy the Wigner surmise. On the other hand, the SFF and the OTOCS show considerable departures from the $g=0$ case. In particular, the ramp region in the SFF becomes shorter with increasing $g$. And in the OTOCS,  the fast scrambling behaviour persists, but with a smaller value for $\lambda_L$, till $g\simeq 6$,  thereafter eventually vanishing for larger  $g$.

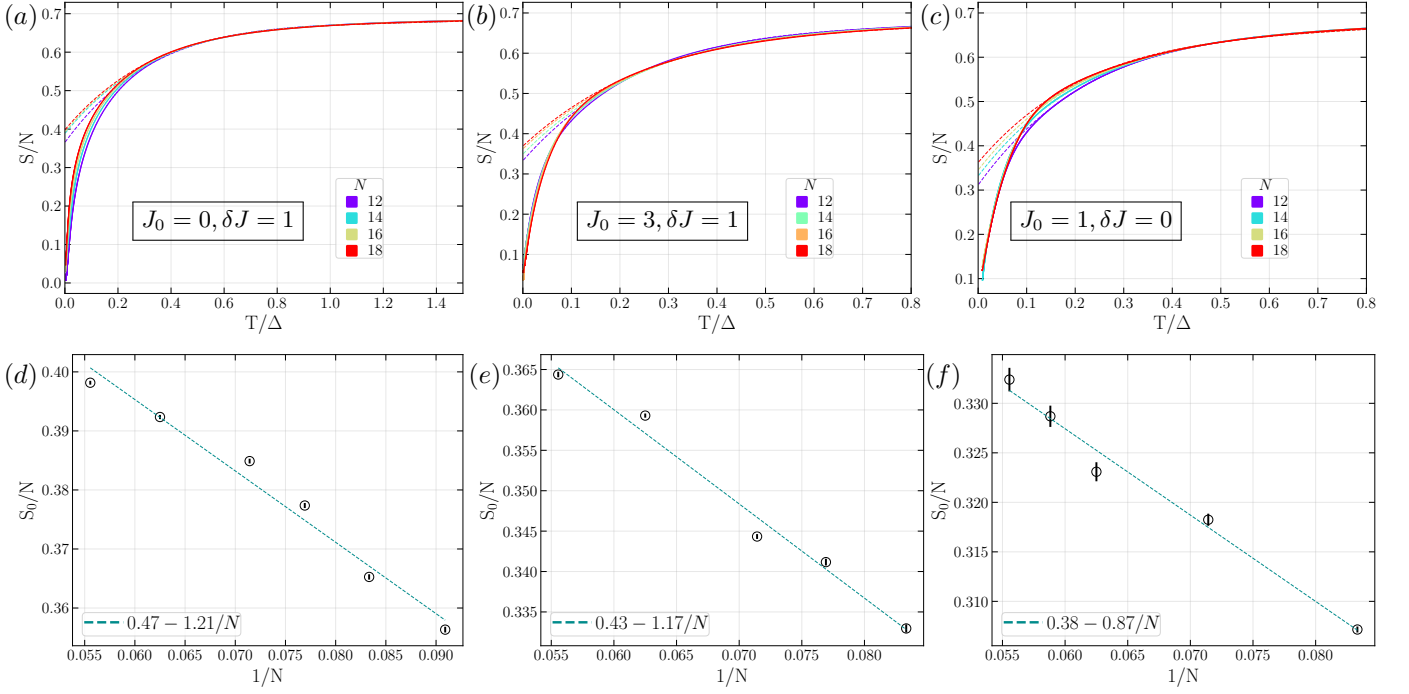
\begin{figure*}
\centering

\setlength{\unitlength}{\linewidth}
\begin{picture}(1, 0.5)
    \put(-0.005, 0.27){\includegraphics[width=0.33\linewidth]{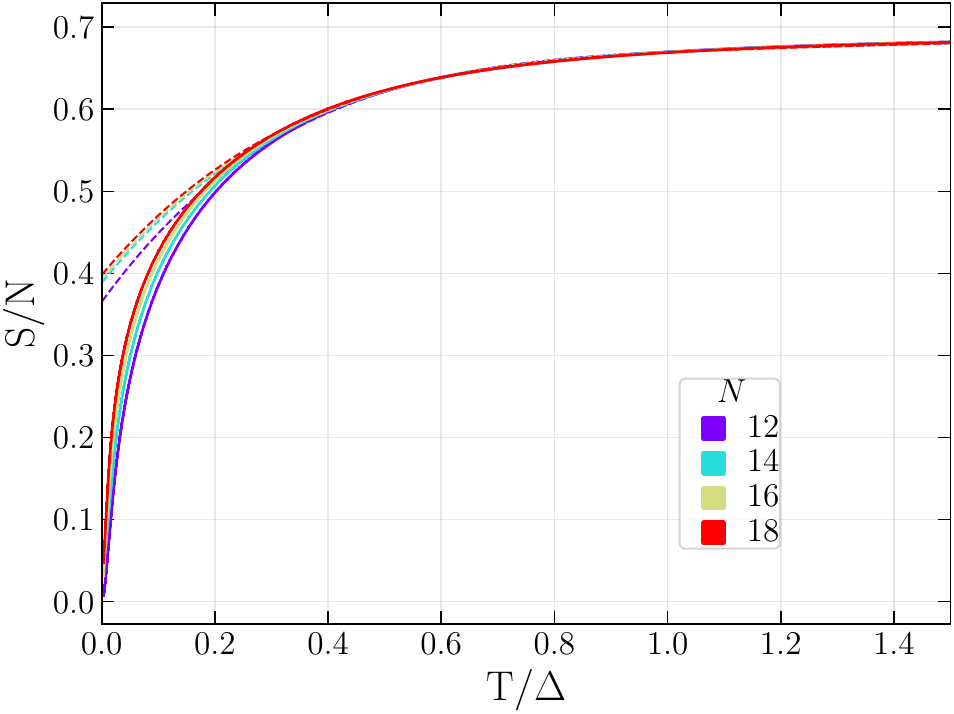}}
    \put(-0.015, 0.5){\normalsize{\textcolor{black}{$(a)$}}}
    \put(0.08, 0.35){\fbox{$J_{0}=0,\delta J=1$}}
    \put(0.333, 0.27){\includegraphics[width=0.33\linewidth]{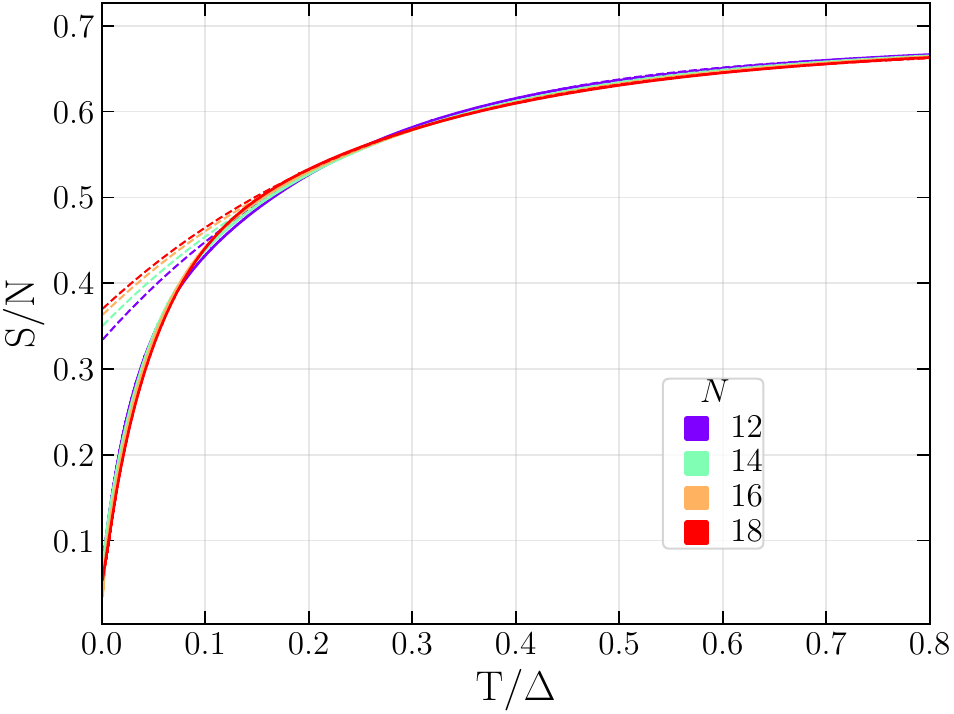}}
    \put(0.326, 0.5){\normalsize{\textcolor{black}{$(b)$}}}
    \put(0.408, 0.35){\fbox{$J_{0}=3,\delta J=1$}}
    \put(0.669, 0.27){\includegraphics[width=0.33\linewidth]{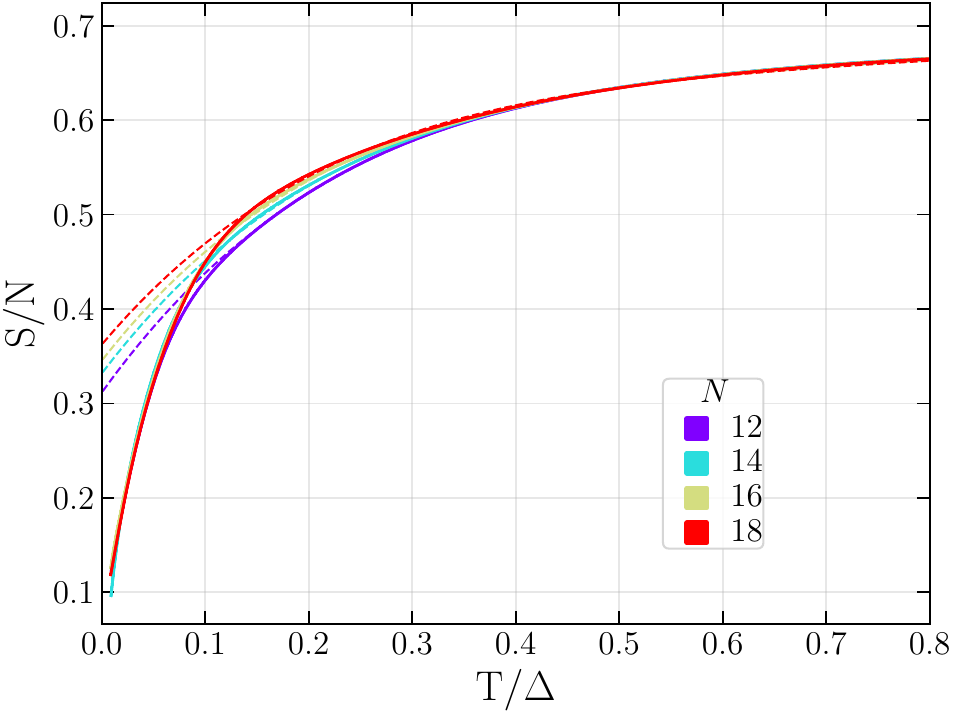}}
    \put(0.661, 0.5){\normalsize{\textcolor{black}{$(c)$}}}
    \put(0.728, 0.35){\fbox{$J_{0}=1,\delta J=0$}}
    \put(-0.005, 0.01){\includegraphics[width=0.33\linewidth]{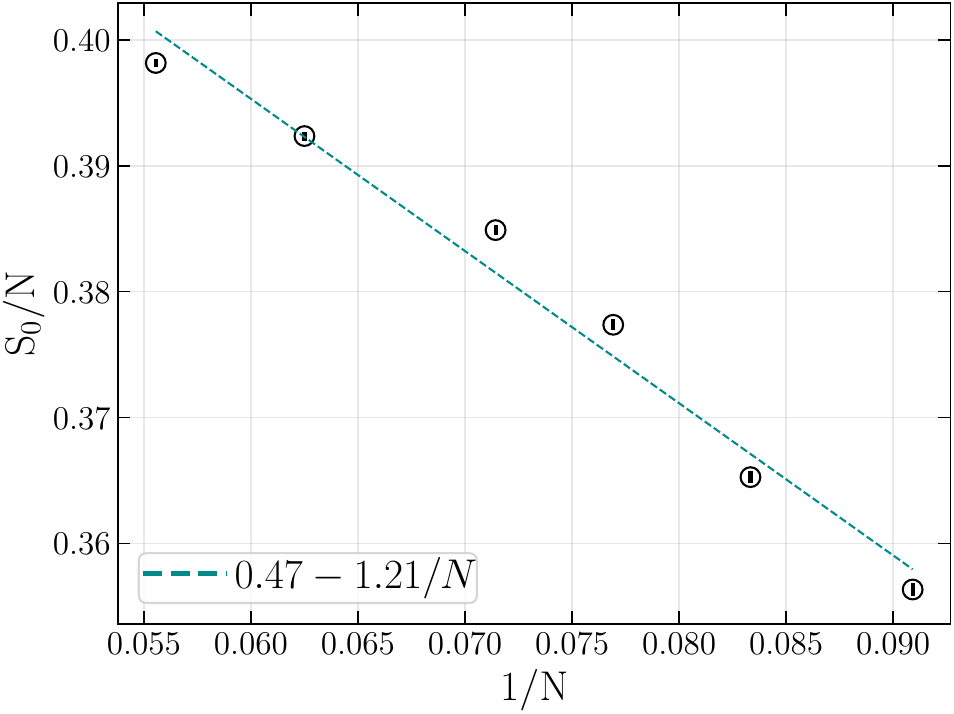}}
    \put(-0.015, 0.235){\normalsize{\textcolor{black}{$(d)$}}}
    \put(0.335, 0.01){\includegraphics[width=0.33\linewidth]{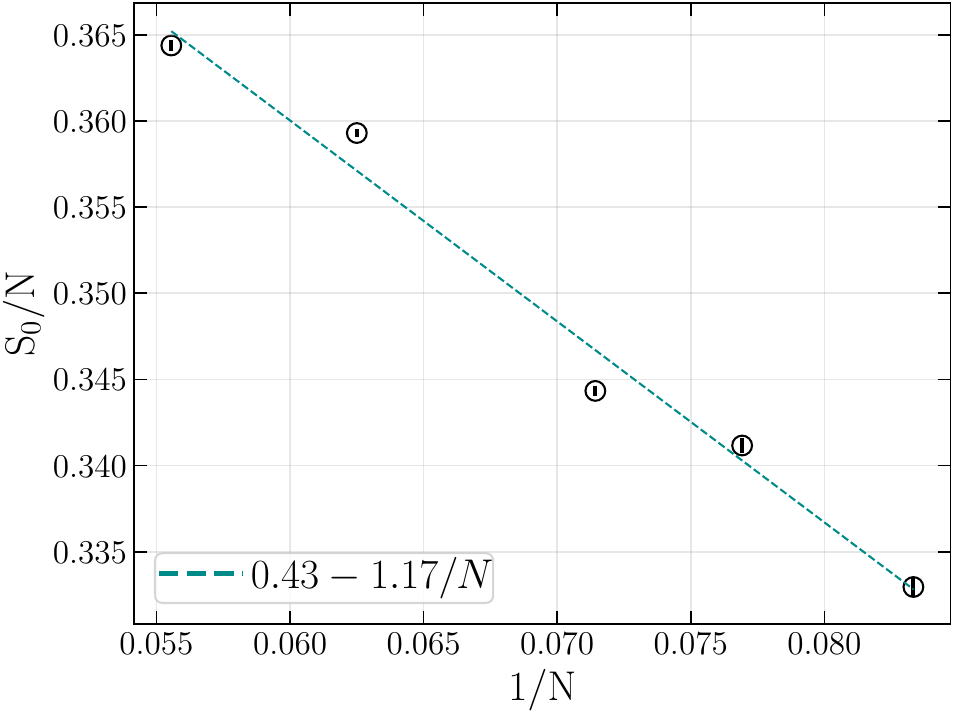}}
    \put(0.328, 0.235){\normalsize{\textcolor{black}{$(e)$}}}
    \put(0.668, 0.01){\includegraphics[width=0.33\linewidth]{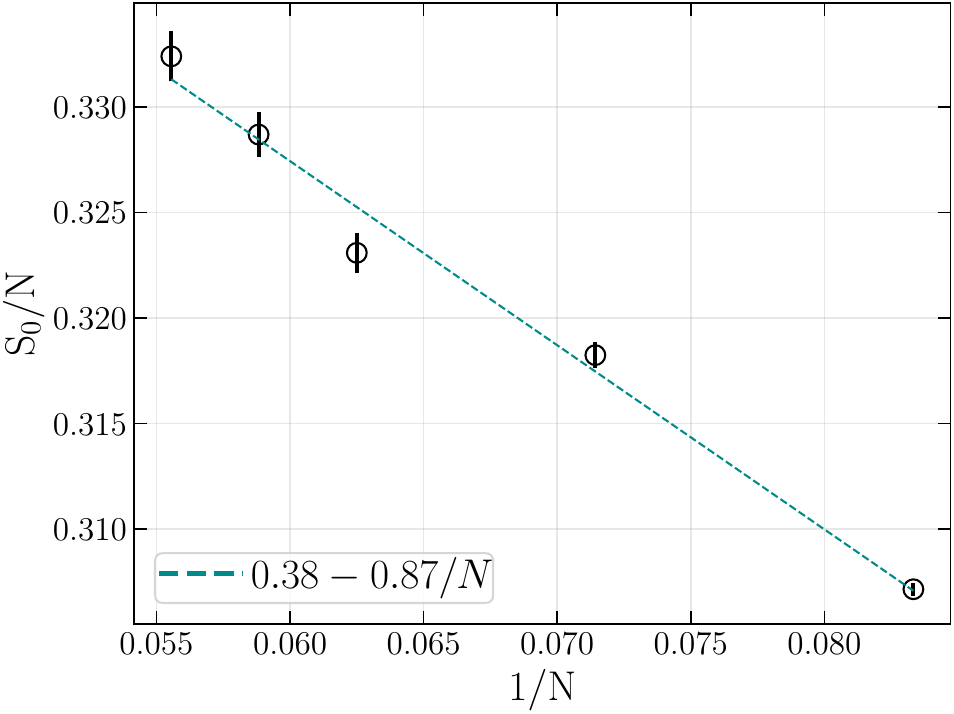}}
    \put(0.665, 0.235){\normalsize{\textcolor{black}{$(f)$}}}
\end{picture}
\vspace{-18pt}
\caption{(a), (b) and (c) are plots for the Entropy per site, $S/N$ as a function of $T$, with varying values of $N$.  (a) corresponds to the zero-mean ($J_0=0, \delta J=1$) case, (b) to the case with ($J_0=3, \delta J=1$), and (c) to the case with ($J_0=1,\delta  J=0$). 
The dashed lines in these plots  indicate a polynomial extrapolation to zero $T$ to obtain the residual entropy density $s_0=S_0(N)/N$ (see main text).  (d) gives a plot of 
$S_0(N)/N$ vs  $1/N$, for the  ($J_0=0, \delta J=1$) case to obtain $s_{0}\approx 0.47$ in the thermodynamic limit. Similarly  (e) and (f) give plots for $S_0(N)/N$ vs  $1/N$, in  the ($J_0=3, \delta J=1$), and
($J_0=1,\delta  J=0$) cases, to obtain $s_{0}\approx 0.43$ and $s_0\approx 0.38$ respectively, in the thermodynamic limit. }
\label{SP_heat}
\end{figure*}
\section{\label{RES_ENT} Residual Entropy at $T=0$}
An unusual feature of the zero-mean SYK model ($J_0=0$) is the zero-temperature residual entropy per site $S_0/N\neq 0$, obtained by taking $T\to 0$ limit after taking the $N\to\infty$, i.e., the large-$N$ limit. This is a key feature in the correspondence between the SYK model and extremal black holes \cite{kitaev2015simple,Gu:2019jub,Maldacena:2016hyu}. The residual entropy implies dense many-body eigenspectrum near the ground state of the SYK model with exponentially small level spacing $\delta(N) \sim \exp{(-Ns_0)}$ for large $N$, where $S_0=Ns_0$ and $s_0$ is the residual entropy density (per site). This distinguishes SYK model from standard quantum many-body models, where many-body level spacing near ground state is of $\mathcal{O}(1)$ or decays as some power of $N$, unlike the exponential decay with $N$ in the SYK model. Here we ask whether the residual entropy persists even for non-zero mean ($J_0\neq 0$). A non-zero $T=0$ residual entropy may imply that the SYK-black hole correspondence continues to hold even away from the large-$N$ solvable limit of $J_0=0$.

In our simulations for this section we only consider the $\mu=0$ case, and consider all $\mathcal{Q}$ charge sectors for the particular $N$ value. In general we consider at least $1000$ disorder realization for $N=12$, $300$ for $N=14$, $100$ for $N=16$ and $5$ for $N=18$.

\medskip

\noindent{\textcolor{blue}{\it { Residual entropy from the extrapolation of thermodynamic entropy.}}} A residual entropy of $S_0=0.46$ has been estimated for the zero-mean SYK model from large-$N$ analytical calculation ~\cite{PhysRevX.5.041025,Gu:2019jub,PhysRevB.94.035135,Maldacena:2015waa,Banerjee:2016ncu}. However, numerical ED calculations, such as ours, can only approach the large-$N$ limit through a different order-of-limit, where $T\to 0$ limit is taken for finite $N$ and $N$ is increased up to system sizes accessible in ED. The thermodynamic entropy per site $S(T)/N$ for any finite $N$ always approaches zero as $T\to 0$. Thus, to estimate $S_0/N$ from such finite-$N$ calculations, one needs to carefully extrapolate $S(T)/N$ from a \emph{suitable} range of temperatures at finite $T$ to $T=0$, as was done in previous numerical studies ~\cite{Gu:2019jub,PhysRevB.94.035135}. A residual entropy of $S_0/N\simeq 0.46$ was estimated in these studies, consistent with the analytical estimate. 
We use extrapolation procedure similar to Refs.~\cite{Gu:2019jub,PhysRevB.94.035135} to estimate $S_0(J_0)$.

To this end, we compute the partition function $Z(T)=\sum_m \exp{(-E_m/T)}$ using the many-body energy eigenvalues from ED for a given disorder realization. The entropy is then extracted directly from
\begin{align}
S(T)&=\left(1+T\frac{\partial}{\partial T}\right)\ln{Z(T)}.
\label{ENTRPY_DEF}
\end{align}
for each realization and then $S(T)$ obtained by averaging over $1000$ disorder realizations. 
To extrapolate $S(T)$ to $T\to 0$ at a finite $N$, we first fix the temperature range of extrapolation to be higher than a temperature $T_l$ for the zero-mean case ($J_0=0$) and fit $S(N,T>T_l)$ with a high-order polynomial of $T$, $S_{pol}(N,T)$.
The polynomial is then extrapolated to $T=0$ to obtain $S_0(N)=S_{pol}(N,T\to 0)$. This extrapolation is shown as  dashed lines, for the different values of $N$, in fig.\ref{SP_heat}(a). 

The $S_0(N)$ extracted in this manner, is plotted in fig.\ref{SP_heat}(d), as a function of $N$, and  then further extrapolated to $1/N\to 0$ to estimate the residual  entropy density, $s_0$. We obtain $s_0\approx 0.47$ with a choice of $T_l=0.4$. 

\begin{figure}[h] 
\includegraphics[width=\linewidth]{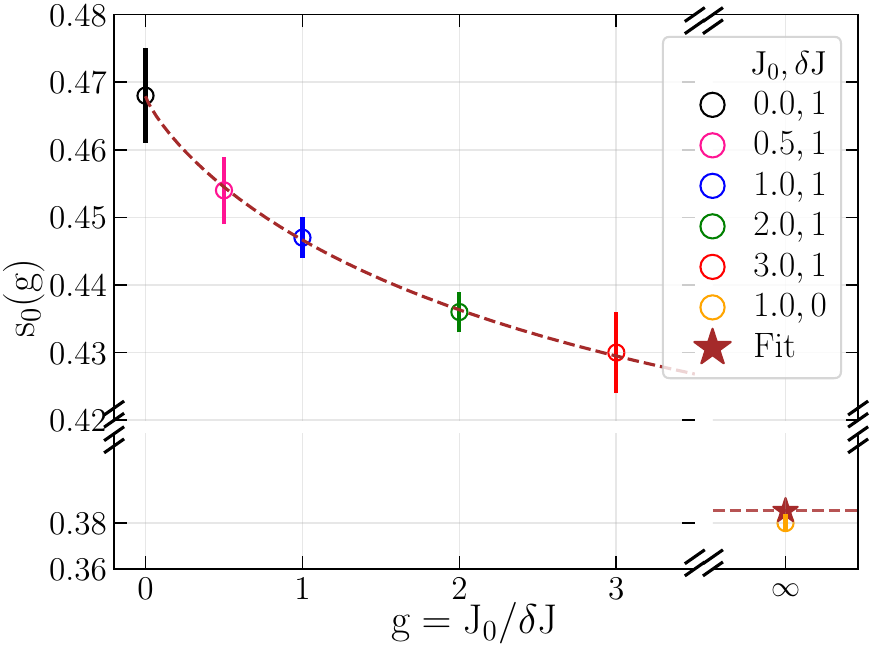}
\caption{The residual entropy for $N\to\infty$,  $s_{0}(g)$,
is  plotted for different values of $g$ and fitted to function  $y(g) = A + \frac{B}{1 + (g/\tau)^\alpha}$  $( A=0.38, B=0.08, \tau=3.53, \alpha=0.83)$, with error bars. We see that there is a 
 finite residual entropy even in the clean limit, $g=\infty$, with the extrapolated value agreeing well with the direct calculation.} 
 \label{RES_ENT_FIT_SYK}
\end{figure}
Next, as demonstrated in figs.\ref{SP_heat}(b), (c), and (e),(f), we repeat this procedure for two other values, ($J_0=3,\delta J=1$) and ($J_0=1,\delta J=0$) respectively. We take the temperature range to be  $T>T_l/\Delta(J_0)$, i.e., appropriately scaled with normalized many-body bandwidth $\Delta(J_0)$ [eq.\eqref{DELTAJ}] with respect to the zero-mean case. 
In the case ($J_0=3,\delta J=1$) we obtain $s_0\approx 0.43$, which is somewhat reduced compared to the value with vanishing mean. Interestingly, even in the case ($J_0=1, \delta J=0$) where there is no disorder, we obtain a non-vanishing value for $s_0\approx 0.38$.


The result that the residual entropy density is non-vanishing, even in the case with no disorder, is quite non-trivial and merits further investigation. We will describe three different types of analysis we have carried out for this purpose. 
All three  give  results in good agreement,  in  particular for   the value of $s_0$ without disorder ( $g=\infty$),  that we have obtained above.

In the rest of this section we describe these three methods of analysis. 
\begin{figure}[h]
  \setlength{\unitlength}{1pt}
  \begin{picture}(\linewidth, 340)
    \put(0,0){\includegraphics[width=0.9\columnwidth]{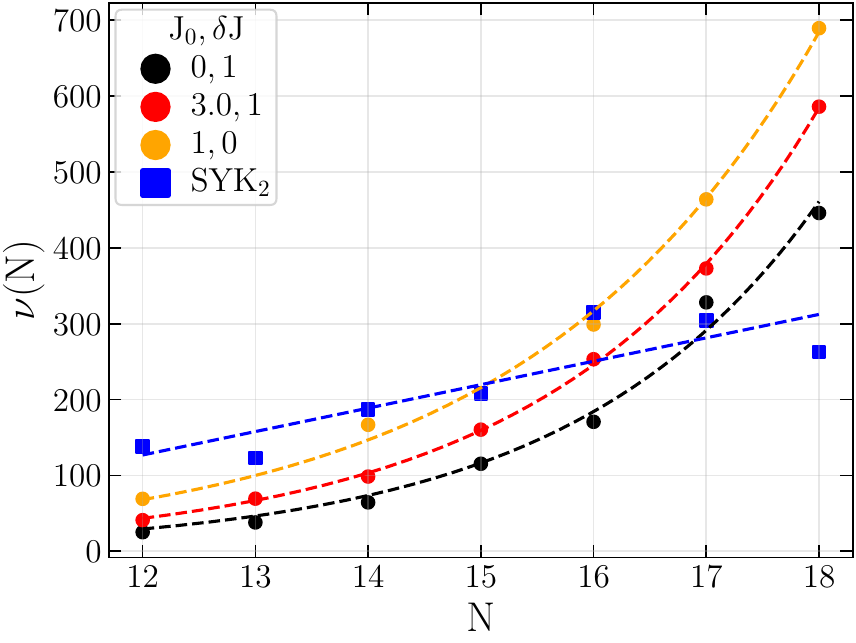}}
    \put(0,155){\normalsize{\textcolor{black}{$(b)$}}}
    \put(0,170){\includegraphics[width=0.9\columnwidth]{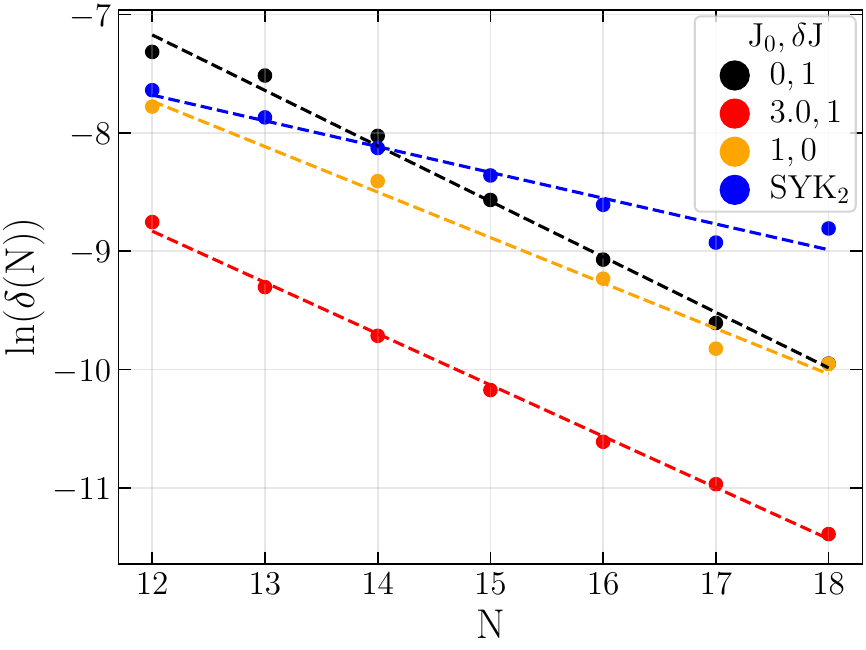}}
    \put(0,325){\normalsize{\textcolor{black}{$(a)$}}}

\end{picture}
\vspace*{-0.15in}
\caption{(a) Disorder-averaged mean level spacing 
$\delta(N)$ from the low-energy eigen spectrum, within an energy window near the ground state, as a function of system size $N$ (see main text). The dashed lines are fits to the data. Different colors denote different values of $(J_0,\delta J)$. 
The blue dots and curve refer to the  $\mathrm{SYK}_2$ model,  after dividing $\delta(N)$ by a factor of $30$. 
(b) The total number of states $\nu(N)$ within the low-energy window above the ground state. The dashed lines are exponential fits. The different colors refer to   different values of $(J_0,\delta J)$. The blue dots and curve refer to the $\mathrm{SYK}_2$ model, after rescaling by a factor of $30$.
}
\label{SP_heat_0}
\end{figure}

I) \noindent{\textcolor{blue}{\it {Additional data, and extrapolation.}}} First, we  obtained additional data for the 
residual entropy density $s_0(g)$ at other values of the coupling $g$, as shown in fig.\ref{RES_ENT_FIT_SYK}, following the same method described above, with $T>T_l/ \Delta$.
Evidently, $s_0(g)$ decreases monotonically with $g$.
The five values obtained for $g\ne \infty$ are  fitted with a  function $y(g)$ as described in the figure caption.  The error bars to the fit are also shown, we see that a good fit is obtained. 
Further we used  this   function to  extrapolate to $g\rightarrow \infty$ giving  rise to the value for $s_0(\infty)$ marked as the star in fig.\ref{RES_ENT_FIT_SYK}. We see that this extrapolated value is close to $s_0=0.38$ which was obtained above in fig.\ref{SP_heat}(f) by direct calculation for  the $g= \infty$ case. For three values of $(J_0,\delta J)$ the resulting values of $s_0$, obtained from this method, are given in the third column in table \ref{tab:S0_LevelSpacing}. 


The other two methods are related to  studying the spacing of low-energy states in the system and its limit as  $N\rightarrow \infty$.


\medskip


II) \noindent{\textcolor{blue}{\it {Level spacing near the ground state.}}} In this analysis we obtain the disorder averaged mean  spacing $\delta(N)$, as a function of $N$, for   the low-energy states with energies $E<E_w$. Here the energy window, $E_w$, is given by 
\begin{equation}
E_w=\langle E\rangle-\sqrt{\langle \Delta E^{2}\rangle}
\label{Ewdef}
\end{equation}
with  $\langle E\rangle$ being the expectation value of the energy at the rescaled temperature $T_l/\Delta$, that was used in fig.\ref{SP_heat}. And  $\sqrt{\langle \Delta E^{2}\rangle}$ being   the root mean square of the energy  fluctuation at $T_l/\Delta$.

 The results for $\ln{\delta}$ vs. $N$ are shown in fig.\ref{SP_heat_0} (a) for the three cases, $(J_0=0,\delta J=1)$, $(J_0=3, \delta J=1)$, and  in the clean limit ($J_0=1$, $\delta J=0$). 
 We see that a good linear fit is obtained and from the slope find the value of $s_0$ in the three cases which are shown in Column 4 of table \ref{tab:S0_LevelSpacing}. 
 As a check we also calculate the disorder average mean spacing for the $\mathrm{SYK}_2$ model with the Hamiltonian 
 \be
\displaystyle
\mathcal{H}=\frac{1}{(2N)^{1/2}}\sum_{i,j}J_{i,j}c_{i}^{\dagger}c_{j}.
\label{SYK2_def}
\ee
where $J_{ij}$ are Gaussian random couplings with zero mean. 
These results are also plotted in fig.\ref{SP_heat_0} (a). In this case, in contrast to our model, and in accord with expectations, the decrease of $\delta(N)$ is consistent with a power-law decay with $N$ implying the absence of residual entropy, $s_0$.

III)\noindent{\textcolor{blue}{\it {Number of states near the ground state.}}} The third method of analysis involved counting the number of states $\nu(N)$ with energy $E\le E_w$, where $E_w$ was defined in eq. \ref{Ewdef} above. This method is closely related to II) above, and in the absence of disorder averaging would have yielded the same result for $s_0$.

The results for the three cases mentioned above are plotted as a function of $N$ in fig.\ref{SP_heat_0}(b). We see that the data fits the  exponential form well in all three cases. From the exponent a value of $s_0(N)$ can be obtained, and these are also summarised in Column 5 of table \ref{tab:S0_LevelSpacing}, and agree well with the values obtained in II). In addition $\nu(N)$ for the $\mathrm{SYK}_2$ model is also plotted in fig.\ref{SP_heat_0}(b). In this case the data  fits well  with a linearly increasing function of $N$. This implies that the residual entropy for the $SYK_2$ model vanishes as $N\rightarrow \infty$, as expected.

\begin{table}[H]
\centering
\begin{tabular}{|c|c|c|c|c|}
       \hline
        $J_{0}$   &  $\delta J$& \begin{tabular}[c]{@{}c@{}}$s_{0}$ from \\[-0.6ex] thermodynamic entropy\end{tabular}& $s_0$ from $\delta(N)$ & $s_0$ from $\nu(N)$ \\
        \hline
        0   & 1 & $0.47\pm 0.007$& $0.47\pm 0.02$ & $0.46\pm0.03$ \\
        \hline
        3   & 1& $0.43\pm 0.003$ & $0.43\pm0.01$ & $0.43\pm0.01$  \\
        \hline
        1   & 0& $0.38\pm 0.004$ &$0.38\pm0.03$ & $0.38\pm0.01$\\
        \hline
\end{tabular}
\caption{The residual entropy density $s_0=S_0/N$ extracted from $\delta(N)$ and $\nu(N)$ in fig.\ref{SP_heat_0}.}
\label{tab:S0_LevelSpacing}
\end{table}
We see from table \ref{tab:S0_LevelSpacing} that the  results for $s_0$ obtained from the three methods described above agree, within error bars.  
This agreement  strongly suggests  that at $T=0$ a non-zero residual entropy $s_0$ is present, even in the model with no randomness, i.e. with  $\delta J=0$.

Note, as was discussed earlier,  that without disorder the OTOCs are non-chaotic and do not exhibit scrambling behaviour, despite the presence of this non-zero $s_0$. 

We close this section with one final comment. Although we used several different methods, one might be  worried about the dependence of our results on the value of $T_l$  we chose(see paragraph after eq. \ref{ENTRPY_DEF} and also related definition of $E_w$, eq. \ref{Ewdef}). In particular, one might worry about whether our conclusions could  change if $T_l$   depends on $N$. To address this worry we have carried out one further  type of  analysis. We took, for the values $(J_0,\delta J)$ in table \ref{tab:S0_LevelSpacing}, an energy  window $E_w$ which is 3\% of the total energy band width above the ground state, i.e. 
\begin{equation}\label{defnew ew}
E_w=E_{min}+ {3\over 100} (E_{max}-E_{min})
\end{equation}
where $E_{max},E_{min}$ are the maximum and minimum values of the energy. Then we calculated in each case the mean spacing between these states and repeated the same analysis for $\delta(N)$, as a function of $N$, as in method II) to obtain the residual entropy. This is described in appendix \ref{app:ENTRP_EXTRA} and gives values for $s_0$, see table \ref{tab:S1_LevelSpacing},  which also agree, within errors with those in table \ref{tab:S0_LevelSpacing}. 






\begin{figure}[h]
\setlength{\unitlength}{1pt}
\begin{picture}(\linewidth,380)
    \put(0,0){\includegraphics[width=\columnwidth]{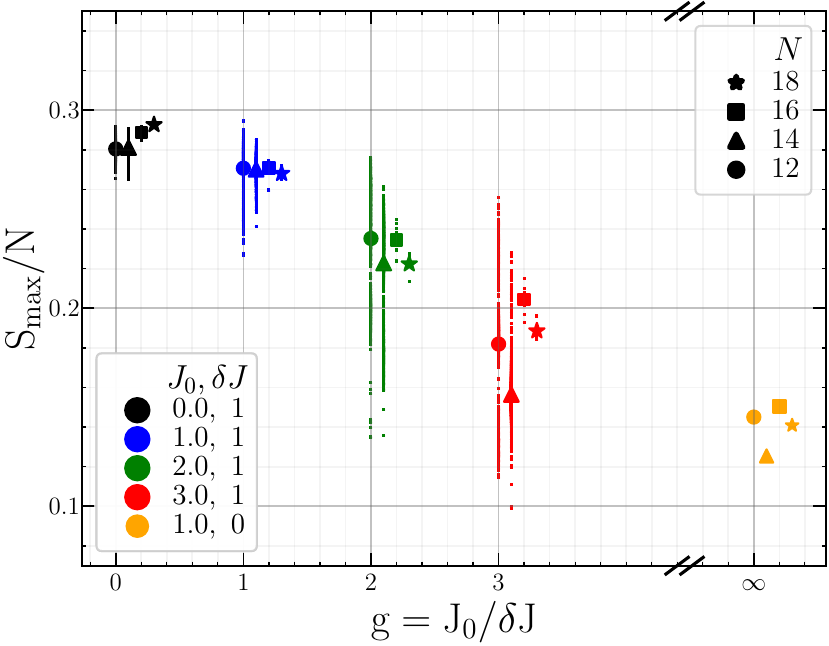}}
    \put(-3,175){\large{\textcolor{black}{$(b)$}}}

    \put(-4,190){\includegraphics[width=\linewidth]{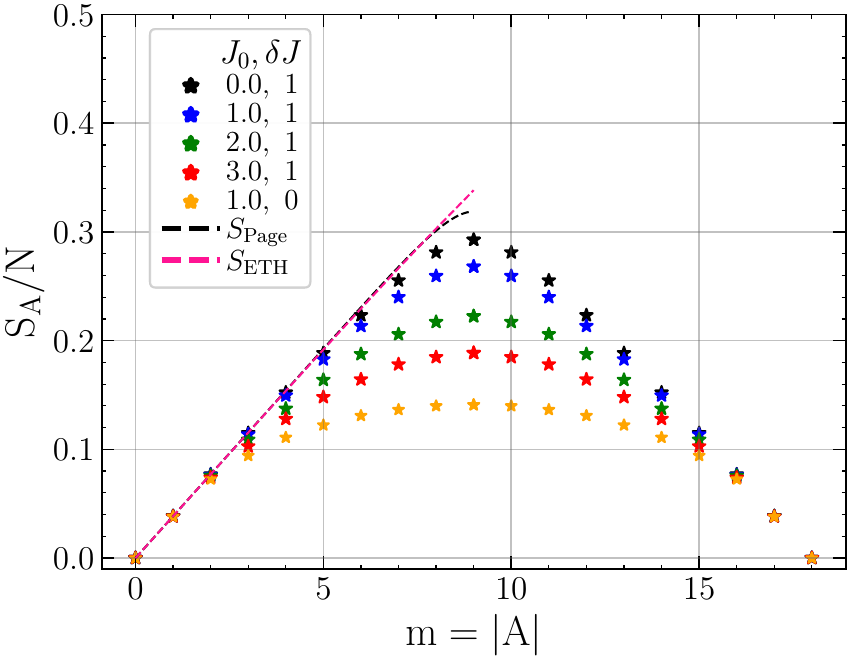}}
    \put(-3,365){\large{\textcolor{black}{$(a)$}}}

\end{picture}
\caption{(a) Ground-state von Neumann entanglement entropy (EE) per site versus subsystem size $m=|A|$ for $N=18$ across different combinations of the mean $J_{0}$ and standard deviation $\delta J$. The parameters are $(J_{0}, \delta J) = (0,1)$ (black), $(1,1)$ (blue), $(2,1)$ (green), $(0,1)$ (red), and $(1,0)$ (orange). The black dotted line indicates the Page value [eq.\eqref{eq:EE_Page}], and the pink dashed line represents the ETH prediction for the EE [eq.\eqref{eq:ETH_EE}]. In all the cases, we observe a gradual decrease in EE as the mean coupling shifts from zero to finite values. (b) Maximum entanglement entropy at $m=N/2$ for various system sizes ($N=12,14,16,18$) and $(J_{0},\delta J)$ combinations, plotted as a function of $g=J_0/\delta J$. The results exhibit finite-$N$ effects, primarily due to the smaller ensemble sizes used for averaging at larger $N$.  
}
\label{SYK_EE_GR}
\end{figure}

\section{ Ground-State Entanglement}\label{GS_EE}
Entanglement entropy, that quantifies quantum entanglement among degrees of freedom, e.g., subsystems, is one of the prominent probes of the quantum correlations for pure quantum many-body states. Given the strongly interacting nature of the SYK model, one expects the NFL ground state to be strongly entangled. Moreover, due to all-to-all interactions in the SYK-type models, the entanglement entropy is expected to grow linearly with number of sites $m$ in the subsystem, i.e., obey a \emph{volume-law} of entanglement. There have been several numerical and analytical studies \cite{PhysRevB.94.035135,Liu2018,PhysRevD.100.041901,Zhang2020,ZhangLiu2020,Haldar2020,Zhang:2022yaw} of entanglement in SYK model in various equilibrium and non-equilibrium situations. Indeed, the NFL ground state in the SYK model has been found to have almost close to the maximal entanglement \cite{PhysRevB.94.035135,Liu2018,PhysRevD.100.041901,Zhang2020,ZhangLiu2020,Haldar2020,Zhang:2022yaw}, i.e., the Page value \cite{PAGE_PAPER} corresponding to the entanglement entropy of completely random states. Here we study how the entanglement entropy of the ground state of the SYK model is affected by the presence of a finite mean.

We compute von Neumann entanglement entropy of a subsystem $A$, consisting of $m\leq N$ sites out of total $N$ sites, in the ground state $|\Psi_g\rangle$ of the SYK model, i.e.,
\begin{align}
S_A=-\mathrm{Tr}_A[\rho_A \ln \rho_A],
\end{align}
where the reduced density matrix $\rho_A$ of the subsystem is given by $\rho_A=\mathrm{Tr}_B[|\Psi_g\rangle\langle \Psi_g|]$, where $B$ denotes the rest of the subsystem with $N-m$ sites. Here the ground state is obtained from the lowest energy eigenstate over all the charge sectors. For all our simulations we set $\mu=0$, and consider all charge sectors to determine the lowest energy state.

In fig.\ref{SYK_EE_GR}(a), we plot the ground-state EE (per site) of a system of size $N=18$ for SYK model with different mean $J_0$, including the clean limit with $J_0=1, \delta J=0$. For comparison, we also plot the Page value,
\begin{align}
S_\mathrm{Page}=m\ln 2-2^{2m-N-1},  \label{eq:EE_Page}
\end{align}
corresponding to subsystem and full-system Hilbert-space dimensions $2^m$ and $2^N$, where $m\leq N/2$. In fig.\ref{SYK_EE_GR}(a), we also compare our results with a bound on EE \cite{PhysRevD.100.041901}, henceforth called the \emph{ETH estimate}. The latter is based on the application of Eigenstate Thermalization Hypothesis (ETH) for low-energy eigenstates, including the ground-state, of the zero-mean SYK model with Majorana fermions. The ETH bound on EE  of the $A$ subsystem with $m$ sites for eigenstate with energy $E$ is given by,
\begin{align}
S_A&\leq S_\mathrm{ETH}=\ln{\rho_m\left[(m/N)^{3/2}(E/N)\right]}  \nonumber\\
&\simeq m\left[\ln{2}-\frac{1}{8}\mathrm{arcsin}^2\left[\left(\frac{m}{N}\right)^{3/2}\frac{E}{E_0}\right]\right].\label{eq:ETH_EE}
\end{align}
The above estimate is valid for large $m$ and $N$ with a finite $m/N$ ratio. We have appropriately adopted the expressions of Ref.\onlinecite{PhysRevD.100.041901}  by translating the number Majorana fermions, $2m,2N\to m,N$, number of complex fermions. Here $\rho_m(\epsilon)$ is the many-body density of states at energy density (energy per site) $\epsilon$ for SYK model with $m$ sites. Due to all-to-all nature of SYK couplings, the energy density $\epsilon$ of the subsystem with $m$ sites is related to the energy density $E/N$ of the SYK model with $N$ sites via $\epsilon=(m/N)^{3/2}(E/N)$. In the second line of eq.\eqref{eq:ETH_EE}, the many-body density of states \cite{Maldacena:2016hyu,GarciaSPHEAT,RevModPhys.94.035004} estimated from the low-energy Schwarzian action has been used. For purposes of comparison with fig.\ref{SYK_EE_GR} we will set $E=E_0$ in eq.(\ref{eq:ETH_EE}). 

We see from  fig.\ref{SYK_EE_GR}(a), that $S_\mathrm{Page}$ is effectively the same as $S_\mathrm{ETH}$, except very close to $m=N/2$. We also see that the numerically computed EE for the SYK model approaches the Page value for small subsystem sizes $m$ for all $(J_0,\delta J)$. For the zero-mean SYK model, the EE falls perceptibly below the Page/ETH value only for $m\simeq N/2$, when two subsystems are of comparable sizes. However, for the finite-mean cases the deviation from the Page value becomes progressively pronounced with increasing $g=J_0/\delta J$, indicating a significant suppression of entanglement relative to both the Page/ETH value and the zero-mean SYK model.

In fig.\ref{SYK_EE_GR} (b)  we show the maximum EE value at $m=N/2$ for $N=12,14,16,18$ and different combinations of mean $J_{0}$ and standard deviation $\delta J$. It shows that as we increase mean-to-standard deviation ratio $g$ from $0\to\infty$, the maximal EE value decreases. Evidently, the minimal EE is attained in the clean limit $J_0=1,\delta J=0$.

\section{Discussion and Implications}\label{discupaper}
In this paper we have studied a deformed version of the SYK model where the quartic couplings are  Gaussian with   a non-zero mean, $J_0$, and standard deviation, $\delta J$. This gives rise to a dimensionless ratio, $g={J_0\over \delta J}$, which can be varied continuously. We  analysed the behaviour of the system for varying values of  $g$, going from the conventional SYK model at $g=0$, to the clean case with no randomness at $g=\infty$.


 In particular, we examined the chaotic behaviour using three different diagnostics of chaos, which probe three different time scales: level spacing statisics  (late time), spectral form factor (intermediate time), and OTOCs/Lyapunov exponent (early time). 
 We find that these decouple as $g$ increases.  

One of our key results   is that the system retains its fast scrambling character even at finite mean, though with a reduced value of the Lyapunov exponent. This provides a rare example of a quantum chaotic system where the strength of chaos can be smoothly varied  without destroying the overall scrambling nature.  The reduction in the Lyapunov coefficient, and the shortening of the ramp in the spectral form factor, reflect a decrease in the degree of chaos at intermediate times. In contrast, the late-time behavior—captured by the level spacing statistics—remains unaffected, and continues to be of RMT type, suggesting that the long-time chaotic properties are robust against the presence of a non-zero mean.

 

Another key result  pertains to the residual ground state entropy, $s_0$. A careful analysis, which we carried out in different ways,
allowed us  to extrapolate the behaviour of our system to $N\rightarrow \infty$. This revealed that   while $s_0$ decreases as $g$ increases, it does not vanish, and has a non-zero value even in the clean case, $g\rightarrow \infty$.

The two key results mentioned above show that  a finite residual entropy does not necessarily imply maximal chaos. This is  contrary to the implicit assumptions in the  holography and condensed matter literature, suggesting that the two are  tied together.
Our analysis also reveals that the subsystem entanglement reduces as $g$ increases, resulting in a growing departure  from the Page curve.

Previous efforts to tune chaos in the SYK model have often involved adding a solvable two-fermion random all-to-all hopping term  to the fundamental four-fermion interactions \cite{Garcia-Garcia:2017bkg}. Any finite two-fermion hopping immediately destroys the SYK NFL state at low temperature and leads to a slow scrambling Fermi liquid (FL) state \cite{Banerjee:2016ncu,Kim2021comment} with $\lambda_\mr{L}\sim T^2$ for $N\to 0$. However, the system can become non-chaotic with $\lambda_\mr{L}=0$ \cite{Garcia-Garcia:2017bkg}, or even undergo many-body localization (MBL) \cite{Altshuler1997} at finite $N$ for sufficiently large strength of the two-fermion hopping $\sim N^2$ \cite{Micklitz2019,Garcia2021}. Similarly, a slow scrambling phase with activated or exponentially suppressed Lyapunov exponent at low temperature can be realized at a non-zero chemical potential, in addition to the random hopping \cite{Samui:2020jli, Sorokhaibam:2019qho}. One can also obtain quantum phase transitions from fastest or fast scrambling NFL phases to slow scrambling FL phases by introducing auxiliary fermions \cite{Banerjee:2016ncu} or by varying the rank of the four-fermion SYK coupling \cite{Kim2020} or by tuning non-interacting band structure in higher dimensions \cite{Haldar2018a}. For example, in Ref.\onlinecite{Banerjee:2016ncu}, a transition between a NFL phase with $\lambda_\mr{L}=2\pi T$ to FL phase with $\lambda_\mr{L}\sim T^2$ is tuned through ratio of the number of flavors of the auxiliary and the SYK fermions. Nevertheless, these models do not consider the consequences of more natural tuning, generic in disordered systems, namely the variation of the randomness of the fundamental SYK four-fermion coupling from the fully random to \emph{ordered} or uniform limit, as done in this work by tuning the mean to standard deviation ratio from $g=0$ to $g\to \infty$.

As was mentioned in the introduction, the SYK model- and related topics-  are  fertile grounds for cross-connections between  the fields of condensed matter physics, quantum information theory, and quantum gravity. 
Broadly speaking, our analysis shows that  different measures of chaos can be independently tuned. This tunability introduces a rich structure not present in the zero-mean case, offering both theoretical  and experimental opportunities for further investigation which would be of interest from these perspectives.  
For example, our findings are likely to have  implications for quantum information dynamics in disordered systems and for constructing tunable chaotic quantum simulators.

From the perspective of quantum gravity, it is now understood that entanglement can give rise to a wormhole geometry, which can sometimes allow communication between its two ends \cite{Gao:2016bin,YANG_MALDACENA}. This process is a version of quantum teleportation, but as becomes evident in the gravitational description due to its geometric nature, it involves a  new type of collective mode which allows for an efficient transfer of information \cite{PRXQuantum.4.010320,PRXQuantum.4.010321}. A tunable model of chaos, and  varying entanglement, such as the one studied here, is particularly interesting in this context, since one can examine when this kind of behaviour arises, and whether it turns off, signaling the closing of the wormhole, as the randomness is decreased.




Most of our analysis was numerical in nature. Several checks were carried out   against known analytical results and found to be in agreement. 
For the complex SYK model, the Hilbert space grows exponentially with fermion flavor number $N$, limiting standard simulations to $N = 16$. By exploiting charge conservation and particle-hole symmetry to block-diagonalize the Hamiltonian, we were  able to push the simulations up to $N = 26$ for smaller charge sectors, though full-spectrum exact diagonalization was  feasible up to $N = 18$ — the current computational frontier. Importantly, by extrapolating to large $N$, we are able to benchmark against established analytical results and confirm a residual entropy of $S_0/N \approx 0.46–0.47$, in the $g=0$ case, correct level spacing statistics, and accurate spectral form factor values, while OTOCs computed via matrix-free methods successfully reproduce the expected Lyapunov exponent $\lambda  = 2\pi/\beta$, for $g=0$. We hope these studies will motivate both analytical results in specific regimes and numerical studies on larger systems.

In a companion paper, now under preparation, \cite{MukherjeeFuture}, we  examine, using more analytical techniques, to understand some limits of the behaviour of the system studied here. In some cases, for suitable couplings,  the analytical approach turns out to be quite tractable,  allowing us to start from a precursor of chaos, in the completely non-chaotic regime and study the growing chaotic nature of the system, and changes in residual entropy.

\begin{acknowledgments}
We are grateful to Vatsal for collaboration at an early stage in  this study. 
A.M and N.T. acknowledge support from NSF-DMR 2138905. Computations were performed at
the Unity cluster of Arts and Science College, Ohio State University. SB acknowledges support from ANRF, DST, India (File No. ANRF/ARG/2025/004045/PS). SPT acknowledge support from Government of India, Department of Atomic Energy, under Project Identification No. RTI 4002 and from the Quantum Space- Time Endowment of the Infosys Science Foundation
\end{acknowledgments}
\bibliography{ALL_REFERENCES}

\providecommand{\noopsort}[1]{}\providecommand{\singleletter}[1]{#1}%
\begin{thebibliography}{73}%
\makeatletter
\providecommand \@ifxundefined [1]{%
 \@ifx{#1\undefined}
}%
\providecommand \@ifnum [1]{%
 \ifnum #1\expandafter \@firstoftwo
 \else \expandafter \@secondoftwo
 \fi
}%
\providecommand \@ifx [1]{%
 \ifx #1\expandafter \@firstoftwo
 \else \expandafter \@secondoftwo
 \fi
}%
\providecommand \natexlab [1]{#1}%
\providecommand \enquote  [1]{``#1''}%
\providecommand \bibnamefont  [1]{#1}%
\providecommand \bibfnamefont [1]{#1}%
\providecommand \citenamefont [1]{#1}%
\providecommand \href@noop [0]{\@secondoftwo}%
\providecommand \href [0]{\begingroup \@sanitize@url \@href}%
\providecommand \@href[1]{\@@startlink{#1}\@@href}%
\providecommand \@@href[1]{\endgroup#1\@@endlink}%
\providecommand \@sanitize@url [0]{\catcode `\\12\catcode `\$12\catcode
  `\&12\catcode `\#12\catcode `\^12\catcode `\_12\catcode `\%12\relax}%
\providecommand \@@startlink[1]{}%
\providecommand \@@endlink[0]{}%
\providecommand \url  [0]{\begingroup\@sanitize@url \@url }%
\providecommand \@url [1]{\endgroup\@href {#1}{\urlprefix }}%
\providecommand \urlprefix  [0]{URL }%
\providecommand \Eprint [0]{\href }%
\providecommand \doibase [0]{https://doi.org/}%
\providecommand \selectlanguage [0]{\@gobble}%
\providecommand \bibinfo  [0]{\@secondoftwo}%
\providecommand \bibfield  [0]{\@secondoftwo}%
\providecommand \translation [1]{[#1]}%
\providecommand \BibitemOpen [0]{}%
\providecommand \bibitemStop [0]{}%
\providecommand \bibitemNoStop [0]{.\EOS\space}%
\providecommand \EOS [0]{\spacefactor3000\relax}%
\providecommand \BibitemShut  [1]{\csname bibitem#1\endcsname}%
\let\auto@bib@innerbib\@empty
\bibitem [{\citenamefont {Kitaev}(2015)}]{kitaev2015simple}%
  \BibitemOpen
  \bibfield  {author} {\bibinfo {author} {\bibfnamefont {A.}~\bibnamefont
  {Kitaev}},\ }\href@noop {} {\bibinfo {title} {A simple model of quantum
  holography}},\ \bibinfo {howpublished} {Lecture, KITP Program: Entanglement
  in Strongly-Correlated Quantum Matter, Santa Barbara} (\bibinfo {year}
  {2015}),\ \bibinfo {note} {2015}\BibitemShut {NoStop}%
\bibitem [{\citenamefont {Maldacena}\ and\ \citenamefont
  {Stanford}(2016)}]{Maldacena:2016hyu}%
  \BibitemOpen
  \bibfield  {author} {\bibinfo {author} {\bibfnamefont {J.}~\bibnamefont
  {Maldacena}}\ and\ \bibinfo {author} {\bibfnamefont {D.}~\bibnamefont
  {Stanford}},\ }\bibfield  {title} {\bibinfo {title} {{Remarks on the
  Sachdev-Ye-Kitaev model}},\ }\href
  {https://doi.org/10.1103/PhysRevD.94.106002} {\bibfield  {journal} {\bibinfo
  {journal} {Phys. Rev. D}\ }\textbf {\bibinfo {volume} {94}},\ \bibinfo
  {pages} {106002} (\bibinfo {year} {2016})},\ \Eprint
  {https://arxiv.org/abs/1604.07818} {arXiv:1604.07818 [hep-th]} \BibitemShut
  {NoStop}%
\bibitem [{\citenamefont {Sachdev}(2015)}]{PhysRevX.5.041025}%
  \BibitemOpen
  \bibfield  {author} {\bibinfo {author} {\bibfnamefont {S.}~\bibnamefont
  {Sachdev}},\ }\bibfield  {title} {\bibinfo {title} {Bekenstein-hawking
  entropy and strange metals},\ }\href
  {https://doi.org/10.1103/PhysRevX.5.041025} {\bibfield  {journal} {\bibinfo
  {journal} {Phys. Rev. X}\ }\textbf {\bibinfo {volume} {5}},\ \bibinfo {pages}
  {041025} (\bibinfo {year} {2015})}\BibitemShut {NoStop}%
\bibitem [{\citenamefont {Sachdev}\ and\ \citenamefont {Ye}(1993)}]{SY1992}%
  \BibitemOpen
  \bibfield  {author} {\bibinfo {author} {\bibfnamefont {S.}~\bibnamefont
  {Sachdev}}\ and\ \bibinfo {author} {\bibfnamefont {J.}~\bibnamefont {Ye}},\
  }\bibfield  {title} {\bibinfo {title} {{Gapless spin-fluid ground state in a
  random quantum Heisenberg magnet}},\ }\href
  {https://doi.org/10.1103/PhysRevLett.70.3339} {\bibfield  {journal} {\bibinfo
   {journal} {Phys. Rev. Lett.}\ }\textbf {\bibinfo {volume} {70}},\ \bibinfo
  {pages} {3339} (\bibinfo {year} {1993})}\BibitemShut {NoStop}%
\bibitem [{\citenamefont {Chowdhury}\ \emph {et~al.}(2022)\citenamefont
  {Chowdhury}, \citenamefont {Georges}, \citenamefont {Parcollet},\ and\
  \citenamefont {Sachdev}}]{RevModPhys.94.035004}%
  \BibitemOpen
  \bibfield  {author} {\bibinfo {author} {\bibfnamefont {D.}~\bibnamefont
  {Chowdhury}}, \bibinfo {author} {\bibfnamefont {A.}~\bibnamefont {Georges}},
  \bibinfo {author} {\bibfnamefont {O.}~\bibnamefont {Parcollet}},\ and\
  \bibinfo {author} {\bibfnamefont {S.}~\bibnamefont {Sachdev}},\ }\bibfield
  {title} {\bibinfo {title} {Sachdev-ye-kitaev models and beyond: Window into
  non-fermi liquids},\ }\href {https://doi.org/10.1103/RevModPhys.94.035004}
  {\bibfield  {journal} {\bibinfo  {journal} {Rev. Mod. Phys.}\ }\textbf
  {\bibinfo {volume} {94}},\ \bibinfo {pages} {035004} (\bibinfo {year}
  {2022})}\BibitemShut {NoStop}%
\bibitem [{\citenamefont {{Gu}}\ \emph {et~al.}(2017)\citenamefont {{Gu}},
  \citenamefont {{Qi}},\ and\ \citenamefont {{Stanford}}}]{Gu2017}%
  \BibitemOpen
  \bibfield  {author} {\bibinfo {author} {\bibfnamefont {Y.}~\bibnamefont
  {{Gu}}}, \bibinfo {author} {\bibfnamefont {X.-L.}\ \bibnamefont {{Qi}}},\
  and\ \bibinfo {author} {\bibfnamefont {D.}~\bibnamefont {{Stanford}}},\
  }\bibfield  {title} {\bibinfo {title} {{Local criticality, diffusion and
  chaos in generalized Sachdev-Ye-Kitaev models}},\ }\href
  {https://doi.org/10.1007/JHEP05(2017)125} {\bibfield  {journal} {\bibinfo
  {journal} {Journal of High Energy Physics}\ }\textbf {\bibinfo {volume}
  {2017}},\ \bibinfo {eid} {125} (\bibinfo {year} {2017})},\ \Eprint
  {https://arxiv.org/abs/1609.07832} {arXiv:1609.07832 [hep-th]} \BibitemShut
  {NoStop}%
\bibitem [{\citenamefont {Banerjee}\ and\ \citenamefont
  {Altman}(2017)}]{Banerjee:2016ncu}%
  \BibitemOpen
  \bibfield  {author} {\bibinfo {author} {\bibfnamefont {S.}~\bibnamefont
  {Banerjee}}\ and\ \bibinfo {author} {\bibfnamefont {E.}~\bibnamefont
  {Altman}},\ }\bibfield  {title} {\bibinfo {title} {{Solvable model for a
  dynamical quantum phase transition from fast to slow scrambling}},\ }\href
  {https://doi.org/10.1103/PhysRevB.95.134302} {\bibfield  {journal} {\bibinfo
  {journal} {Phys. Rev. B}\ }\textbf {\bibinfo {volume} {95}},\ \bibinfo
  {pages} {134302} (\bibinfo {year} {2017})},\ \Eprint
  {https://arxiv.org/abs/1610.04619} {arXiv:1610.04619 [cond-mat.str-el]}
  \BibitemShut {NoStop}%
\bibitem [{\citenamefont {Jian}\ \emph {et~al.}(2017)\citenamefont {Jian},
  \citenamefont {Bi},\ and\ \citenamefont {Xu}}]{Jian2017}%
  \BibitemOpen
  \bibfield  {author} {\bibinfo {author} {\bibfnamefont {C.-M.}\ \bibnamefont
  {Jian}}, \bibinfo {author} {\bibfnamefont {Z.}~\bibnamefont {Bi}},\ and\
  \bibinfo {author} {\bibfnamefont {C.}~\bibnamefont {Xu}},\ }\bibfield
  {title} {\bibinfo {title} {Model for continuous thermal metal to insulator
  transition},\ }\href {https://doi.org/10.1103/PhysRevB.96.115122} {\bibfield
  {journal} {\bibinfo  {journal} {Phys. Rev. B}\ }\textbf {\bibinfo {volume}
  {96}},\ \bibinfo {pages} {115122} (\bibinfo {year} {2017})}\BibitemShut
  {NoStop}%
\bibitem [{\citenamefont {Song}\ \emph {et~al.}(2017)\citenamefont {Song},
  \citenamefont {Jian},\ and\ \citenamefont {Balents}}]{Song2017}%
  \BibitemOpen
  \bibfield  {author} {\bibinfo {author} {\bibfnamefont {X.-Y.}\ \bibnamefont
  {Song}}, \bibinfo {author} {\bibfnamefont {C.-M.}\ \bibnamefont {Jian}},\
  and\ \bibinfo {author} {\bibfnamefont {L.}~\bibnamefont {Balents}},\
  }\bibfield  {title} {\bibinfo {title} {Strongly correlated metal built from
  sachdev-ye-kitaev models},\ }\href
  {https://doi.org/10.1103/PhysRevLett.119.216601} {\bibfield  {journal}
  {\bibinfo  {journal} {Phys. Rev. Lett.}\ }\textbf {\bibinfo {volume} {119}},\
  \bibinfo {pages} {216601} (\bibinfo {year} {2017})}\BibitemShut {NoStop}%
\bibitem [{\citenamefont {Davison}\ \emph {et~al.}(2017)\citenamefont
  {Davison}, \citenamefont {Fu}, \citenamefont {Georges}, \citenamefont {Gu},
  \citenamefont {Jensen},\ and\ \citenamefont {Sachdev}}]{Davison2017}%
  \BibitemOpen
  \bibfield  {author} {\bibinfo {author} {\bibfnamefont {R.~A.}\ \bibnamefont
  {Davison}}, \bibinfo {author} {\bibfnamefont {W.}~\bibnamefont {Fu}},
  \bibinfo {author} {\bibfnamefont {A.}~\bibnamefont {Georges}}, \bibinfo
  {author} {\bibfnamefont {Y.}~\bibnamefont {Gu}}, \bibinfo {author}
  {\bibfnamefont {K.}~\bibnamefont {Jensen}},\ and\ \bibinfo {author}
  {\bibfnamefont {S.}~\bibnamefont {Sachdev}},\ }\bibfield  {title} {\bibinfo
  {title} {Thermoelectric transport in disordered metals without
  quasiparticles: The sachdev-ye-kitaev models and holography},\ }\href
  {https://doi.org/10.1103/PhysRevB.95.155131} {\bibfield  {journal} {\bibinfo
  {journal} {Phys. Rev. B}\ }\textbf {\bibinfo {volume} {95}},\ \bibinfo
  {pages} {155131} (\bibinfo {year} {2017})}\BibitemShut {NoStop}%
\bibitem [{\citenamefont {Zhang}(2017)}]{Zhang2017}%
  \BibitemOpen
  \bibfield  {author} {\bibinfo {author} {\bibfnamefont {P.}~\bibnamefont
  {Zhang}},\ }\bibfield  {title} {\bibinfo {title} {Dispersive
  sachdev-ye-kitaev model: Band structure and quantum chaos},\ }\href
  {https://doi.org/10.1103/PhysRevB.96.205138} {\bibfield  {journal} {\bibinfo
  {journal} {Phys. Rev. B}\ }\textbf {\bibinfo {volume} {96}},\ \bibinfo
  {pages} {205138} (\bibinfo {year} {2017})}\BibitemShut {NoStop}%
\bibitem [{\citenamefont {Chowdhury}\ \emph {et~al.}(2018)\citenamefont
  {Chowdhury}, \citenamefont {Werman}, \citenamefont {Berg},\ and\
  \citenamefont {Senthil}}]{Chowdhury2018}%
  \BibitemOpen
  \bibfield  {author} {\bibinfo {author} {\bibfnamefont {D.}~\bibnamefont
  {Chowdhury}}, \bibinfo {author} {\bibfnamefont {Y.}~\bibnamefont {Werman}},
  \bibinfo {author} {\bibfnamefont {E.}~\bibnamefont {Berg}},\ and\ \bibinfo
  {author} {\bibfnamefont {T.}~\bibnamefont {Senthil}},\ }\bibfield  {title}
  {\bibinfo {title} {Translationally invariant non-fermi-liquid metals with
  critical fermi surfaces: Solvable models},\ }\href
  {https://doi.org/10.1103/PhysRevX.8.031024} {\bibfield  {journal} {\bibinfo
  {journal} {Phys. Rev. X}\ }\textbf {\bibinfo {volume} {8}},\ \bibinfo {pages}
  {031024} (\bibinfo {year} {2018})}\BibitemShut {NoStop}%
\bibitem [{\citenamefont {Haldar}\ and\ \citenamefont
  {Shenoy}(2018)}]{Haldar2018}%
  \BibitemOpen
  \bibfield  {author} {\bibinfo {author} {\bibfnamefont {A.}~\bibnamefont
  {Haldar}}\ and\ \bibinfo {author} {\bibfnamefont {V.~B.}\ \bibnamefont
  {Shenoy}},\ }\bibfield  {title} {\bibinfo {title} {Strange half-metals and
  mott insulators in sachdev-ye-kitaev models},\ }\href
  {https://doi.org/10.1103/PhysRevB.98.165135} {\bibfield  {journal} {\bibinfo
  {journal} {Phys. Rev. B}\ }\textbf {\bibinfo {volume} {98}},\ \bibinfo
  {pages} {165135} (\bibinfo {year} {2018})}\BibitemShut {NoStop}%
\bibitem [{\citenamefont {Haldar}\ \emph {et~al.}(2018)\citenamefont {Haldar},
  \citenamefont {Banerjee},\ and\ \citenamefont {Shenoy}}]{Haldar2018a}%
  \BibitemOpen
  \bibfield  {author} {\bibinfo {author} {\bibfnamefont {A.}~\bibnamefont
  {Haldar}}, \bibinfo {author} {\bibfnamefont {S.}~\bibnamefont {Banerjee}},\
  and\ \bibinfo {author} {\bibfnamefont {V.~B.}\ \bibnamefont {Shenoy}},\
  }\bibfield  {title} {\bibinfo {title} {Higher-dimensional sachdev-ye-kitaev
  non-fermi liquids at lifshitz transitions},\ }\href
  {https://doi.org/10.1103/PhysRevB.97.241106} {\bibfield  {journal} {\bibinfo
  {journal} {Phys. Rev. B}\ }\textbf {\bibinfo {volume} {97}},\ \bibinfo
  {pages} {241106} (\bibinfo {year} {2018})}\BibitemShut {NoStop}%
\bibitem [{\citenamefont {Jian}\ \emph {et~al.}(2018)\citenamefont {Jian},
  \citenamefont {Xian},\ and\ \citenamefont {Yao}}]{Jian2018}%
  \BibitemOpen
  \bibfield  {author} {\bibinfo {author} {\bibfnamefont {S.-K.}\ \bibnamefont
  {Jian}}, \bibinfo {author} {\bibfnamefont {Z.-Y.}\ \bibnamefont {Xian}},\
  and\ \bibinfo {author} {\bibfnamefont {H.}~\bibnamefont {Yao}},\ }\bibfield
  {title} {\bibinfo {title} {Quantum criticality and duality in the
  $\text{Sachdev-Ye-Kitaev}/{\mathrm{ads}}_{2}$ chain},\ }\href
  {https://doi.org/10.1103/PhysRevB.97.205141} {\bibfield  {journal} {\bibinfo
  {journal} {Phys. Rev. B}\ }\textbf {\bibinfo {volume} {97}},\ \bibinfo
  {pages} {205141} (\bibinfo {year} {2018})}\BibitemShut {NoStop}%
\bibitem [{\citenamefont {Esterlis}\ and\ \citenamefont
  {Schmalian}(2019)}]{Esterlis2019}%
  \BibitemOpen
  \bibfield  {author} {\bibinfo {author} {\bibfnamefont {I.}~\bibnamefont
  {Esterlis}}\ and\ \bibinfo {author} {\bibfnamefont {J.}~\bibnamefont
  {Schmalian}},\ }\bibfield  {title} {\bibinfo {title} {Cooper pairing of
  incoherent electrons: An electron-phonon version of the sachdev-ye-kitaev
  model},\ }\href {https://doi.org/10.1103/PhysRevB.100.115132} {\bibfield
  {journal} {\bibinfo  {journal} {Phys. Rev. B}\ }\textbf {\bibinfo {volume}
  {100}},\ \bibinfo {pages} {115132} (\bibinfo {year} {2019})}\BibitemShut
  {NoStop}%
\bibitem [{\citenamefont {Kim}\ \emph {et~al.}(2021)\citenamefont {Kim},
  \citenamefont {Altman},\ and\ \citenamefont {Cao}}]{Kim2021}%
  \BibitemOpen
  \bibfield  {author} {\bibinfo {author} {\bibfnamefont {J.}~\bibnamefont
  {Kim}}, \bibinfo {author} {\bibfnamefont {E.}~\bibnamefont {Altman}},\ and\
  \bibinfo {author} {\bibfnamefont {X.}~\bibnamefont {Cao}},\ }\bibfield
  {title} {\bibinfo {title} {Dirac fast scramblers},\ }\href
  {https://doi.org/10.1103/PhysRevB.103.L081113} {\bibfield  {journal}
  {\bibinfo  {journal} {Phys. Rev. B}\ }\textbf {\bibinfo {volume} {103}},\
  \bibinfo {pages} {L081113} (\bibinfo {year} {2021})}\BibitemShut {NoStop}%
\bibitem [{\citenamefont {{Patel}}\ \emph {et~al.}(2023)\citenamefont
  {{Patel}}, \citenamefont {{Guo}}, \citenamefont {{Esterlis}},\ and\
  \citenamefont {{Sachdev}}}]{Patel2023}%
  \BibitemOpen
  \bibfield  {author} {\bibinfo {author} {\bibfnamefont {A.~A.}\ \bibnamefont
  {{Patel}}}, \bibinfo {author} {\bibfnamefont {H.}~\bibnamefont {{Guo}}},
  \bibinfo {author} {\bibfnamefont {I.}~\bibnamefont {{Esterlis}}},\ and\
  \bibinfo {author} {\bibfnamefont {S.}~\bibnamefont {{Sachdev}}},\ }\bibfield
  {title} {\bibinfo {title} {{Universal theory of strange metals from spatially
  random interactions}},\ }\href {https://doi.org/10.1126/science.abq6011}
  {\bibfield  {journal} {\bibinfo  {journal} {Science}\ }\textbf {\bibinfo
  {volume} {381}},\ \bibinfo {pages} {790} (\bibinfo {year} {2023})},\ \Eprint
  {https://arxiv.org/abs/2203.04990} {arXiv:2203.04990 [cond-mat.str-el]}
  \BibitemShut {NoStop}%
\bibitem [{\citenamefont {Kim}\ \emph {et~al.}(2019)\citenamefont {Kim},
  \citenamefont {Klebanov}, \citenamefont {Tarnopolsky},\ and\ \citenamefont
  {Zhao}}]{PhysRevX.9.021043}%
  \BibitemOpen
  \bibfield  {author} {\bibinfo {author} {\bibfnamefont {J.}~\bibnamefont
  {Kim}}, \bibinfo {author} {\bibfnamefont {I.~R.}\ \bibnamefont {Klebanov}},
  \bibinfo {author} {\bibfnamefont {G.}~\bibnamefont {Tarnopolsky}},\ and\
  \bibinfo {author} {\bibfnamefont {W.}~\bibnamefont {Zhao}},\ }\bibfield
  {title} {\bibinfo {title} {Symmetry breaking in coupled syk or tensor
  models},\ }\href {https://doi.org/10.1103/PhysRevX.9.021043} {\bibfield
  {journal} {\bibinfo  {journal} {Phys. Rev. X}\ }\textbf {\bibinfo {volume}
  {9}},\ \bibinfo {pages} {021043} (\bibinfo {year} {2019})}\BibitemShut
  {NoStop}%
\bibitem [{\citenamefont {Maldacena}\ \emph
  {et~al.}(2016{\natexlab{a}})\citenamefont {Maldacena}, \citenamefont
  {Shenker},\ and\ \citenamefont {Stanford}}]{Maldacena:2015waa}%
  \BibitemOpen
  \bibfield  {author} {\bibinfo {author} {\bibfnamefont {J.}~\bibnamefont
  {Maldacena}}, \bibinfo {author} {\bibfnamefont {S.~H.}\ \bibnamefont
  {Shenker}},\ and\ \bibinfo {author} {\bibfnamefont {D.}~\bibnamefont
  {Stanford}},\ }\bibfield  {title} {\bibinfo {title} {{A bound on chaos}},\
  }\href {https://doi.org/10.1007/JHEP08(2016)106} {\bibfield  {journal}
  {\bibinfo  {journal} {JHEP}\ }\textbf {\bibinfo {volume} {08}},\ \bibinfo
  {pages} {106}},\ \Eprint {https://arxiv.org/abs/1503.01409} {arXiv:1503.01409
  [hep-th]} \BibitemShut {NoStop}%
\bibitem [{\citenamefont {Maldacena}\ \emph
  {et~al.}(2016{\natexlab{b}})\citenamefont {Maldacena}, \citenamefont
  {Stanford},\ and\ \citenamefont {Yang}}]{Maldacena:2016upp}%
  \BibitemOpen
  \bibfield  {author} {\bibinfo {author} {\bibfnamefont {J.}~\bibnamefont
  {Maldacena}}, \bibinfo {author} {\bibfnamefont {D.}~\bibnamefont
  {Stanford}},\ and\ \bibinfo {author} {\bibfnamefont {Z.}~\bibnamefont
  {Yang}},\ }\bibfield  {title} {\bibinfo {title} {{Conformal symmetry and its
  breaking in two dimensional Nearly Anti-de-Sitter space}},\ }\href
  {https://doi.org/10.1093/ptep/ptw124} {\bibfield  {journal} {\bibinfo
  {journal} {PTEP}\ }\textbf {\bibinfo {volume} {2016}},\ \bibinfo {pages}
  {12C104} (\bibinfo {year} {2016}{\natexlab{b}})},\ \Eprint
  {https://arxiv.org/abs/1606.01857} {arXiv:1606.01857 [hep-th]} \BibitemShut
  {NoStop}%
\bibitem [{\citenamefont {Gu}\ \emph {et~al.}(2020)\citenamefont {Gu},
  \citenamefont {Kitaev}, \citenamefont {Sachdev},\ and\ \citenamefont
  {Tarnopolsky}}]{Gu:2019jub}%
  \BibitemOpen
  \bibfield  {author} {\bibinfo {author} {\bibfnamefont {Y.}~\bibnamefont
  {Gu}}, \bibinfo {author} {\bibfnamefont {A.}~\bibnamefont {Kitaev}}, \bibinfo
  {author} {\bibfnamefont {S.}~\bibnamefont {Sachdev}},\ and\ \bibinfo {author}
  {\bibfnamefont {G.}~\bibnamefont {Tarnopolsky}},\ }\bibfield  {title}
  {\bibinfo {title} {{Notes on the complex Sachdev-Ye-Kitaev model}},\ }\href
  {https://doi.org/10.1007/JHEP02(2020)157} {\bibfield  {journal} {\bibinfo
  {journal} {JHEP}\ }\textbf {\bibinfo {volume} {02}},\ \bibinfo {pages}
  {157}},\ \Eprint {https://arxiv.org/abs/1910.14099} {arXiv:1910.14099
  [hep-th]} \BibitemShut {NoStop}%
\bibitem [{\citenamefont {Nayak}\ \emph {et~al.}(2018)\citenamefont {Nayak},
  \citenamefont {Shukla}, \citenamefont {Soni}, \citenamefont {Trivedi},\ and\
  \citenamefont {Vishal}}]{Nayak:2018qej}%
  \BibitemOpen
  \bibfield  {author} {\bibinfo {author} {\bibfnamefont {P.}~\bibnamefont
  {Nayak}}, \bibinfo {author} {\bibfnamefont {A.}~\bibnamefont {Shukla}},
  \bibinfo {author} {\bibfnamefont {R.~M.}\ \bibnamefont {Soni}}, \bibinfo
  {author} {\bibfnamefont {S.~P.}\ \bibnamefont {Trivedi}},\ and\ \bibinfo
  {author} {\bibfnamefont {V.}~\bibnamefont {Vishal}},\ }\bibfield  {title}
  {\bibinfo {title} {{On the Dynamics of Near-Extremal Black Holes}},\ }\href
  {https://doi.org/10.1007/JHEP09(2018)048} {\bibfield  {journal} {\bibinfo
  {journal} {JHEP}\ }\textbf {\bibinfo {volume} {09}},\ \bibinfo {pages}
  {048}},\ \Eprint {https://arxiv.org/abs/1802.09547} {arXiv:1802.09547
  [hep-th]} \BibitemShut {NoStop}%
\bibitem [{\citenamefont {Moitra}\ \emph
  {et~al.}(2019{\natexlab{a}})\citenamefont {Moitra}, \citenamefont {Trivedi},\
  and\ \citenamefont {Vishal}}]{Moitra:2018jqs}%
  \BibitemOpen
  \bibfield  {author} {\bibinfo {author} {\bibfnamefont {U.}~\bibnamefont
  {Moitra}}, \bibinfo {author} {\bibfnamefont {S.~P.}\ \bibnamefont
  {Trivedi}},\ and\ \bibinfo {author} {\bibfnamefont {V.}~\bibnamefont
  {Vishal}},\ }\bibfield  {title} {\bibinfo {title} {{Extremal and
  near-extremal black holes and near-CFT$_{1}$}},\ }\href
  {https://doi.org/10.1007/JHEP07(2019)055} {\bibfield  {journal} {\bibinfo
  {journal} {JHEP}\ }\textbf {\bibinfo {volume} {07}},\ \bibinfo {pages}
  {055}},\ \Eprint {https://arxiv.org/abs/1808.08239} {arXiv:1808.08239
  [hep-th]} \BibitemShut {NoStop}%
\bibitem [{\citenamefont {Sachdev}(2019)}]{Sachdev:2019bjn}%
  \BibitemOpen
  \bibfield  {author} {\bibinfo {author} {\bibfnamefont {S.}~\bibnamefont
  {Sachdev}},\ }\bibfield  {title} {\bibinfo {title} {Universal low temperature
  theory of charged black holes with ads2 horizons},\ }\href
  {https://doi.org/10.1063/1.5092726} {\bibfield  {journal} {\bibinfo
  {journal} {Journal of Mathematical Physics}\ }\textbf {\bibinfo {volume}
  {60}},\ \bibinfo {pages} {052303} (\bibinfo {year} {2019})},\ \Eprint
  {https://arxiv.org/abs/https://pubs.aip.org/aip/jmp/article-pdf/doi/10.1063/1.5092726/16034331/052303\_1\_online.pdf}
  {https://pubs.aip.org/aip/jmp/article-pdf/doi/10.1063/1.5092726/16034331/052303\_1\_online.pdf}
  \BibitemShut {NoStop}%
\bibitem [{\citenamefont {Moitra}\ \emph
  {et~al.}(2019{\natexlab{b}})\citenamefont {Moitra}, \citenamefont {Sake},
  \citenamefont {Trivedi},\ and\ \citenamefont {Vishal}}]{Moitra:2019bub}%
  \BibitemOpen
  \bibfield  {author} {\bibinfo {author} {\bibfnamefont {U.}~\bibnamefont
  {Moitra}}, \bibinfo {author} {\bibfnamefont {S.~K.}\ \bibnamefont {Sake}},
  \bibinfo {author} {\bibfnamefont {S.~P.}\ \bibnamefont {Trivedi}},\ and\
  \bibinfo {author} {\bibfnamefont {V.}~\bibnamefont {Vishal}},\ }\bibfield
  {title} {\bibinfo {title} {{Jackiw-Teitelboim Gravity and Rotating Black
  Holes}},\ }\href {https://doi.org/10.1007/JHEP11(2019)047} {\bibfield
  {journal} {\bibinfo  {journal} {JHEP}\ }\textbf {\bibinfo {volume} {11}},\
  \bibinfo {pages} {047}},\ \Eprint {https://arxiv.org/abs/1905.10378}
  {arXiv:1905.10378 [hep-th]} \BibitemShut {NoStop}%
\bibitem [{\citenamefont {Fu}\ and\ \citenamefont
  {Sachdev}(2016)}]{PhysRevB.94.035135}%
  \BibitemOpen
  \bibfield  {author} {\bibinfo {author} {\bibfnamefont {W.}~\bibnamefont
  {Fu}}\ and\ \bibinfo {author} {\bibfnamefont {S.}~\bibnamefont {Sachdev}},\
  }\bibfield  {title} {\bibinfo {title} {Numerical study of fermion and boson
  models with infinite-range random interactions},\ }\href
  {https://doi.org/10.1103/PhysRevB.94.035135} {\bibfield  {journal} {\bibinfo
  {journal} {Phys. Rev. B}\ }\textbf {\bibinfo {volume} {94}},\ \bibinfo
  {pages} {035135} (\bibinfo {year} {2016})}\BibitemShut {NoStop}%
\bibitem [{\citenamefont {Tikhanovskaya}\ \emph {et~al.}(2021)\citenamefont
  {Tikhanovskaya}, \citenamefont {Guo}, \citenamefont {Sachdev},\ and\
  \citenamefont {Tarnopolsky}}]{PhysRevB.103.075141}%
  \BibitemOpen
  \bibfield  {author} {\bibinfo {author} {\bibfnamefont {M.}~\bibnamefont
  {Tikhanovskaya}}, \bibinfo {author} {\bibfnamefont {H.}~\bibnamefont {Guo}},
  \bibinfo {author} {\bibfnamefont {S.}~\bibnamefont {Sachdev}},\ and\ \bibinfo
  {author} {\bibfnamefont {G.}~\bibnamefont {Tarnopolsky}},\ }\bibfield
  {title} {\bibinfo {title} {Excitation spectra of quantum matter without
  quasiparticles. i. sachdev-ye-kitaev models},\ }\href
  {https://doi.org/10.1103/PhysRevB.103.075141} {\bibfield  {journal} {\bibinfo
   {journal} {Phys. Rev. B}\ }\textbf {\bibinfo {volume} {103}},\ \bibinfo
  {pages} {075141} (\bibinfo {year} {2021})}\BibitemShut {NoStop}%
\bibitem [{\citenamefont {Chew}\ \emph {et~al.}(2017)\citenamefont {Chew},
  \citenamefont {Essin},\ and\ \citenamefont {Alicea}}]{PhysRevB.96.121119}%
  \BibitemOpen
  \bibfield  {author} {\bibinfo {author} {\bibfnamefont {A.}~\bibnamefont
  {Chew}}, \bibinfo {author} {\bibfnamefont {A.}~\bibnamefont {Essin}},\ and\
  \bibinfo {author} {\bibfnamefont {J.}~\bibnamefont {Alicea}},\ }\bibfield
  {title} {\bibinfo {title} {Approximating the sachdev-ye-kitaev model with
  majorana wires},\ }\href {https://doi.org/10.1103/PhysRevB.96.121119}
  {\bibfield  {journal} {\bibinfo  {journal} {Phys. Rev. B}\ }\textbf {\bibinfo
  {volume} {96}},\ \bibinfo {pages} {121119} (\bibinfo {year}
  {2017})}\BibitemShut {NoStop}%
\bibitem [{\citenamefont {Pikulin}\ and\ \citenamefont
  {Franz}(2017)}]{PhysRevX.7.031006}%
  \BibitemOpen
  \bibfield  {author} {\bibinfo {author} {\bibfnamefont {D.~I.}\ \bibnamefont
  {Pikulin}}\ and\ \bibinfo {author} {\bibfnamefont {M.}~\bibnamefont
  {Franz}},\ }\bibfield  {title} {\bibinfo {title} {Black hole on a chip:
  Proposal for a physical realization of the sachdev-ye-kitaev model in a
  solid-state system},\ }\href {https://doi.org/10.1103/PhysRevX.7.031006}
  {\bibfield  {journal} {\bibinfo  {journal} {Phys. Rev. X}\ }\textbf {\bibinfo
  {volume} {7}},\ \bibinfo {pages} {031006} (\bibinfo {year}
  {2017})}\BibitemShut {NoStop}%
\bibitem [{\citenamefont {Franz}\ and\ \citenamefont
  {Rozali}(2018)}]{Franz:2018cqi}%
  \BibitemOpen
  \bibfield  {author} {\bibinfo {author} {\bibfnamefont {M.}~\bibnamefont
  {Franz}}\ and\ \bibinfo {author} {\bibfnamefont {M.}~\bibnamefont {Rozali}},\
  }\bibfield  {title} {\bibinfo {title} {{Mimicking black hole event horizons
  in atomic and solid-state systems}},\ }\href
  {https://doi.org/10.1038/s41578-018-0058-z} {\bibfield  {journal} {\bibinfo
  {journal} {Nature Rev. Mater.}\ }\textbf {\bibinfo {volume} {3}},\ \bibinfo
  {pages} {491} (\bibinfo {year} {2018})},\ \Eprint
  {https://arxiv.org/abs/1808.00541} {arXiv:1808.00541 [cond-mat.str-el]}
  \BibitemShut {NoStop}%
\bibitem [{\citenamefont {Chen}\ \emph {et~al.}(2018)\citenamefont {Chen},
  \citenamefont {Ilan}, \citenamefont {de~Juan}, \citenamefont {Pikulin},\ and\
  \citenamefont {Franz}}]{PhysRevLett.121.036403}%
  \BibitemOpen
  \bibfield  {author} {\bibinfo {author} {\bibfnamefont {A.}~\bibnamefont
  {Chen}}, \bibinfo {author} {\bibfnamefont {R.}~\bibnamefont {Ilan}}, \bibinfo
  {author} {\bibfnamefont {F.}~\bibnamefont {de~Juan}}, \bibinfo {author}
  {\bibfnamefont {D.~I.}\ \bibnamefont {Pikulin}},\ and\ \bibinfo {author}
  {\bibfnamefont {M.}~\bibnamefont {Franz}},\ }\bibfield  {title} {\bibinfo
  {title} {Quantum holography in a graphene flake with an irregular boundary},\
  }\href {https://doi.org/10.1103/PhysRevLett.121.036403} {\bibfield  {journal}
  {\bibinfo  {journal} {Phys. Rev. Lett.}\ }\textbf {\bibinfo {volume} {121}},\
  \bibinfo {pages} {036403} (\bibinfo {year} {2018})}\BibitemShut {NoStop}%
\bibitem [{\citenamefont {Danshita}\ \emph {et~al.}(2017)\citenamefont
  {Danshita}, \citenamefont {Hanada},\ and\ \citenamefont
  {Tezuka}}]{Danshita:2016xbo}%
  \BibitemOpen
  \bibfield  {author} {\bibinfo {author} {\bibfnamefont {I.}~\bibnamefont
  {Danshita}}, \bibinfo {author} {\bibfnamefont {M.}~\bibnamefont {Hanada}},\
  and\ \bibinfo {author} {\bibfnamefont {M.}~\bibnamefont {Tezuka}},\
  }\bibfield  {title} {\bibinfo {title} {{Creating and probing the
  Sachdev-Ye-Kitaev model with ultracold gases: Towards experimental studies of
  quantum gravity}},\ }\href {https://doi.org/10.1093/ptep/ptx108} {\bibfield
  {journal} {\bibinfo  {journal} {PTEP}\ }\textbf {\bibinfo {volume} {2017}},\
  \bibinfo {pages} {083I01} (\bibinfo {year} {2017})},\ \Eprint
  {https://arxiv.org/abs/1606.02454} {arXiv:1606.02454 [cond-mat.quant-gas]}
  \BibitemShut {NoStop}%
\bibitem [{\citenamefont {Lantagne-Hurtubise}\ \emph
  {et~al.}(2020)\citenamefont {Lantagne-Hurtubise}, \citenamefont {Plugge},
  \citenamefont {Can},\ and\ \citenamefont {Franz}}]{PhysRevResearch.2.013254}%
  \BibitemOpen
  \bibfield  {author} {\bibinfo {author} {\bibfnamefont {E.}~\bibnamefont
  {Lantagne-Hurtubise}}, \bibinfo {author} {\bibfnamefont {S.}~\bibnamefont
  {Plugge}}, \bibinfo {author} {\bibfnamefont {O.}~\bibnamefont {Can}},\ and\
  \bibinfo {author} {\bibfnamefont {M.}~\bibnamefont {Franz}},\ }\bibfield
  {title} {\bibinfo {title} {Diagnosing quantum chaos in many-body systems
  using entanglement as a resource},\ }\href
  {https://doi.org/10.1103/PhysRevResearch.2.013254} {\bibfield  {journal}
  {\bibinfo  {journal} {Phys. Rev. Res.}\ }\textbf {\bibinfo {volume} {2}},\
  \bibinfo {pages} {013254} (\bibinfo {year} {2020})}\BibitemShut {NoStop}%
\bibitem [{\citenamefont {Brzezi\ifmmode~\acute{n}\else \'{n}\fi{}ska}\ \emph
  {et~al.}(2023)\citenamefont {Brzezi\ifmmode~\acute{n}\else \'{n}\fi{}ska},
  \citenamefont {Guan}, \citenamefont {Yazyev}, \citenamefont {Sachdev},\ and\
  \citenamefont {Kruchkov}}]{PhysRevLett.131.036503}%
  \BibitemOpen
  \bibfield  {author} {\bibinfo {author} {\bibfnamefont {M.}~\bibnamefont
  {Brzezi\ifmmode~\acute{n}\else \'{n}\fi{}ska}}, \bibinfo {author}
  {\bibfnamefont {Y.}~\bibnamefont {Guan}}, \bibinfo {author} {\bibfnamefont
  {O.~V.}\ \bibnamefont {Yazyev}}, \bibinfo {author} {\bibfnamefont
  {S.}~\bibnamefont {Sachdev}},\ and\ \bibinfo {author} {\bibfnamefont
  {A.}~\bibnamefont {Kruchkov}},\ }\bibfield  {title} {\bibinfo {title}
  {Engineering syk interactions in disordered graphene flakes under realistic
  experimental conditions},\ }\href
  {https://doi.org/10.1103/PhysRevLett.131.036503} {\bibfield  {journal}
  {\bibinfo  {journal} {Phys. Rev. Lett.}\ }\textbf {\bibinfo {volume} {131}},\
  \bibinfo {pages} {036503} (\bibinfo {year} {2023})}\BibitemShut {NoStop}%
\bibitem [{\citenamefont {Altshuler}\ \emph {et~al.}(1997)\citenamefont
  {Altshuler}, \citenamefont {Gefen}, \citenamefont {Kamenev},\ and\
  \citenamefont {Levitov}}]{Altshuler1997}%
  \BibitemOpen
  \bibfield  {author} {\bibinfo {author} {\bibfnamefont {B.~L.}\ \bibnamefont
  {Altshuler}}, \bibinfo {author} {\bibfnamefont {Y.}~\bibnamefont {Gefen}},
  \bibinfo {author} {\bibfnamefont {A.}~\bibnamefont {Kamenev}},\ and\ \bibinfo
  {author} {\bibfnamefont {L.~S.}\ \bibnamefont {Levitov}},\ }\bibfield
  {title} {\bibinfo {title} {Quasiparticle lifetime in a finite system: A
  nonperturbative approach},\ }\href
  {https://doi.org/10.1103/PhysRevLett.78.2803} {\bibfield  {journal} {\bibinfo
   {journal} {Phys. Rev. Lett.}\ }\textbf {\bibinfo {volume} {78}},\ \bibinfo
  {pages} {2803} (\bibinfo {year} {1997})}\BibitemShut {NoStop}%
\bibitem [{\citenamefont {Micklitz}\ \emph {et~al.}(2019)\citenamefont
  {Micklitz}, \citenamefont {Monteiro},\ and\ \citenamefont
  {Altland}}]{Micklitz2019}%
  \BibitemOpen
  \bibfield  {author} {\bibinfo {author} {\bibfnamefont {T.}~\bibnamefont
  {Micklitz}}, \bibinfo {author} {\bibfnamefont {F.}~\bibnamefont {Monteiro}},\
  and\ \bibinfo {author} {\bibfnamefont {A.}~\bibnamefont {Altland}},\
  }\bibfield  {title} {\bibinfo {title} {Nonergodic extended states in the
  sachdev-ye-kitaev model},\ }\href
  {https://doi.org/10.1103/PhysRevLett.123.125701} {\bibfield  {journal}
  {\bibinfo  {journal} {Phys. Rev. Lett.}\ }\textbf {\bibinfo {volume} {123}},\
  \bibinfo {pages} {125701} (\bibinfo {year} {2019})}\BibitemShut {NoStop}%
\bibitem [{Note1()}]{Note1}%
  \BibitemOpen
  \bibinfo {note} {For a precise definition of mid-spectrum states see section
  \ref {sec:level2}.}\BibitemShut {Stop}%
\bibitem [{\citenamefont {Mukherjee}\ \emph {et~al.}()\citenamefont
  {Mukherjee}, \citenamefont {Trivedi}, \citenamefont {Banerjee},\ and\
  \citenamefont {Trivedi}}]{MukherjeeFuture}%
  \BibitemOpen
  \bibfield  {author} {\bibinfo {author} {\bibfnamefont {A.}~\bibnamefont
  {Mukherjee}}, \bibinfo {author} {\bibfnamefont {S.}~\bibnamefont {Trivedi}},
  \bibinfo {author} {\bibfnamefont {S.}~\bibnamefont {Banerjee}},\ and\
  \bibinfo {author} {\bibfnamefont {N.}~\bibnamefont {Trivedi}},\ }\bibfield
  {title} {\bibinfo {title} {{Does Chaos in SYK-model survive Without
  disorder}},\ }\href@noop {} {\bibinfo  {journal} {In Preparations}\
  }\BibitemShut {NoStop}%
\bibitem [{\citenamefont {Bohigas}\ \emph {et~al.}(1984)\citenamefont
  {Bohigas}, \citenamefont {Giannoni},\ and\ \citenamefont
  {Schmit}}]{Bohigas1984}%
  \BibitemOpen
\bibfield  {journal} {  }\bibfield  {author} {\bibinfo {author} {\bibfnamefont
  {O.}~\bibnamefont {Bohigas}}, \bibinfo {author} {\bibfnamefont {M.~J.}\
  \bibnamefont {Giannoni}},\ and\ \bibinfo {author} {\bibfnamefont
  {C.}~\bibnamefont {Schmit}},\ }\bibfield  {title} {\bibinfo {title}
  {Characterization of chaotic quantum spectra and universality of level
  fluctuation laws},\ }\href {https://doi.org/10.1103/PhysRevLett.52.1}
  {\bibfield  {journal} {\bibinfo  {journal} {Phys. Rev. Lett.}\ }\textbf
  {\bibinfo {volume} {52}},\ \bibinfo {pages} {1} (\bibinfo {year}
  {1984})}\BibitemShut {NoStop}%
\bibitem [{\citenamefont {Behrends}\ and\ \citenamefont
  {B\'eri}(2020)}]{PhysRevD.101.066017}%
  \BibitemOpen
  \bibfield  {author} {\bibinfo {author} {\bibfnamefont {J.}~\bibnamefont
  {Behrends}}\ and\ \bibinfo {author} {\bibfnamefont {B.}~\bibnamefont
  {B\'eri}},\ }\bibfield  {title} {\bibinfo {title} {Symmetry classes,
  many-body zero modes, and supersymmetry in the complex sachdev-ye-kitaev
  model},\ }\href {https://doi.org/10.1103/PhysRevD.101.066017} {\bibfield
  {journal} {\bibinfo  {journal} {Phys. Rev. D}\ }\textbf {\bibinfo {volume}
  {101}},\ \bibinfo {pages} {066017} (\bibinfo {year} {2020})}\BibitemShut
  {NoStop}%
\bibitem [{\citenamefont {You}\ \emph {et~al.}(2017)\citenamefont {You},
  \citenamefont {Ludwig},\ and\ \citenamefont {Xu}}]{you2017sachdev}%
  \BibitemOpen
  \bibfield  {author} {\bibinfo {author} {\bibfnamefont {Y.-Z.}\ \bibnamefont
  {You}}, \bibinfo {author} {\bibfnamefont {A.~W.~W.}\ \bibnamefont {Ludwig}},\
  and\ \bibinfo {author} {\bibfnamefont {C.}~\bibnamefont {Xu}},\ }\bibfield
  {title} {\bibinfo {title} {Sachdev-ye-kitaev model and thermalization on the
  boundary of many-body localized fermionic symmetry-protected topological
  states},\ }\href {https://doi.org/10.1103/PhysRevB.95.115150} {\bibfield
  {journal} {\bibinfo  {journal} {Phys. Rev. B}\ }\textbf {\bibinfo {volume}
  {95}},\ \bibinfo {pages} {115150} (\bibinfo {year} {2017})}\BibitemShut
  {NoStop}%
\bibitem [{\citenamefont {Garc\'{\i}a-Garc\'{\i}a}\ \emph
  {et~al.}(2022)\citenamefont {Garc\'{\i}a-Garc\'{\i}a}, \citenamefont {S\'a},\
  and\ \citenamefont {Verbaarschot}}]{PhysRevX.12.021040}%
  \BibitemOpen
  \bibfield  {author} {\bibinfo {author} {\bibfnamefont {A.~M.}\ \bibnamefont
  {Garc\'{\i}a-Garc\'{\i}a}}, \bibinfo {author} {\bibfnamefont
  {L.}~\bibnamefont {S\'a}},\ and\ \bibinfo {author} {\bibfnamefont {J.~J.~M.}\
  \bibnamefont {Verbaarschot}},\ }\bibfield  {title} {\bibinfo {title}
  {Symmetry classification and universality in non-hermitian many-body quantum
  chaos by the sachdev-ye-kitaev model},\ }\href
  {https://doi.org/10.1103/PhysRevX.12.021040} {\bibfield  {journal} {\bibinfo
  {journal} {Phys. Rev. X}\ }\textbf {\bibinfo {volume} {12}},\ \bibinfo
  {pages} {021040} (\bibinfo {year} {2022})}\BibitemShut {NoStop}%
\bibitem [{\citenamefont {Cotler}\ \emph {et~al.}(2017)\citenamefont {Cotler},
  \citenamefont {Gur-Ari}, \citenamefont {Hanada}, \citenamefont {Polchinski},
  \citenamefont {Saad}, \citenamefont {Shenker}, \citenamefont {Stanford},
  \citenamefont {Streicher},\ and\ \citenamefont {Tezuka}}]{Cotler:2016fpe}%
  \BibitemOpen
  \bibfield  {author} {\bibinfo {author} {\bibfnamefont {J.~S.}\ \bibnamefont
  {Cotler}}, \bibinfo {author} {\bibfnamefont {G.}~\bibnamefont {Gur-Ari}},
  \bibinfo {author} {\bibfnamefont {M.}~\bibnamefont {Hanada}}, \bibinfo
  {author} {\bibfnamefont {J.}~\bibnamefont {Polchinski}}, \bibinfo {author}
  {\bibfnamefont {P.}~\bibnamefont {Saad}}, \bibinfo {author} {\bibfnamefont
  {S.~H.}\ \bibnamefont {Shenker}}, \bibinfo {author} {\bibfnamefont
  {D.}~\bibnamefont {Stanford}}, \bibinfo {author} {\bibfnamefont
  {A.}~\bibnamefont {Streicher}},\ and\ \bibinfo {author} {\bibfnamefont
  {M.}~\bibnamefont {Tezuka}},\ }\bibfield  {title} {\bibinfo {title} {{Black
  Holes and Random Matrices}},\ }\href
  {https://doi.org/10.1007/JHEP05(2017)118} {\bibfield  {journal} {\bibinfo
  {journal} {JHEP}\ }\textbf {\bibinfo {volume} {05}},\ \bibinfo {pages}
  {118}},\ \bibinfo {note} {[Erratum: JHEP 09, 002 (2018)]},\ \Eprint
  {https://arxiv.org/abs/1611.04650} {arXiv:1611.04650 [hep-th]} \BibitemShut
  {NoStop}%
\bibitem [{\citenamefont {{Jackiw}}(1985)}]{1985NuPhB.252..343J}%
  \BibitemOpen
  \bibfield  {author} {\bibinfo {author} {\bibfnamefont {R.}~\bibnamefont
  {{Jackiw}}},\ }\bibfield  {title} {\bibinfo {title} {{Lower dimensional
  gravity}},\ }\href {https://doi.org/10.1016/0550-3213(85)90448-1} {\bibfield
  {journal} {\bibinfo  {journal} {Nuclear Physics B}\ }\textbf {\bibinfo
  {volume} {252}},\ \bibinfo {pages} {343} (\bibinfo {year}
  {1985})}\BibitemShut {NoStop}%
\bibitem [{\citenamefont {{Regge}}\ and\ \citenamefont
  {{Teitelboim}}(1974)}]{1974AnPhy..88..286R}%
  \BibitemOpen
  \bibfield  {author} {\bibinfo {author} {\bibfnamefont {T.}~\bibnamefont
  {{Regge}}}\ and\ \bibinfo {author} {\bibfnamefont {C.}~\bibnamefont
  {{Teitelboim}}},\ }\bibfield  {title} {\bibinfo {title} {{Role of surface
  integrals in the Hamiltonian formulation of general relativity}},\ }\href
  {https://doi.org/10.1016/0003-4916(74)90404-7} {\bibfield  {journal}
  {\bibinfo  {journal} {Annals of Physics}\ }\textbf {\bibinfo {volume} {88}},\
  \bibinfo {pages} {286} (\bibinfo {year} {1974})}\BibitemShut {NoStop}%
\bibitem [{\citenamefont {Saad}\ \emph {et~al.}(2019)\citenamefont {Saad},
  \citenamefont {Shenker},\ and\ \citenamefont {Stanford}}]{Saad:2019lba}%
  \BibitemOpen
  \bibfield  {author} {\bibinfo {author} {\bibfnamefont {P.}~\bibnamefont
  {Saad}}, \bibinfo {author} {\bibfnamefont {S.~H.}\ \bibnamefont {Shenker}},\
  and\ \bibinfo {author} {\bibfnamefont {D.}~\bibnamefont {Stanford}},\
  }\bibfield  {title} {\bibinfo {title} {{JT gravity as a matrix integral}},\
  }\href@noop {} {\  (\bibinfo {year} {2019})},\ \Eprint
  {https://arxiv.org/abs/1903.11115} {arXiv:1903.11115 [hep-th]} \BibitemShut
  {NoStop}%
\bibitem [{\citenamefont {Altland}\ and\ \citenamefont
  {Zirnbauer}(1997)}]{altland1997nonstandard}%
  \BibitemOpen
  \bibfield  {author} {\bibinfo {author} {\bibfnamefont {A.}~\bibnamefont
  {Altland}}\ and\ \bibinfo {author} {\bibfnamefont {M.~R.}\ \bibnamefont
  {Zirnbauer}},\ }\bibfield  {title} {\bibinfo {title} {Nonstandard symmetry
  classes in mesoscopic normal-superconducting hybrid structures},\ }\href
  {https://doi.org/10.1103/PhysRevB.55.1142} {\bibfield  {journal} {\bibinfo
  {journal} {Phys. Rev. B}\ }\textbf {\bibinfo {volume} {55}},\ \bibinfo
  {pages} {1142} (\bibinfo {year} {1997})}\BibitemShut {NoStop}%
\bibitem [{Note2()}]{Note2}%
  \BibitemOpen
  \bibinfo {note} {Alternatively one can consider the nearest neighbour spacing
  itself, rather than the ratio $r$, but in that case one has to `unfold' the
  spectrum, \cite {PhysRevD.101.066017}}\BibitemShut {NoStop}%
\bibitem [{\citenamefont {Atas}\ \emph {et~al.}(2013)\citenamefont {Atas},
  \citenamefont {Bogomolny}, \citenamefont {Giraud},\ and\ \citenamefont
  {Roux}}]{PhysRevLett.110.084101}%
  \BibitemOpen
  \bibfield  {author} {\bibinfo {author} {\bibfnamefont {Y.~Y.}\ \bibnamefont
  {Atas}}, \bibinfo {author} {\bibfnamefont {E.}~\bibnamefont {Bogomolny}},
  \bibinfo {author} {\bibfnamefont {O.}~\bibnamefont {Giraud}},\ and\ \bibinfo
  {author} {\bibfnamefont {G.}~\bibnamefont {Roux}},\ }\bibfield  {title}
  {\bibinfo {title} {Distribution of the ratio of consecutive level spacings in
  random matrix ensembles},\ }\href
  {https://doi.org/10.1103/PhysRevLett.110.084101} {\bibfield  {journal}
  {\bibinfo  {journal} {Phys. Rev. Lett.}\ }\textbf {\bibinfo {volume} {110}},\
  \bibinfo {pages} {084101} (\bibinfo {year} {2013})}\BibitemShut {NoStop}%
\bibitem [{\citenamefont {Garc\'{\i}a-Garc\'{\i}a}\ and\ \citenamefont
  {Verbaarschot}(2016{\natexlab{a}})}]{PhysRevD.94.126010}%
  \BibitemOpen
  \bibfield  {author} {\bibinfo {author} {\bibfnamefont {A.~M.}\ \bibnamefont
  {Garc\'{\i}a-Garc\'{\i}a}}\ and\ \bibinfo {author} {\bibfnamefont {J.~J.~M.}\
  \bibnamefont {Verbaarschot}},\ }\bibfield  {title} {\bibinfo {title}
  {Spectral and thermodynamic properties of the sachdev-ye-kitaev model},\
  }\href {https://doi.org/10.1103/PhysRevD.94.126010} {\bibfield  {journal}
  {\bibinfo  {journal} {Phys. Rev. D}\ }\textbf {\bibinfo {volume} {94}},\
  \bibinfo {pages} {126010} (\bibinfo {year} {2016}{\natexlab{a}})}\BibitemShut
  {NoStop}%
\bibitem [{\citenamefont {Kobrin}\ \emph {et~al.}(2021)\citenamefont {Kobrin},
  \citenamefont {Yang}, \citenamefont {Kahanamoku-Meyer}, \citenamefont
  {Olund}, \citenamefont {Moore}, \citenamefont {Stanford},\ and\ \citenamefont
  {Yao}}]{PhysRevLett.126.030602}%
  \BibitemOpen
  \bibfield  {author} {\bibinfo {author} {\bibfnamefont {B.}~\bibnamefont
  {Kobrin}}, \bibinfo {author} {\bibfnamefont {Z.}~\bibnamefont {Yang}},
  \bibinfo {author} {\bibfnamefont {G.~D.}\ \bibnamefont {Kahanamoku-Meyer}},
  \bibinfo {author} {\bibfnamefont {C.~T.}\ \bibnamefont {Olund}}, \bibinfo
  {author} {\bibfnamefont {J.~E.}\ \bibnamefont {Moore}}, \bibinfo {author}
  {\bibfnamefont {D.}~\bibnamefont {Stanford}},\ and\ \bibinfo {author}
  {\bibfnamefont {N.~Y.}\ \bibnamefont {Yao}},\ }\bibfield  {title} {\bibinfo
  {title} {Many-body chaos in the sachdev-ye-kitaev model},\ }\href
  {https://doi.org/10.1103/PhysRevLett.126.030602} {\bibfield  {journal}
  {\bibinfo  {journal} {Phys. Rev. Lett.}\ }\textbf {\bibinfo {volume} {126}},\
  \bibinfo {pages} {030602} (\bibinfo {year} {2021})}\BibitemShut {NoStop}%
\bibitem [{\citenamefont {Anegawa}\ \emph {et~al.}(2023)\citenamefont
  {Anegawa}, \citenamefont {Iizuka}, \citenamefont {Mukherjee}, \citenamefont
  {Sake},\ and\ \citenamefont {Trivedi}}]{Anegawa:2023vxq}%
  \BibitemOpen
  \bibfield  {author} {\bibinfo {author} {\bibfnamefont {T.}~\bibnamefont
  {Anegawa}}, \bibinfo {author} {\bibfnamefont {N.}~\bibnamefont {Iizuka}},
  \bibinfo {author} {\bibfnamefont {A.}~\bibnamefont {Mukherjee}}, \bibinfo
  {author} {\bibfnamefont {S.~K.}\ \bibnamefont {Sake}},\ and\ \bibinfo
  {author} {\bibfnamefont {S.~P.}\ \bibnamefont {Trivedi}},\ }\bibfield
  {title} {\bibinfo {title} {{Sparse random matrices and Gaussian ensembles
  with varying randomness}},\ }\href {https://doi.org/10.1007/JHEP11(2023)234}
  {\bibfield  {journal} {\bibinfo  {journal} {JHEP}\ }\textbf {\bibinfo
  {volume} {11}},\ \bibinfo {pages} {234}},\ \Eprint
  {https://arxiv.org/abs/2305.07505} {arXiv:2305.07505 [hep-th]} \BibitemShut
  {NoStop}%
\bibitem [{\citenamefont {Liu}\ \emph {et~al.}(2018)\citenamefont {Liu},
  \citenamefont {Chen},\ and\ \citenamefont {Balents}}]{Liu2018}%
  \BibitemOpen
  \bibfield  {author} {\bibinfo {author} {\bibfnamefont {C.}~\bibnamefont
  {Liu}}, \bibinfo {author} {\bibfnamefont {X.}~\bibnamefont {Chen}},\ and\
  \bibinfo {author} {\bibfnamefont {L.}~\bibnamefont {Balents}},\ }\bibfield
  {title} {\bibinfo {title} {Quantum entanglement of the sachdev-ye-kitaev
  models},\ }\href {https://doi.org/10.1103/PhysRevB.97.245126} {\bibfield
  {journal} {\bibinfo  {journal} {Phys. Rev. B}\ }\textbf {\bibinfo {volume}
  {97}},\ \bibinfo {pages} {245126} (\bibinfo {year} {2018})}\BibitemShut
  {NoStop}%
\bibitem [{\citenamefont {Huang}\ and\ \citenamefont
  {Gu}(2019)}]{PhysRevD.100.041901}%
  \BibitemOpen
  \bibfield  {author} {\bibinfo {author} {\bibfnamefont {Y.}~\bibnamefont
  {Huang}}\ and\ \bibinfo {author} {\bibfnamefont {Y.}~\bibnamefont {Gu}},\
  }\bibfield  {title} {\bibinfo {title} {Eigenstate entanglement in the
  sachdev-ye-kitaev model},\ }\href
  {https://doi.org/10.1103/PhysRevD.100.041901} {\bibfield  {journal} {\bibinfo
   {journal} {Phys. Rev. D}\ }\textbf {\bibinfo {volume} {100}},\ \bibinfo
  {pages} {041901} (\bibinfo {year} {2019})}\BibitemShut {NoStop}%
\bibitem [{\citenamefont {{Zhang}}(2020)}]{Zhang2020}%
  \BibitemOpen
  \bibfield  {author} {\bibinfo {author} {\bibfnamefont {P.}~\bibnamefont
  {{Zhang}}},\ }\bibfield  {title} {\bibinfo {title} {{Entanglement entropy and
  its quench dynamics for pure states of the Sachdev-Ye-Kitaev model}},\ }\href
  {https://doi.org/10.1007/JHEP06(2020)143} {\bibfield  {journal} {\bibinfo
  {journal} {Journal of High Energy Physics}\ }\textbf {\bibinfo {volume}
  {2020}},\ \bibinfo {eid} {143} (\bibinfo {year} {2020})},\ \Eprint
  {https://arxiv.org/abs/2004.05339} {arXiv:2004.05339 [hep-th]} \BibitemShut
  {NoStop}%
\bibitem [{\citenamefont {Zhang}\ \emph {et~al.}(2020)\citenamefont {Zhang},
  \citenamefont {Liu},\ and\ \citenamefont {Chen}}]{ZhangLiu2020}%
  \BibitemOpen
  \bibfield  {author} {\bibinfo {author} {\bibfnamefont {P.}~\bibnamefont
  {Zhang}}, \bibinfo {author} {\bibfnamefont {C.}~\bibnamefont {Liu}},\ and\
  \bibinfo {author} {\bibfnamefont {X.}~\bibnamefont {Chen}},\ }\bibfield
  {title} {\bibinfo {title} {{Subsystem Rényi entropy of thermal ensembles for
  SYK-like models}},\ }\href {https://doi.org/10.21468/SciPostPhys.8.6.094}
  {\bibfield  {journal} {\bibinfo  {journal} {SciPost Phys.}\ }\textbf
  {\bibinfo {volume} {8}},\ \bibinfo {pages} {094} (\bibinfo {year}
  {2020})}\BibitemShut {NoStop}%
\bibitem [{\citenamefont {Haldar}\ \emph {et~al.}(2020)\citenamefont {Haldar},
  \citenamefont {Bera},\ and\ \citenamefont {Banerjee}}]{Haldar2020}%
  \BibitemOpen
  \bibfield  {author} {\bibinfo {author} {\bibfnamefont {A.}~\bibnamefont
  {Haldar}}, \bibinfo {author} {\bibfnamefont {S.}~\bibnamefont {Bera}},\ and\
  \bibinfo {author} {\bibfnamefont {S.}~\bibnamefont {Banerjee}},\ }\bibfield
  {title} {\bibinfo {title} {R\'enyi entanglement entropy of fermi and
  non-fermi liquids: Sachdev-ye-kitaev model and dynamical mean field
  theories},\ }\href {https://doi.org/10.1103/PhysRevResearch.2.033505}
  {\bibfield  {journal} {\bibinfo  {journal} {Phys. Rev. Res.}\ }\textbf
  {\bibinfo {volume} {2}},\ \bibinfo {pages} {033505} (\bibinfo {year}
  {2020})}\BibitemShut {NoStop}%
\bibitem [{\citenamefont {Zhang}(2022)}]{Zhang:2022yaw}%
  \BibitemOpen
  \bibfield  {author} {\bibinfo {author} {\bibfnamefont {P.}~\bibnamefont
  {Zhang}},\ }\bibfield  {title} {\bibinfo {title} {{Quantum entanglement in
  the Sachdev\textemdash{}Ye\textemdash{}Kitaev model and its
  generalizations}},\ }\href {https://doi.org/10.1007/s11467-022-1162-5}
  {\bibfield  {journal} {\bibinfo  {journal} {Front. Phys. (Beijing)}\ }\textbf
  {\bibinfo {volume} {17}},\ \bibinfo {pages} {43201} (\bibinfo {year}
  {2022})},\ \Eprint {https://arxiv.org/abs/2203.01513} {arXiv:2203.01513
  [cond-mat.str-el]} \BibitemShut {NoStop}%
\bibitem [{\citenamefont {Page}(1993)}]{PAGE_PAPER}%
  \BibitemOpen
  \bibfield  {author} {\bibinfo {author} {\bibfnamefont {D.~N.}\ \bibnamefont
  {Page}},\ }\bibfield  {title} {\bibinfo {title} {Average entropy of a
  subsystem},\ }\href {https://doi.org/10.1103/PhysRevLett.71.1291} {\bibfield
  {journal} {\bibinfo  {journal} {Phys. Rev. Lett.}\ }\textbf {\bibinfo
  {volume} {71}},\ \bibinfo {pages} {1291} (\bibinfo {year}
  {1993})}\BibitemShut {NoStop}%
\bibitem [{\citenamefont {Garc\'{\i}a-Garc\'{\i}a}\ and\ \citenamefont
  {Verbaarschot}(2016{\natexlab{b}})}]{GarciaSPHEAT}%
  \BibitemOpen
  \bibfield  {author} {\bibinfo {author} {\bibfnamefont {A.~M.}\ \bibnamefont
  {Garc\'{\i}a-Garc\'{\i}a}}\ and\ \bibinfo {author} {\bibfnamefont {J.~J.~M.}\
  \bibnamefont {Verbaarschot}},\ }\bibfield  {title} {\bibinfo {title}
  {Spectral and thermodynamic properties of the sachdev-ye-kitaev model},\
  }\href {https://doi.org/10.1103/PhysRevD.94.126010} {\bibfield  {journal}
  {\bibinfo  {journal} {Phys. Rev. D}\ }\textbf {\bibinfo {volume} {94}},\
  \bibinfo {pages} {126010} (\bibinfo {year} {2016}{\natexlab{b}})}\BibitemShut
  {NoStop}%
\bibitem [{\citenamefont {Garc{\'\i}a-Garc{\'\i}a}\ \emph
  {et~al.}(2018)\citenamefont {Garc{\'\i}a-Garc{\'\i}a}, \citenamefont
  {Loureiro}, \citenamefont {Romero-Berm{\'u}dez},\ and\ \citenamefont
  {Tezuka}}]{Garcia-Garcia:2017bkg}%
  \BibitemOpen
  \bibfield  {author} {\bibinfo {author} {\bibfnamefont {A.~M.}\ \bibnamefont
  {Garc{\'\i}a-Garc{\'\i}a}}, \bibinfo {author} {\bibfnamefont
  {B.}~\bibnamefont {Loureiro}}, \bibinfo {author} {\bibfnamefont
  {A.}~\bibnamefont {Romero-Berm{\'u}dez}},\ and\ \bibinfo {author}
  {\bibfnamefont {M.}~\bibnamefont {Tezuka}},\ }\bibfield  {title} {\bibinfo
  {title} {{Chaotic-Integrable Transition in the Sachdev-Ye-Kitaev Model}},\
  }\href {https://doi.org/10.1103/PhysRevLett.120.241603} {\bibfield  {journal}
  {\bibinfo  {journal} {Phys. Rev. Lett.}\ }\textbf {\bibinfo {volume} {120}},\
  \bibinfo {pages} {241603} (\bibinfo {year} {2018})},\ \Eprint
  {https://arxiv.org/abs/1707.02197} {arXiv:1707.02197 [hep-th]} \BibitemShut
  {NoStop}%
\bibitem [{\citenamefont {Kim}\ and\ \citenamefont
  {Cao}(2021)}]{Kim2021comment}%
  \BibitemOpen
  \bibfield  {author} {\bibinfo {author} {\bibfnamefont {J.}~\bibnamefont
  {Kim}}\ and\ \bibinfo {author} {\bibfnamefont {X.}~\bibnamefont {Cao}},\
  }\bibfield  {title} {\bibinfo {title} {Comment on ``chaotic-integrable
  transition in the sachdev-ye-kitaev model''},\ }\href
  {https://doi.org/10.1103/PhysRevLett.126.109101} {\bibfield  {journal}
  {\bibinfo  {journal} {Phys. Rev. Lett.}\ }\textbf {\bibinfo {volume} {126}},\
  \bibinfo {pages} {109101} (\bibinfo {year} {2021})}\BibitemShut {NoStop}%
\bibitem [{\citenamefont {Garc\'{\i}a-Garc\'{\i}a}\ \emph
  {et~al.}(2021)\citenamefont {Garc\'{\i}a-Garc\'{\i}a}, \citenamefont
  {Loureiro}, \citenamefont {Romero-Berm\'udez},\ and\ \citenamefont
  {Tezuka}}]{Garcia2021}%
  \BibitemOpen
  \bibfield  {author} {\bibinfo {author} {\bibfnamefont {A.~M.}\ \bibnamefont
  {Garc\'{\i}a-Garc\'{\i}a}}, \bibinfo {author} {\bibfnamefont
  {B.}~\bibnamefont {Loureiro}}, \bibinfo {author} {\bibfnamefont
  {A.}~\bibnamefont {Romero-Berm\'udez}},\ and\ \bibinfo {author}
  {\bibfnamefont {M.}~\bibnamefont {Tezuka}},\ }\bibfield  {title} {\bibinfo
  {title} {Garc\'{\i}a-garc\'{\i}a et al. reply:},\ }\href
  {https://doi.org/10.1103/PhysRevLett.126.109102} {\bibfield  {journal}
  {\bibinfo  {journal} {Phys. Rev. Lett.}\ }\textbf {\bibinfo {volume} {126}},\
  \bibinfo {pages} {109102} (\bibinfo {year} {2021})}\BibitemShut {NoStop}%
\bibitem [{\citenamefont {Samui}\ and\ \citenamefont
  {Sorokhaibam}(2021)}]{Samui:2020jli}%
  \BibitemOpen
  \bibfield  {author} {\bibinfo {author} {\bibfnamefont {T.}~\bibnamefont
  {Samui}}\ and\ \bibinfo {author} {\bibfnamefont {N.}~\bibnamefont
  {Sorokhaibam}},\ }\bibfield  {title} {\bibinfo {title} {{Thermalization in
  different phases of charged SYK model}},\ }\href
  {https://doi.org/10.1007/JHEP04(2021)157} {\bibfield  {journal} {\bibinfo
  {journal} {JHEP}\ }\textbf {\bibinfo {volume} {04}},\ \bibinfo {pages}
  {157}},\ \Eprint {https://arxiv.org/abs/2004.14376} {arXiv:2004.14376
  [hep-th]} \BibitemShut {NoStop}%
\bibitem [{\citenamefont {Sorokhaibam}(2020)}]{Sorokhaibam:2019qho}%
  \BibitemOpen
  \bibfield  {author} {\bibinfo {author} {\bibfnamefont {N.}~\bibnamefont
  {Sorokhaibam}},\ }\bibfield  {title} {\bibinfo {title} {{Phase transition and
  chaos in charged SYK model}},\ }\href
  {https://doi.org/10.1007/JHEP07(2020)055} {\bibfield  {journal} {\bibinfo
  {journal} {JHEP}\ }\textbf {\bibinfo {volume} {07}},\ \bibinfo {pages}
  {055}},\ \Eprint {https://arxiv.org/abs/1912.04326} {arXiv:1912.04326
  [hep-th]} \BibitemShut {NoStop}%
\bibitem [{\citenamefont {Kim}\ \emph {et~al.}(2020)\citenamefont {Kim},
  \citenamefont {Cao},\ and\ \citenamefont {Altman}}]{Kim2020}%
  \BibitemOpen
  \bibfield  {author} {\bibinfo {author} {\bibfnamefont {J.}~\bibnamefont
  {Kim}}, \bibinfo {author} {\bibfnamefont {X.}~\bibnamefont {Cao}},\ and\
  \bibinfo {author} {\bibfnamefont {E.}~\bibnamefont {Altman}},\ }\bibfield
  {title} {\bibinfo {title} {Low-rank sachdev-ye-kitaev models},\ }\href
  {https://doi.org/10.1103/PhysRevB.101.125112} {\bibfield  {journal} {\bibinfo
   {journal} {Phys. Rev. B}\ }\textbf {\bibinfo {volume} {101}},\ \bibinfo
  {pages} {125112} (\bibinfo {year} {2020})}\BibitemShut {NoStop}%
\bibitem [{\citenamefont {Gao}\ \emph {et~al.}(2017)\citenamefont {Gao},
  \citenamefont {Jafferis},\ and\ \citenamefont {Wall}}]{Gao:2016bin}%
  \BibitemOpen
  \bibfield  {author} {\bibinfo {author} {\bibfnamefont {P.}~\bibnamefont
  {Gao}}, \bibinfo {author} {\bibfnamefont {D.~L.}\ \bibnamefont {Jafferis}},\
  and\ \bibinfo {author} {\bibfnamefont {A.~C.}\ \bibnamefont {Wall}},\
  }\bibfield  {title} {\bibinfo {title} {{Traversable Wormholes via a Double
  Trace Deformation}},\ }\href {https://doi.org/10.1007/JHEP12(2017)151}
  {\bibfield  {journal} {\bibinfo  {journal} {JHEP}\ }\textbf {\bibinfo
  {volume} {12}},\ \bibinfo {pages} {151}},\ \Eprint
  {https://arxiv.org/abs/1608.05687} {arXiv:1608.05687 [hep-th]} \BibitemShut
  {NoStop}%
\bibitem [{\citenamefont {Maldacena}\ \emph {et~al.}(2017)\citenamefont
  {Maldacena}, \citenamefont {Stanford},\ and\ \citenamefont
  {Yang}}]{YANG_MALDACENA}%
  \BibitemOpen
  \bibfield  {author} {\bibinfo {author} {\bibfnamefont {J.}~\bibnamefont
  {Maldacena}}, \bibinfo {author} {\bibfnamefont {D.}~\bibnamefont
  {Stanford}},\ and\ \bibinfo {author} {\bibfnamefont {Z.}~\bibnamefont
  {Yang}},\ }\bibfield  {title} {\bibinfo {title} {Diving into traversable
  wormholes},\ }\href {https://doi.org/https://doi.org/10.1002/prop.201700034}
  {\bibfield  {journal} {\bibinfo  {journal} {Fortschritte der Physik}\
  }\textbf {\bibinfo {volume} {65}},\ \bibinfo {pages} {1700034} (\bibinfo
  {year} {2017})},\ \Eprint
  {https://arxiv.org/abs/https://onlinelibrary.wiley.com/doi/pdf/10.1002/prop.201700034}
  {https://onlinelibrary.wiley.com/doi/pdf/10.1002/prop.201700034} \BibitemShut
  {NoStop}%
\bibitem [{\citenamefont {Brown}\ \emph {et~al.}(2023)\citenamefont {Brown},
  \citenamefont {Gharibyan}, \citenamefont {Leichenauer}, \citenamefont {Lin},
  \citenamefont {Nezami}, \citenamefont {Salton}, \citenamefont {Susskind},
  \citenamefont {Swingle},\ and\ \citenamefont {Walter}}]{PRXQuantum.4.010320}%
  \BibitemOpen
  \bibfield  {author} {\bibinfo {author} {\bibfnamefont {A.~R.}\ \bibnamefont
  {Brown}}, \bibinfo {author} {\bibfnamefont {H.}~\bibnamefont {Gharibyan}},
  \bibinfo {author} {\bibfnamefont {S.}~\bibnamefont {Leichenauer}}, \bibinfo
  {author} {\bibfnamefont {H.~W.}\ \bibnamefont {Lin}}, \bibinfo {author}
  {\bibfnamefont {S.}~\bibnamefont {Nezami}}, \bibinfo {author} {\bibfnamefont
  {G.}~\bibnamefont {Salton}}, \bibinfo {author} {\bibfnamefont
  {L.}~\bibnamefont {Susskind}}, \bibinfo {author} {\bibfnamefont
  {B.}~\bibnamefont {Swingle}},\ and\ \bibinfo {author} {\bibfnamefont
  {M.}~\bibnamefont {Walter}},\ }\bibfield  {title} {\bibinfo {title} {Quantum
  gravity in the lab. i. teleportation by size and traversable wormholes},\
  }\href {https://doi.org/10.1103/PRXQuantum.4.010320} {\bibfield  {journal}
  {\bibinfo  {journal} {PRX Quantum}\ }\textbf {\bibinfo {volume} {4}},\
  \bibinfo {pages} {010320} (\bibinfo {year} {2023})}\BibitemShut {NoStop}%
\bibitem [{\citenamefont {Nezami}\ \emph {et~al.}(2023)\citenamefont {Nezami},
  \citenamefont {Lin}, \citenamefont {Brown}, \citenamefont {Gharibyan},
  \citenamefont {Leichenauer}, \citenamefont {Salton}, \citenamefont
  {Susskind}, \citenamefont {Swingle},\ and\ \citenamefont
  {Walter}}]{PRXQuantum.4.010321}%
  \BibitemOpen
  \bibfield  {author} {\bibinfo {author} {\bibfnamefont {S.}~\bibnamefont
  {Nezami}}, \bibinfo {author} {\bibfnamefont {H.~W.}\ \bibnamefont {Lin}},
  \bibinfo {author} {\bibfnamefont {A.~R.}\ \bibnamefont {Brown}}, \bibinfo
  {author} {\bibfnamefont {H.}~\bibnamefont {Gharibyan}}, \bibinfo {author}
  {\bibfnamefont {S.}~\bibnamefont {Leichenauer}}, \bibinfo {author}
  {\bibfnamefont {G.}~\bibnamefont {Salton}}, \bibinfo {author} {\bibfnamefont
  {L.}~\bibnamefont {Susskind}}, \bibinfo {author} {\bibfnamefont
  {B.}~\bibnamefont {Swingle}},\ and\ \bibinfo {author} {\bibfnamefont
  {M.}~\bibnamefont {Walter}},\ }\bibfield  {title} {\bibinfo {title} {Quantum
  gravity in the lab. ii. teleportation by size and traversable wormholes},\
  }\href {https://doi.org/10.1103/PRXQuantum.4.010321} {\bibfield  {journal}
  {\bibinfo  {journal} {PRX Quantum}\ }\textbf {\bibinfo {volume} {4}},\
  \bibinfo {pages} {010321} (\bibinfo {year} {2023})}\BibitemShut {NoStop}%
\bibitem [{\citenamefont {{Gur-Ari}}\ \emph {et~al.}(2018)\citenamefont
  {{Gur-Ari}}, \citenamefont {{Mahajan}},\ and\ \citenamefont
  {{Vaezi}}}]{Gurari2018}%
  \BibitemOpen
  \bibfield  {author} {\bibinfo {author} {\bibfnamefont {G.}~\bibnamefont
  {{Gur-Ari}}}, \bibinfo {author} {\bibfnamefont {R.}~\bibnamefont
  {{Mahajan}}},\ and\ \bibinfo {author} {\bibfnamefont {A.}~\bibnamefont
  {{Vaezi}}},\ }\bibfield  {title} {\bibinfo {title} {{Does the SYK model have
  a spin glass phase?}},\ }\href {https://doi.org/10.1007/JHEP11(2018)070}
  {\bibfield  {journal} {\bibinfo  {journal} {Journal of High Energy Physics}\
  }\textbf {\bibinfo {volume} {2018}},\ \bibinfo {eid} {70} (\bibinfo {year}
  {2018})},\ \Eprint {https://arxiv.org/abs/1806.10145} {arXiv:1806.10145
  [hep-th]} \BibitemShut {NoStop}%
\bibitem [{\citenamefont {{Wang}}\ \emph {et~al.}(2019)\citenamefont {{Wang}},
  \citenamefont {{Bagrets}}, \citenamefont {{Chudnovskiy}},\ and\ \citenamefont
  {{Kamenev}}}]{Wang2019}%
  \BibitemOpen
  \bibfield  {author} {\bibinfo {author} {\bibfnamefont {H.}~\bibnamefont
  {{Wang}}}, \bibinfo {author} {\bibfnamefont {D.}~\bibnamefont {{Bagrets}}},
  \bibinfo {author} {\bibfnamefont {A.~L.}\ \bibnamefont {{Chudnovskiy}}},\
  and\ \bibinfo {author} {\bibfnamefont {A.}~\bibnamefont {{Kamenev}}},\
  }\bibfield  {title} {\bibinfo {title} {{On the replica structure of
  Sachdev-Ye-Kitaev model}},\ }\href {https://doi.org/10.1007/JHEP09(2019)057}
  {\bibfield  {journal} {\bibinfo  {journal} {Journal of High Energy Physics}\
  }\textbf {\bibinfo {volume} {2019}},\ \bibinfo {eid} {57} (\bibinfo {year}
  {2019})},\ \Eprint {https://arxiv.org/abs/1812.02666} {arXiv:1812.02666
  [hep-th]} \BibitemShut {NoStop}%
\end{thebibliography}%

\appendix

\section{Symmetries in Different Charge Sectors}
\label{SYMM_LEVEL}
 Now we discuss symmetries of different charge sectors for complex SYK model and explain the classification shown in eq. \ref{RMTCLASS}. Firstly particle-hole symmetry operator $S=\mathcal{U}\mathcal{K}$ can be written in terms of $\mathcal{U}$, an unitary operator and $\mathcal{K}$ the charge conjugation operator. In spin operator representation $c_{i}$ is real, so $\mathcal{K}$ has no effect, and we get the following relation, which we can use to express the unitary operator $\mathcal{U}$ in terms of spin-operators.
\bef
\displaystyle 
~~~~~~~~~c^{\dagger}_{i}=\mathcal{S} c_{i}\mathcal{S}^{-1}=\mathcal{U} c_{i}\mathcal{U}^{\dagger}
\\
\displaystyle
\mathcal{U}= \begin{cases}\sigma_1^x \sigma_2^y \cdots \sigma_{N-1}^x \sigma_N^y & \text { for even } N \\ \sigma_1^x \sigma_2^y \cdots \sigma_{N-1}^y \sigma_N^x & \text { for odd } N\end{cases}
\eef
By counting the number of $\sigma_{y}$ present in $\mathcal{U}$ we can determine the $\mathcal{S}^{2}=\mathcal{U} \mathcal{U}^*$ value from
\bef
\begin{tabular}{ |c|c|c|c|c| } 
 \hline
N mod 4 & ~~~~0 & ~~~~1 &~~~~ 2 &~~~~ 3  \\ 
 \hline
$\mathcal{S}^2$ &~~~~1&~~~~ 1 &~~~~ -1 &~~~~ -1  \\
 \hline
\end{tabular}
\label{PHS_OP}
\eef
Now for $\mathcal{Q}=0$ sector $\mathcal{S}$ maps it to itself
\bef
\displaystyle
\mathcal{S} H_{\mathcal{Q}=0} \mathcal{S}^{-1}=\mathcal{U} H_{\mathcal{Q}=0}^* \mathcal{U}^{-1}=H_{\mathcal{Q}=0} .
\label{SYM_EQ_1}
\eef
When $N \bmod 4=0$ we have $\mathcal{U} \mathcal{U}^*=+I$, and so one can choose a basis of the many-body Fock space in which $\mathcal{U}=I$; and from eq. \ref{SYM_EQ_1}; $H_{\mathcal{Q}=0} \in \mathbb{R}$ is a real symmetric matrix, which should exhibit GOE level statistics. \newline
For $N \bmod 4=2$ we have $\mathcal{U} \mathcal{U}^*=-I$, and so one can choose a basis of the many-body Fock space in which $\mathcal{U}=\begin{pmatrix}
0 & +I \\
-I & 0 
\end{pmatrix}$, and $H_{\mathcal{Q}=0} \in \mathbb{H}$ is a quaternion Hermitian matrix, which should exhibit GSE level statistics. 
\newline
Next, for non-zero charge sectors, $\mathcal{Q}\neq 0$ 
\bef
\displaystyle
\mathcal{S} H_{\mathcal{Q}} \mathcal{S}^{-1}=\mathcal{U} H_{\mathcal{Q}}^* \mathcal{U}^{-1}=H_{-\mathcal{Q}} .
\label{SYM_EQ_2}
\eef
which gives $H_{\mathcal{Q}}^{*}=H_{-\mathcal{Q}}$, with no extra condition. Then the only constraint on $H_{\mathcal{Q}\neq 0}$ is being Hermitian. So its elements can be complex numbers, and it should follow GUE statistics. Based on this we get the classification based on $N$ and charge sector value.

\begin{figure}[h]
  \setlength{\unitlength}{1pt}
  \begin{picture}(\linewidth, 340)
    \put(0,0){\includegraphics[width=0.9\columnwidth]{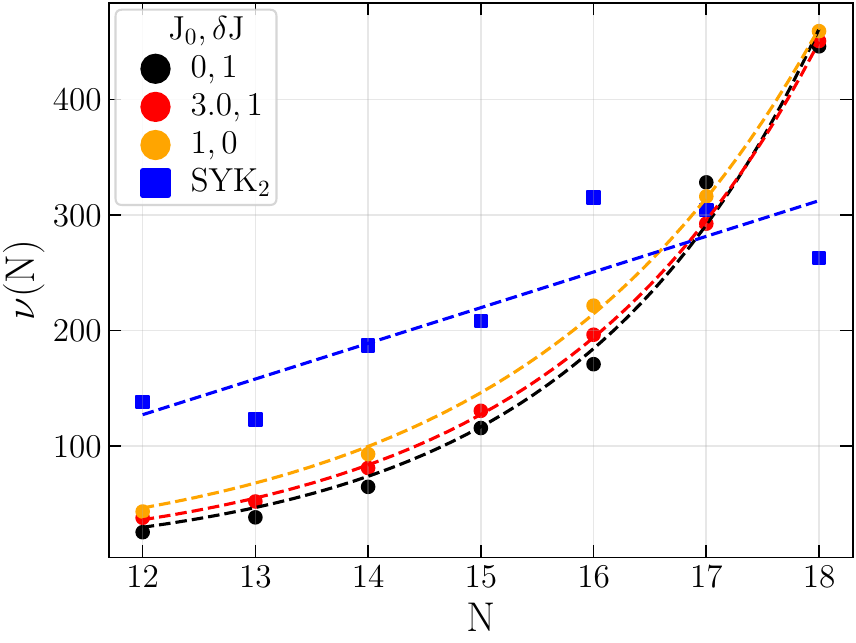}}
    \put(0,155){\normalsize{\textcolor{black}{$(b)$}}}
    \put(0,170){\includegraphics[width=0.9\columnwidth]{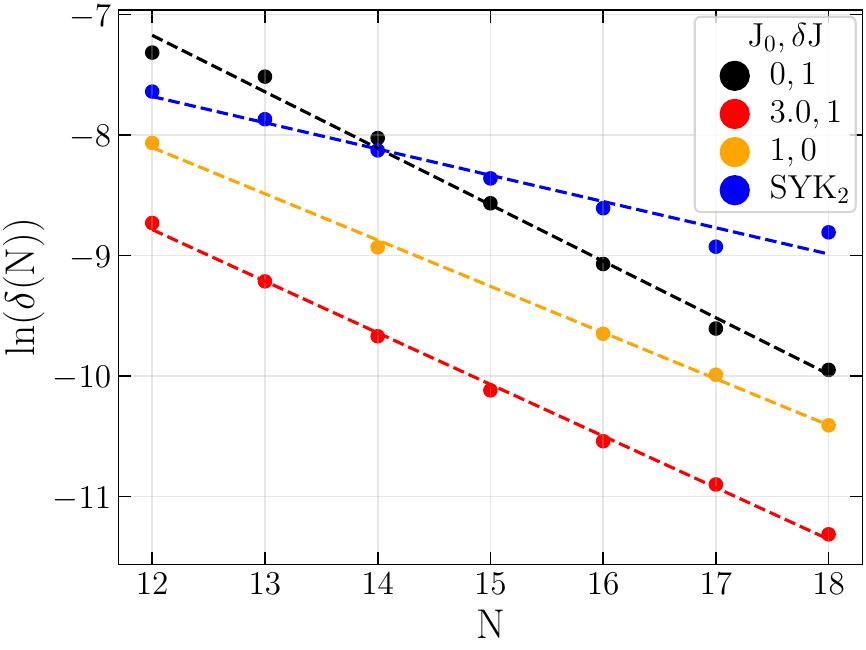}}
    \put(0,325){\normalsize{\textcolor{black}{$(a)$}}}

\end{picture}
\vspace*{-0.15in}
\caption{(a) Disorder-averaged mean level spacing $\delta(N)$ from 3\% of the eigen spectrum near the ground state as a function of system size $N$. The dashed lines are fits to the data, detailed in table. \ref{tab:S1_LevelSpacing}, for different values of $(J_0,\delta J)$. The blue dots and curve refer to the  $\mathrm{SYK}_2$ model,  after dividing $\delta(N)$ by a factor of $30$. 
(b) The total number of states $\nu(N)$  within 5\% of the many-body bandwidth above the ground state. The dashed lines are exponential fits. The different colors refer to different values of $(J_0,\delta J)$. The blue dots and curve refer to the $\mathrm{SYK}_2$ model, after rescaling by a factor of $30$.
}
\label{SP_heat_2}
\end{figure}
\section{Residual entropy Analysis}\label{app:ENTRP_EXTRA}
Here, we explore an alternative method for obtaining the residual entropy from the average level spacing and the number of states near the ground state. In Section \ref{RES_ENT}, the disorder-averaged mean level spacing was calculated using an energy window (eq. \ref{Ewdef}) derived from the $T_{l}$ value (see the paragraph following eq. \ref{ENTRPY_DEF}) used in the $T\to 0$ thermodynamic entropy extrapolation. As noted at the end of that section, this approach raises potential concerns regarding the sensitivity of the analysis to the choice of $T_{l}$ and its possible dependence on $N$. To address this, we perform a similar analysis here but select an $E_{w}$ that is entirely independent of $T_l$. Specifically, we define the energy window as the lowest $3\%$ of the total many-body energy bandwidth above the ground state.
\begin{equation}\label{defnew ew}
E_w=E_{min}+ {3\over 100} (E_{max}-E_{min})
\end{equation}
We obtain the disorder averaged mean level spacing $\delta(N)$ for states within this window $E_{w}$ for different values of  $J_0, \delta J$. The results for $\ln{\delta}$ vs. $N$ are shown in fig.\ref{SP_heat_2} (a). We also show the corresponding result for the non-interacting $\mathrm{SYK}_2$ model eq. \ref{SYK2_def}. For $\mathrm{SYK}_2$ case we have divided $\delta(N)$ by $30$ to plot it in the same scale. For all the interacting cases, the decrease of level spacing with $N$ is consistent with an exponential $\sim \exp[-N s_0]$ throughout the range of $N$ studied, suggesting a non-zero residual entropy upon extrapolation to $N\to \infty$ limit. Obtained values of $s_0$ in each cases is mentioned in table \ref{tab:S1_LevelSpacing}, and matches with the results obtained in $T_l$ dependent analysis in table \ref{tab:S0_LevelSpacing}.

\begin{table}[h]
\centering
\begin{tabular}{|c|c|c|c|}
       \hline
        $J_{0}$   &  $\delta J$& $s_0$ from $\delta(N)$ & $s_0$ from $\nu(N)$ \\
        \hline
        0   & 1 & $0.468\pm 0.019$ & $0.46\pm0.032$ \\
        \hline
        3   & 1& $0.428\pm0.008$ & $0.42\pm0.004$ \\
        \hline
        1   & 0& $0.383\pm0.01$ & $0.384\pm0.01$\\
        \hline
\end{tabular}
\caption{The residual entropy density $s_0=S_0/N$ extracted from $\delta(N)$ and $\nu(N)$ in fig.\ref{SP_heat_1}.}
\label{tab:S1_LevelSpacing}
\end{table}

Next we analyze the exponentially dense spectrum near the ground state. We obtain the total number of eigenstates $\nu(N)$ over an energy window $E_{w}$ ($5\%$ of the many-body band width) above the ground state. Fig.\ref{SP_heat_0}(b) is for $\nu(N)$ for different values of $J_{0},\delta J$. Since the number of states within an energy window is inversely proportional to the level spacing, $\nu(N)$ is expected increases as $\sim \exp{Ns_0}$ with $N$. This is indeed the case for all the values of $J_0,\delta J$ in fig.\ref{SP_heat_0}(b). In contrast, $\nu(N)$ increases linearly with $N$ for $\mathrm{SYK}_2$ model. For $\mathrm{SYK}_2$ case we have scaled $\nu(N)$ by a factor of $30$. The coefficient extracted from the exponential growth of $\nu$ with $N$ is mentioned in table \ref{tab:S1_LevelSpacing}, and matches well with all previous analysis.

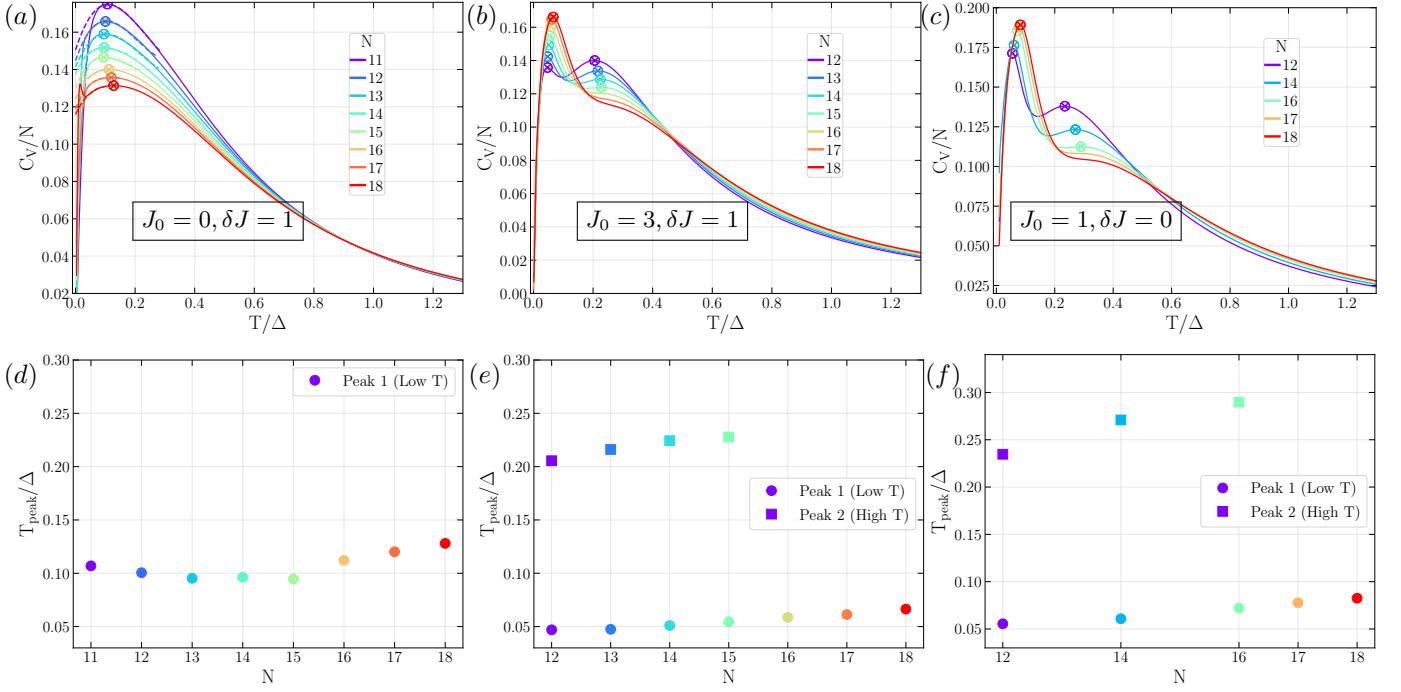
\begin{figure*}
\centering

\setlength{\unitlength}{\linewidth}
\begin{picture}(1, 0.5)
    \put(-0.005, 0.27){\includegraphics[width=0.33\linewidth]{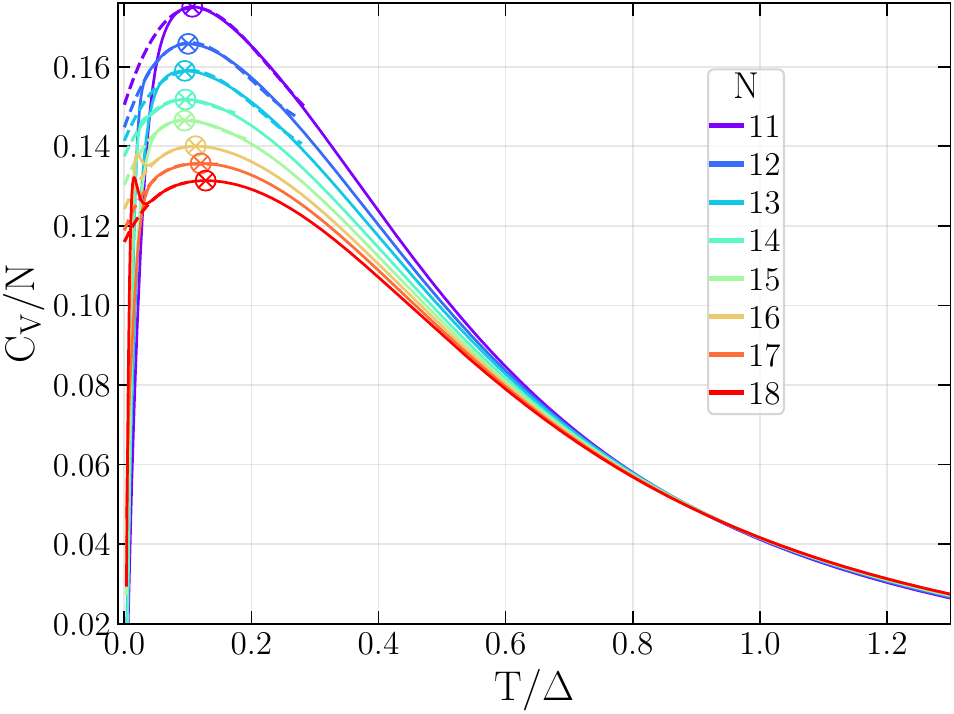}}
    \put(-0.015, 0.5){\normalsize{\textcolor{black}{$(a)$}}}
    \put(0.08, 0.35){\fbox{$J_{0}=0,\delta J=1$}}
    \put(0.333, 0.27){\includegraphics[width=0.33\linewidth]{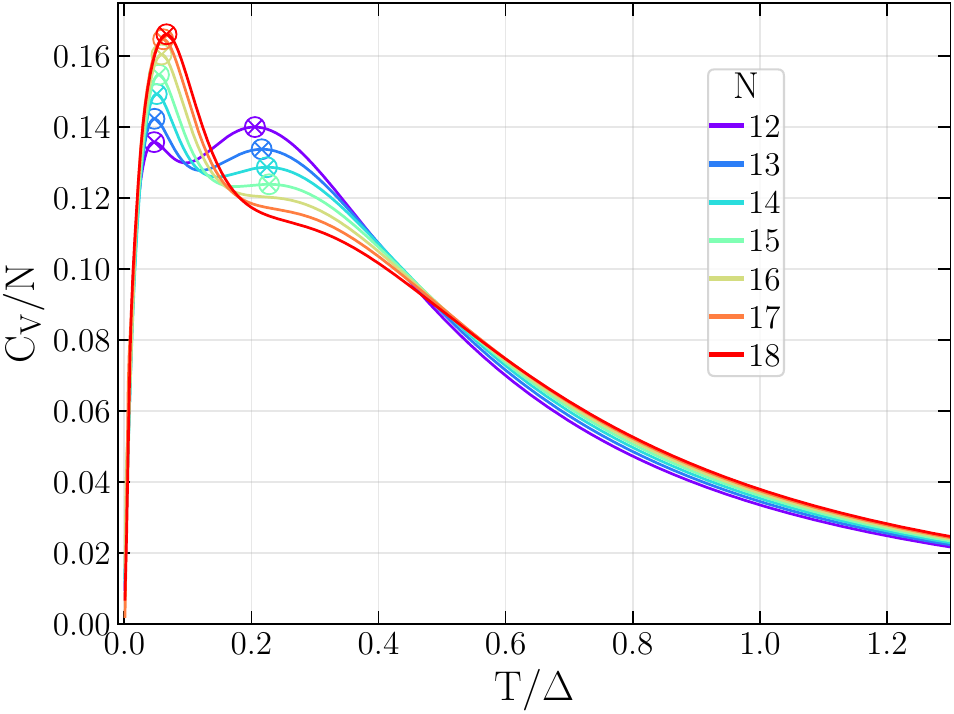}}
    \put(0.326, 0.5){\normalsize{\textcolor{black}{$(b)$}}}
    \put(0.408, 0.35){\fbox{$J_{0}=3,\delta J=1$}}
    \put(0.669, 0.27){\includegraphics[width=0.33\linewidth]{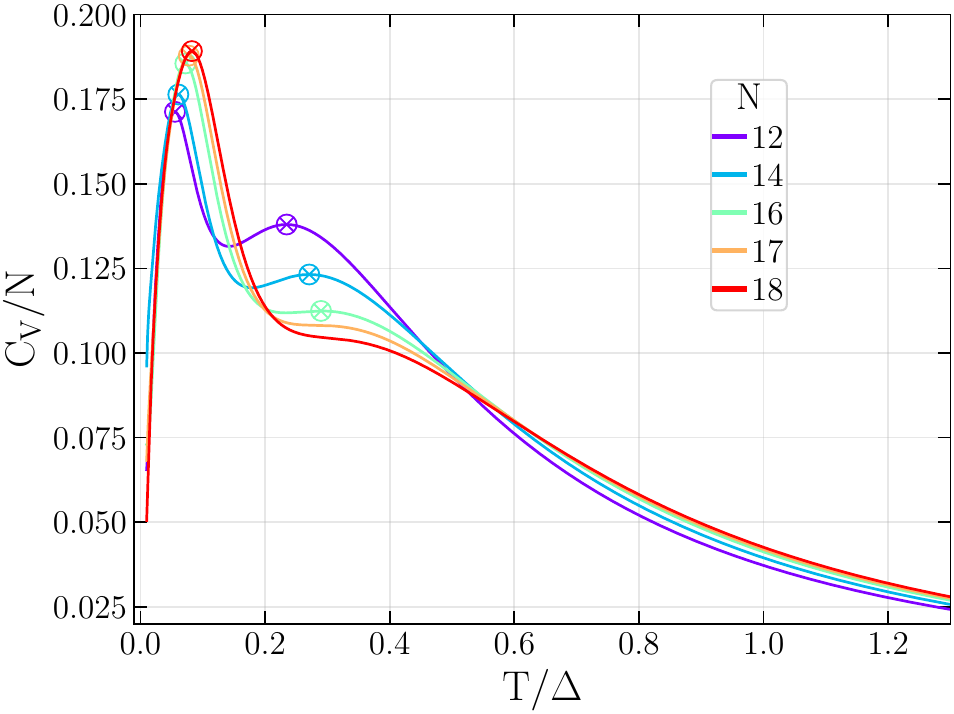}}
    \put(0.661, 0.5){\normalsize{\textcolor{black}{$(c)$}}}
    \put(0.728, 0.35){\fbox{$J_{0}=1,\delta J=0$}}
    \put(-0.005, 0.01){\includegraphics[width=0.33\linewidth]{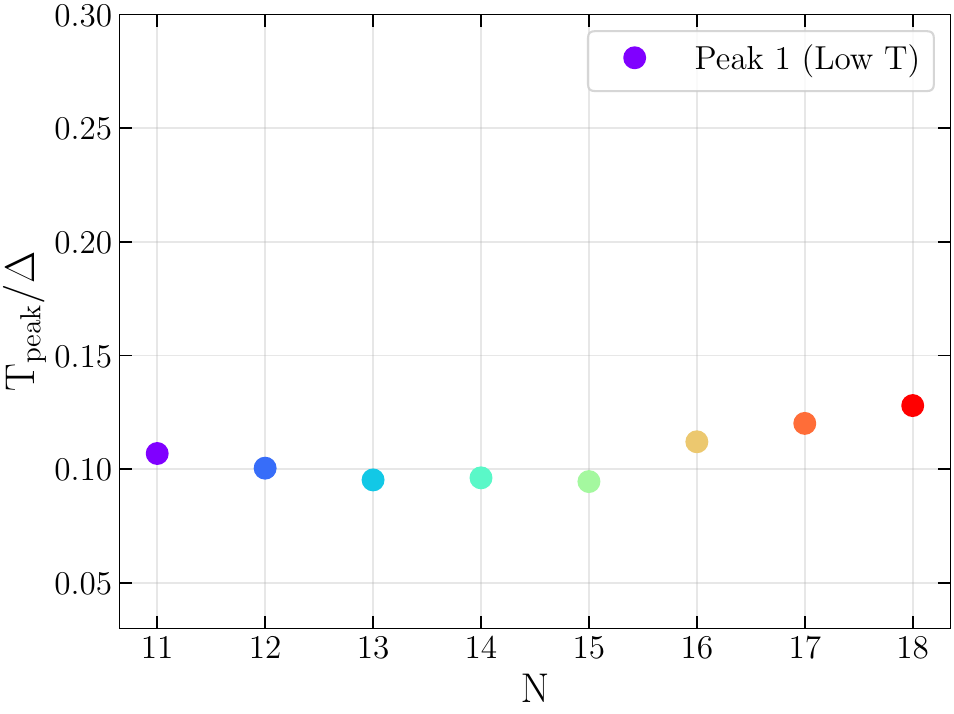}}
    \put(-0.015, 0.235){\normalsize{\textcolor{black}{$(d)$}}}
    \put(0.335, 0.01){\includegraphics[width=0.33\linewidth]{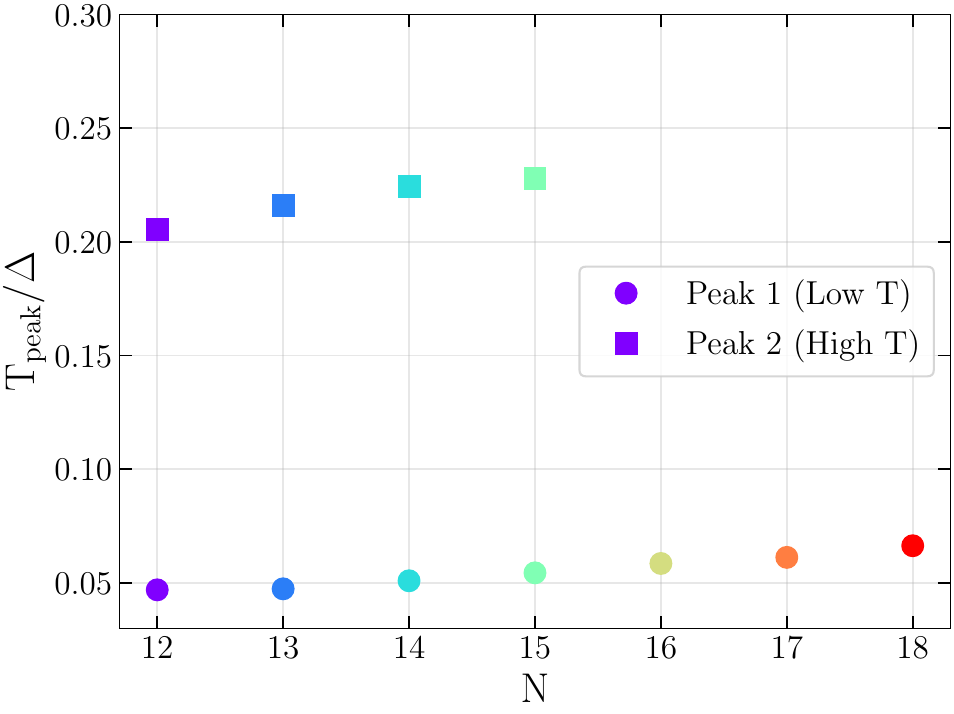}}
    \put(0.328, 0.235){\normalsize{\textcolor{black}{$(e)$}}}
    \put(0.668, 0.01){\includegraphics[width=0.33\linewidth]{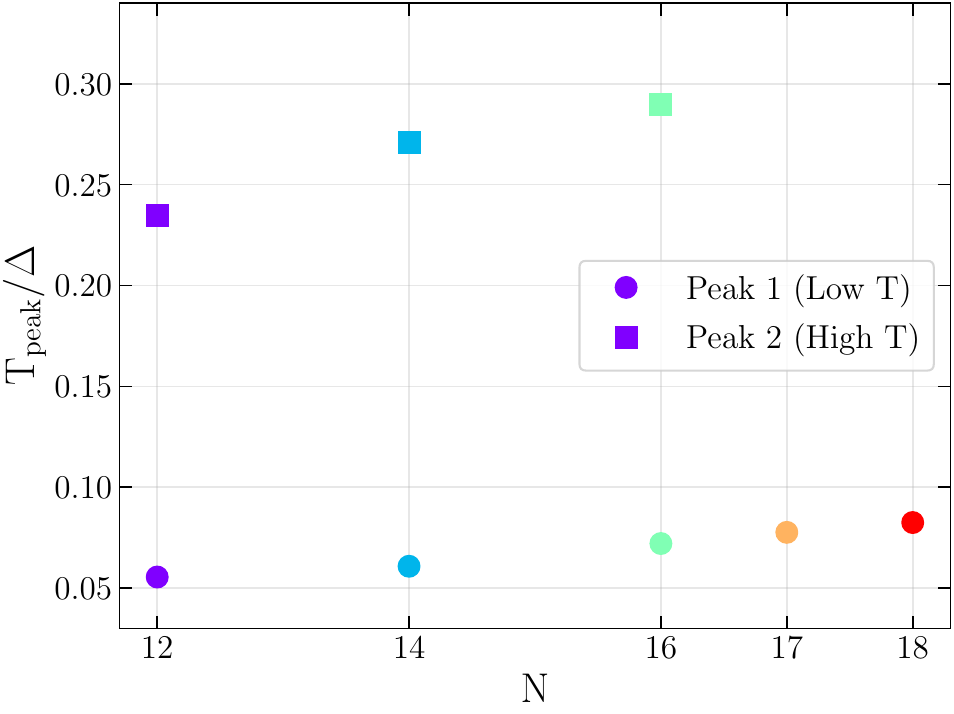}}
    \put(0.665, 0.235){\normalsize{\textcolor{black}{$(f)$}}}
\end{picture}
\caption{Specific heat (per site) for the zero mean ($J_0=0, \delta J=1$) SYK model for different N. Corresponding peak for the specific heat is plotted in (d) for each $N$. (b)  Specific heat for $J_0=3, \delta J=1$ for different N, with peaks for each cases plotted in (e). (c) Specific heat  for $J_{0}=1, \delta J=0$ for different $N$ and peak position in each cases are plotted in (f). 
}
\label{SP_heat_1}
\end{figure*}

\section{Specific Heat} \label{app:SpHeat}
In this appendix, we show that the temperature dependence of low-temperature specific heat $C_v(T)$, obtained from the ED of the finite-mean SYK model, is consistent a linear-$T$ behavior like the standard SYK NFL \cite{Maldacena:2016hyu,Gu:2019jub,GarciaSPHEAT}. 
The disorder-averaged specific heat can be either obtained from the temperature derivative of the disorder averaged internal energy or energy fluctuation, i.e.,
\bef
\displaystyle
\frac{C_{v}}{N}=\frac{1}{N}\frac{d\langle E\rangle}{dT}=\frac{\langle \Delta E^{2}\rangle}{N T^{2}},
\label{CV_EQN}
\eef
where $\langle E\rangle =-d\ln{Z}(\beta)/d\beta$ is the internal energy and $Z(\beta)$ is the partition function.

In the SYK model ($J_0=0$), the partition function can be obtained as $Z(\beta)\sim (1/\beta^q) \exp{[-\beta E_0+S_0+(\gamma/2\beta)]}$ at low temperature, where $E_0$ ($\propto N$) is the ground-state energy, $S_0$ the residual entropy and $\gamma$ ($\propto N$) is the linear-$T$ specific heat coefficient \cite{Maldacena:2016hyu,Gu:2019jub,GarciaSPHEAT,RevModPhys.94.035004}. The pre-exponential $T$-dependent factor with a universal $q=3$ can be obtained from one-loop quantum corrections or from a low-energy soft-mode action. The latter involves emergent Schwarzian and $U(1)$ gauge fields soft modes at low energies around the scaling solution of the large-$N$ saddle point for the complex SYK model. The soft-mode action also leads to the linear-$T$ specific heat for the SYK NFL, i.e., from $Z(\beta)$ one obtains
\begin{align}
\frac{C_v}{N}&=\frac{q}{N}+\frac{\gamma}{N}T+\mathcal{O}(T^2). \label{eq:LowT_Cv}
\end{align}
Here we ask whether such linear-$T$ specific heat persists for the SYK model even with finite mean $J_0\neq 0$.

In out ED computations, we obtain the specific heat from the energy fluctuations, as in eq.\eqref{CV_EQN}. However, due to large sample-to-sample fluctuations at finite $N$, following Ref.\onlinecite{GarciaSPHEAT}, we use
\begin{subequations}
\begin{align}
\frac{C_v}{N}&=\frac{1}{N\overline{Z}} \sum_{m,p} \frac{\left(E_{m,p}-\langle E \rangle_p\right)^2}{T^2}e^{-\beta E_{m,p}}, \\
\overline{Z}&=\sum_{m,p} e^{-\beta E_{m,p}}.
\end{align}
\end{subequations}
Here $E_{m,p}$ is the $m$-th energy eigenvalue for $p$-th disorder realization and $\langle E\rangle_p=(\sum_{m} E_{m,p}e^{-\beta E_{m,p}})/(\sum_m e^{-\beta E_{m,p}})$ is the internal energy for the $p$-th realization. We have used $300$ realizations. The above is different from a \emph{quenched averaging}, where the expression of eq.\eqref{CV_EQN}
is averaged over disorder realization. Instead an \emph{annealed} partition function $\overline{Z}$ is used here. This procedure reduces the strong finite-size sample-to-sample fluctuations and allows us to compute smooth $C_v(T)$ curves. Due to self-averaging property \cite{PhysRevB.94.035135,RevModPhys.94.035004} of the SYK model and the absence of replica-symmetry breaking spin-glass order \cite{Gurari2018,Wang2019}, the quenched and annealed averaging are expected to give same result in the thermodynamic limit. 

We show the specific heat $C_v(T)$, computed from ED at low temperatures, in fig.\ref{SP_heat_1}(a), for $J_0=0, \delta J=1$. The specific heat coefficient $\gamma$ cannot be calculated directly from the ED data for $T\to 0$, since $C_V\sim e^{-\delta(N)/T}$ for $T\to 0$ at finite $N$ due to finite level spacing $\delta(N)$ above the ground state. Hence, following Ref.\cite{GarciaSPHEAT}, we perform a cubic extrapolation of the low-temperature data at given $N$ for $0.04\leq T\leq 0.1$ to obtain $q(N)$ and $\gamma(N)/N$ [eq.\eqref{eq:LowT_Cv}], as shown in fig. \ref{SP_heat_1}. Further, we extrapolate $q(N)$ and $\gamma(N)$ to $1/N\to 0$ to get a matching estimate of $q\simeq 2.7$ and $\gamma/N\simeq 0.4$ for the zero-mean SYK model ($J_0=0, \delta J=1$). 

Applying this procedure to the finite mean case does not work due to the absence of an observable linear behavior of the specific heat in the low-temperature limit. In figs.~\ref{SP_heat_1} (b) and (c), we can observe a pronounced peak near $T\sim 0.05$, along with a secondary peak near $T\sim 0.2$. The second peak vanishes with increasing $N$, which suggests that it may be a finite-size effect. We track the numerical values of the first peak in fig.\ref{SP_heat_1} (d), (e), (f). If we treat these temperature values as corresponding energy scales and try to find these exact positions in the density of states plots in fig.~\ref{DOS_PLOT_0}, we make the following observations:
\begin{itemize}[noitemsep]
\item For $J_{0}=0, \delta J=1$, the peak from the specific heat corresponds to the left shoulder in the density of states (DOS) in fig.~\ref{DOS_PLOT_0} (a) near $\mathcal{E}/\Delta\sim -0.05$. At this point, the DOS changes its nature from convex to concave as well.
\item For $J_{0}=3.0, \delta J=1$, the specific heat peak occurs around $T/\Delta\sim 0.05$, and tracking that in the density of states plot in fig.~\ref{DOS_PLOT_0} (b), it appears to be near the left shoulder at around $\mathcal{E}/\Delta\sim -0.07$. Exactly near this position, the DOS changes its derivative and shows a huge increase.
\item For $J_{0}=1, \delta J=0$, the specific heat peak occurs around $T/\Delta\sim 0.08$, and tracking that in the density of states plot in fig.~\ref{DOS_PLOT_0} (c), it appears to be near the left shoulder at around $\mathcal{E}/\Delta\sim -0.1$. Exactly near this position, the DOS shows a clear dip followed by a huge increase.
\end{itemize}
From this analysis, it appears that the peak features in the specific heat come from specific features of the density of states in the low-energy region. Due to this effect in low temperature limit, the linear nature of the specific heat remains masked.

\section{Finite N Simulations}\label{app:FINN}
For standard SYK model, creation and annihilation operators, $c^{\dagger}$ and $c$, can be represented in terms of Pauli matrices as follows:
\bef
\displaystyle
c^{\dagger}_{i}=\sigma_{1}^{z}\otimes \sigma_{2}^{z} .......\sigma_{i-1}^{z}\otimes \sigma^{+}_{i}\otimes \mathrm{I}...\\
\displaystyle
c_{i}=\sigma_{1}^{z}\otimes \sigma_{2}^{z} .......\sigma_{i-1}^{z}\otimes \sigma^{-}_{i}\otimes \mathrm{I}... 
\eef
For a system with $N$ flavor of fermions this generates a Hilbert space of size $2^{N}$. Because of the all-to-all interactions, the Hamiltonian matrix is relatively dense. This exponential growth in basis size typically restricts simulations of the standard SYK model to a maximum of $N=16$.

In the complex SYK model, charge operator defined in eq. \ref{charge_op} commutes with the Hamiltonian, $[{\mathcal Q},H]=0$. This conservation law allows us to block-diagonalize the Hilbert space into distinct $\mathcal{Q}$ sectors. Furthermore, the particle-hole symmetry defined in eq. \ref{SYM_CASE} relates the positive and negative charge sectors, ensuring they possess identical eigenvalues. Exploiting these symmetries, we perform our simulations within the Fock space for each individual $\mathcal{Q}$ sector, the largest of which has a dimension of $\displaystyle \binom N {N/2}$. While this reduction enables simulations up to $N=26$ for the smaller charge sectors, the necessity of obtaining the full energy spectrum across multiple disorder realizations limits our maximum system size to $N=18$. Notably, achieving full-spectrum exact diagonalization at $N=18$ represents the current computational cutting edge for the complex SYK model.
To ensure robust statistics across these system sizes, we consider at least 1000 disorder realizations for $N=12$, 300 for $N=14$, 100 for $N=16$, and 5 for $N=18$. Simulations for odd values of $N$ employ a comparable order of disorder realizations to their even counterparts.

To benchmark our zero-mean model computations, we verified several key observables against established literature. We extracted the residual entropy, $S_{0}/N$, utilizing two distinct methods. First, we analyzed the number of states and the level spacing near the ground state to capture the exponential growth of the density of states. Second, we evaluated the thermodynamic entropy for each finite $N$ and performed a $1/N$ scaling extrapolation. Both approaches consistently yield $S_{0}/N \approx 0.46 \sim 0.47$, which perfectly matches the large-$N$ analytical derivations \cite{PhysRevB.94.035135,Gu:2019jub}. Moreover, our level spacing statistics for the zero-mean case align precisely with previous findings \cite{you2017sachdev}.

For computations of the spectral form factor (SFF), the plateau region is inherently highly noisy. To mitigate this and significantly reduce the noise, we prioritize a larger ensemble average of approximately 3000 realizations at $N=12$ rather than simulating larger $N$ systems. Using this approach, our SFF plateau value for the zero-mean case at $\beta=0$ matches the theoretical predictions of \cite{Cotler:2016fpe}.

Evaluating out-of-time-order correlators (OTOCs) requires the time evolution of large, dense matrices. To accelerate this process and operate in a matrix-free Krylov methods, we utilize the \texttt{Dynamite} package in conjunction with \texttt{PETSc} and \texttt{SLEPc}, following the methodology described in ~\cite{PhysRevLett.126.030602}. Although the results in ~\cite{PhysRevLett.126.030602} encompass much larger values of $N$ and permit thermodynamic extrapolation ($N \rightarrow \infty$), we fix our system size at $N=12$ and perform averaging over both the flavor index and a smaller set of disorder realizations (around 5). This combined averaging significantly reduces statistical errors, allowing us to effectively benchmark our zero-mean OTOC results (confirming the Lyapunov exponent $\lambda=2\pi/\beta$) with considerably lower computational resources.



\end{document}